\documentclass[11pt,aps,prd,amsmath,amssymb,superscriptaddress,floatfix,nofootinbib,notitlepage]{revtex4-1}
\usepackage{amsmath,amssymb}
\usepackage{epsfig}
\usepackage{color}
\usepackage{bm}
\usepackage{epstopdf}

\usepackage[left=2.5cm,right=2.5cm,top=2cm,bottom=2cm]{geometry}
\usepackage[colorlinks=true,hyperfootnotes=false]{hyperref}
\newcommand{\X}{X(3872)}
\newcommand{\Tr}{\text{Tr}}
\makeatletter
\def\slashchar#1{{\mathpalette\c@ncel{#1}}} % TeXbook, bottom of p360
\makeatother
\def\vsl{\slashchar{v}}

\begin{document}
\title{\boldmath Decay widths of the spin-2 partners of the $X(3872)$}
\author{Miguel Albaladejo}
\email{Miguel.Albaladejo@ific.uv.es}
\affiliation{Instituto de F\'isica Corpuscular (IFIC),
             Centro Mixto CSIC-Universidad de Valencia,
             Institutos de Investigaci\'on de Paterna,
             Aptd. 22085, E-46071 Valencia, Spain}
\author{Feng-Kun Guo}
\email{fkguo@hiskp.uni-bonn.de}
\affiliation{Helmholtz-Institut f\"ur Strahlen- und
             Kernphysik and Bethe Center for Theoretical Physics, \\
             Universit\"at Bonn,  D-53115 Bonn, Germany}
\author{Carlos Hidalgo-Duque}
\email{carloshd@ific.uv.es}
\affiliation{Instituto de F\'isica Corpuscular (IFIC),
             Centro Mixto CSIC-Universidad de Valencia,
             Institutos de Investigaci\'on de Paterna,
             Aptd. 22085, E-46071 Valencia, Spain}
\author{Juan Nieves}
\email{jmnieves@ific.uv.es}
\affiliation{Instituto de F\'isica Corpuscular (IFIC),
             Centro Mixto CSIC-Universidad de Valencia,
             Institutos de Investigaci\'on de Paterna,
             Aptd. 22085, E-46071 Valencia, Spain}
\author{Manuel Pav\'on Valderrama}
\email{pavonvalderrama@ipno.in2p3.fr}
\affiliation{Institut de Physique Nucl\'eaire,
             Universit\'e Paris-Sud, IN2P3/CNRS,
             F-91406 Orsay Cedex, France}

\begin{abstract}

We consider the $X(3872)$ resonance as a $J^{PC}=1^{++}$ $D\bar D^*$
hadronic molecule. According to heavy quark spin symmetry, there will
exist a partner with  quantum numbers $2^{++}$, $X_{2}$, which would be
a $D^*\bar D^*$ loosely bound state. The $X_{2}$ is expected to decay
dominantly into $D\bar D$, $D\bar D^*$ and $\bar D D^*$ in
$d$-wave. In this work, we calculate the decay widths of the $X_{2}$
resonance into the above channels, as well as those of its bottom
partner, $X_{b2}$, the mass of which comes from assuming heavy flavor symmetry for the contact terms. We find partial widths of the $X_{2}$ and $X_{b2}$ of
the order of a few MeV. Finally, we also study the radiative
$X_2\to D\bar D^{*}\gamma$ and $X_{b2} \to \bar B B^{*}\gamma$
decays. These decay modes are more sensitive to the long-distance
structure of the resonances and to the $D\bar D^{*}$ or $B\bar
B^{*}$ final state interaction.

\end{abstract}

\maketitle

\newpage

\section{Introduction}

In the infinite quark mass limit, heavy quark spin-flavor symmetry (HQSFS)
implies that the dynamics involving heavy quarks are independent of their spin
or flavor. In this way, charm and bottom spectra can be related, up to
corrections suppressed by $1/m_Q$ with $m_Q$ the heavy quark mass. It
should also be possible, in a given heavy flavor sector, to relate states with different
spins. These relations are very useful in the study of composites that mix heavy
and light quarks. In this work, we focus on hadronic molecular states composed
by a heavy-light meson and a heavy-light antimeson ($P^{(*)}\bar P^{(*)},\,
P=D,\bar B$). These molecular states were predicted in the mid
70s~\cite{Voloshin:1976ap,De Rujula:1976qd}. So far, the best experimental
candidate to fit this molecular description is the $X(3872)$ resonance, first
observed by the Belle Collaboration in 2003 \cite{Choi:2003ue}, that can be
thought as a $D\bar{D}^{*}$ bound state with $J^{PC} = 1^{++}$ (quantum numbers
confirmed later on in Ref.~\cite{Aaij:2013zoa}). Since then, many other new
$XYZ$ states which are good candidates to be {\em exotic} hadrons have been
experimentally observed \cite{Agashe:2014kda,Brambilla:2010cs}.

Within the molecular description of the $X(3872)$, the existence of a $X_{2}$
[$J^{PC} = 2^{++}$]  $s$-wave $D^{*}\bar{D}^{*}$ bound state was predicted in
the effective field theory (EFT) approach of
Refs.~\cite{Nieves:2012tt,HidalgoDuque:2012pq}.  As a result of the heavy quark
spin symmetry (HQSS), the binding energy of the $X_2$ resonance was found to be
similar to that of the $X(3872)$, {\it i.e.},
\begin{equation}
  M_{X_2} -  M_{X(3872)} \approx M_{D^*} - M_{D} \approx 140~\text{MeV}.
  \label{eq:mx2}
\end{equation}
The existence of such a state was also suggested  in
Refs.~\cite{Tornqvist:1993ng,Hidalgo-Duque:2013pva,Molina:2009ct,Liang:2009sp,
Swanson:2005tn,Sun:2012zzd}. Both the $X(3872)$ and the $X_2$, to be denoted by
$X_2(4013)$ in what follows, have  partners in the bottom
sector~\cite{Guo:2013sya},\footnote{In Ref.~\cite{Guo:2013sya}, the bottom and charm sectors are connected by assuming the bare couplings in the interaction Lagrangian, see Eq.~\eqref{eq:LaLO}, to be independent of the heavy quark mass. This assumption will also be used throughout this work.} which we will call $X_b$ and $X_{b2}$, respectively, with masses approximately related by
\begin{equation}
  M_{X_{b2}} - M_{X_b} \approx M_{B^*} - M_{B} \approx 46~\text{MeV}.\label{eq:mbx2}
\end{equation}
It is worthwhile to mention that states with $2^{++}$ quantum numbers
exist as well as spin partners of the $1^{++}$ states in the spectra
of the conventional heavy quarkonia and tetraquarks. However, the mass
splittings would only accidentally be the same as the fine splitting
between the vector and pseudoscalar charmed mesons, see
Eq.~\eqref{eq:mx2}.\footnote{Were these states due to threshold cusps, the splittings would be the same as those of the hadronic molecules. However, it was shown in Ref.~\cite{Guo:2014iya} that narrow near threshold peaks in the elastic channel cannot be produced by threshold cusps.}  For instance, the mass splitting between the
first radially excited charmonia with $2^{++}$ and $1^{++}$ in the
well-known Godfrey--Isgur quark model is
$30$~MeV~\cite{Godfrey:1985xj}, which is much smaller than the value
in Eq.~\eqref{eq:mx2}. In a quark model calculation with screened
potential, the $2^{++}-1^{++}$ mass splitting for the $2P$ charmonia
is around 40~MeV~\cite{Li:2009ad}. As for the tetraquark states, the
corresponding mass splitting predicted in Ref.~\cite{Maiani:2004vq} is
70~MeV, which is again much smaller than $M_{D^*} - M_{D}$. Notice
that it is generally believed that the $\chi_{c2}(2P)$ has been
discovered~\cite{Uehara:2005qd,Aubert:2010ab}, and its
mass is much lower than $2M_{D^*}$. Therefore, we conclude that a
possible discovery of a $2^{++}$ charmonium-like state with a mass
around 4013~MeV as a consequence of HQSS~\cite{Guo:2009id} would
provide a strong support for the interpretation that the $X(3872)$ is
dominantly a $D\bar D^*$ hadronic molecule. It is thus very important
to search for such a tensor resonance, as well as the bottom
analogues, in various experiments and in lattice QCD (LQCD)
simulations.

Some exotic hidden charm sectors on the lattice have been recently
studied~\cite{Liu:2012ze,Prelovsek:2013cra,Prelovsek:2013xba,Prelovsek:2014swa,Padmanath:2015era},
and evidence for the $X(3872)$ from $D\bar D^*$ scattering on the
lattice has been found~\cite{Prelovsek:2013cra}, while the quark mass
dependence of the $X(3872)$
binding energy was discussed in Refs.~\cite{Baru:2013rta,Jansen:2013cba}.  The
$2^{++}$ sector has not been exhaustively addressed yet in LQCD, though a
state with these quantum numbers and a mass of $(m_{\eta_c}+1041\pm 12)
~\rm{MeV}$= $(4025 \pm 12)~\text{MeV}$, close to the value predicted in
Refs.~\cite{Nieves:2012tt,HidalgoDuque:2012pq}, was reported in
Ref.~\cite{Liu:2012ze}. The simulation used dynamical fermions, novel
computational techniques and the variational method with a large basis
of operators. The calculations were performed on two lattice
volumes with pion mass $\simeq 400$ MeV.  There
exists also a feasibility study~\cite{Albaladejo:2013aka} of future LQCD
simulations, where the EFT approach of
Refs.~\cite{Nieves:2012tt,HidalgoDuque:2012pq} was formulated in a
finite box.

On the other hand, despite  the theoretical predictions on their existence, none of these
hypothetical particles has been observed so far. Nevertheless, they are being and will be
searched for in current and future experiments such as BESIII, LHCb, CMS,
Belle-II and PANDA. It is thus of paramount importance to provide theoretical
estimates on their production rates in various experiments, as well as the
dominant decay modes and widths.\footnote{If a resonance is too broad, say
$\Gamma\gtrsim 200~\text{MeV}$, it would be very difficult to be identified since it
is highly nontrivial to distinguish the signal for a broad resonance
from various backgrounds.}~The production of these states in hadron colliders and electron--positron collisions has been studied in
Refs.~\cite{Guo:2014sca,Guo:2014ura}. In this work, we will investigate the
dominant decay modes of the spin-2 partners of the $X(3872)$, i.e. the
$X_2(4013)$ and $X_{b2}$, and provide an estimate of their decay
widths.

Besides, we will also discuss the radiative $X_2\to D\bar D^{*}\gamma$
and $X_{b2} \to \bar B B^{*}\gamma$ transitions. These decay modes are more
sensitive to the long-distance structure of the resonances and might
provide valuable details on their wave-functions. The situation is
similar to that of the $X(3872)\to D^0 \bar D^0 \pi^0$ decay studied
in Refs.~\cite{Guo:2014hqa,Guo:2014cpb}. Also here, the widths will be affected by
the $D\bar D^{*}$ or $B\bar B^{*}$ final state interaction (FSI).
FSI effects are expected to be large because they should be enhanced
by the presence of the isovector $Z_c(3900)$ and $Z_b(10610)$ resonances located
near the $D \bar D^{*}$ and $\bar B B^{*}$ thresholds,
respectively.  Besides, FSI corrections will be also sensitive to the
negative $C$-parity isoscalar  $D \bar D^*$ or $\bar B B^{*}$
interaction. Eventually, precise measurements of these radiative decay widths might
provide valuable information on the interaction strength in this sector,  which would be
important in understanding the
$P^{(*)}\bar P^{(*)}$ system and other exotic systems related to it through heavy quark
symmetries~\cite{Guo:2013sya,Guo:2013xga}.

The structure of the paper is as follows. First in Sect.~\ref{sec:X2th}, we
briefly discuss the relation of the  charm and bottom $2^{++}$ states with the
$X(3872)$ resonance, and in Sect. \ref{section:hadron-decays} we present our
predictions for the  $X_{2} \to D\bar{D}, D\bar{D}^{*}$ hadron  decays and the
$X_{b2} \to B\bar{B}, B\bar{B}^{*}$ ones in the bottom sector. In
Sect.~\ref{sec:radia}, the $X_{2}$ and $X_{b2}$ radiative decays are
investigated, paying special attention to the loop mechanisms responsible for
the FSI contributions. The conclusions of this work are outlined in Sect.
\ref{section:Conclusions} and in addition, there are three Appendices. In the
first one (Appendix~\ref{sec:4Hpigamma}), we collect different heavy meson
Lagrangians used through this work, while the validity of the
perturbative treatment of the $D\bar D$ for the $X_2$ is discussed in the second one
(Appendix~\ref{app:pc}). Finally, in
Appendix~\ref{app:three-loopfunction}, we give some details on the evaluation of
different three-point loop functions that appear in the computation of the
hadronic and radiative decays.

\section{\boldmath HQSFS, the $X(3872)$ resonance and the charm and bottom $X_2$ states}
\label{sec:X2th}

\subsection{\boldmath $X(3872)$}

As mentioned, we start assuming the $X(3872)$ to be a
positive $C$-parity $D\bar D^*$ bound state, with quantum numbers
$J^{PC}$ = $1^{++}$. At very low energies, the leading order~(LO)
interaction between pseudoscalar and vector charmed ($D^0, D^+,
D^{*0}, D^{*+}$) and anti-charmed ($\bar D^0, D^-, \bar D^{*0},
D^{*-}$) mesons can be described just in terms of a contact-range
potential, which is constrained by
HQSS~\cite{Nieves:2012tt,HidalgoDuque:2012pq,Guo:2013sya}. Pion
exchange and particle coupled-channel\footnote{We do not refer to
  charge channels, but rather to the mixing among the $D\bar D$,
  $D\bar D^*$, $D^*\bar D^*$ pairs in a given $IJC$ (isospin, spin and
  charge conjugation) sector.} effects turn out to be
sub-leading~\cite{Nieves:2012tt,Valderrama:2012jv}.  For the case of
the $X(3872)$, isospin breaking is important~\cite{Gamermann:2009uq}
as this bound state is especially shallow.  The energy gap between the
$D^0 \bar D^{*0}$ and $D^+ D^{*-}$ channels is around 8~MeV, which
is much larger than the $X(3872)$ binding energy with respect to the $D^0 \bar D^{*0}$ threshold. As a consequence,
the neutral ($D^0 \bar D^{*0}$) and charged ($D^+ D^{*-})$
channels should be treated independently. The
coupled-channel\footnote{Actually, positive $C$-parity combinations in
  both the neutral $D^0 \bar D^{*0}$ and charged $D^+\bar
  D^{*-}$ channels are being considered.}  contact potential in the $1^{++}$
sector is given
by~\cite{HidalgoDuque:2012pq} (see also Appendix~\ref{sec:4Hpigamma-4H})
\begin{eqnarray}
V_{X(3872)} = \frac{1}{2} \left( \begin{array}{cc}  C_{0X} + C_{1X} & C_{0X} -
C_{1X} \\ C_{0X} - C_{1X} & C_{0X} + C_{1X}     \end{array} \right), \label{eq:VLO}
\end{eqnarray}
with $C_{0X}$ and $C_{1X}$ low energy constants (LECs) that need to be fixed
from some input. This interaction is used as kernel of the Lippmann--Schwinger
equation (LSE) in the coupled channel space in the $1^{++}$ sector,
\begin{equation}
\label{eq:lse}
    T(E; \vec{p}\,',\vec{p}\,) = V(\vec{p}\,',\vec{p}\,) + \int
\frac{d^3\vec{q}}{(2\pi)^3} V(\vec{p}\,',\vec{q}\,)
\frac{1}{E-\vec{q}^{\,2}/2\mu_{12}-M_1-M_2 +  i \varepsilon} \,
T(E; \vec{q},\vec{p}\,)~,
\end{equation}
with $M_1$ and $M_2$ the masses of the involved mesons,
$\mu_{12}^{-1}=M^{-1}_1+M^{-1}_2$, $E$ the center of mass~(c.m.)
energy of the system and $\vec{p}$ ($\vec{p}^{\, \prime}$) the initial
(final) relative three momentum of the $D\bar D^*$ pair in the
c.m. frame. The used  normalization   is such that
above threshold $\left[E> (M_1+M_2)\right]$, the single channel
elastic unitary condition is ${\rm Im}\, T^{-1}(E) = \mu_{12}k/(2\pi)$ with  $k=
\sqrt{2\mu_{12}\left(E-M_1-M_2\right)}$.
The discussion is similar for any other
$J^{PC}$ sector. Due to the use of contact interactions, the LSE
shows an ill-defined ultraviolet (UV) behavior, and requires a
regularization and renormalization procedure. We employ a standard
Gaussian regulator (see, {\it e.g.} ~\cite{Epelbaum:2008ga})
\begin{equation}
\left<\vec{p}\,|V|\vec{p}\,'\right> = C_{IX}
~e^{-\vec{p}\,^{2}/\Lambda^{2}}~e^{-\vec{p}\,'^{2}/\Lambda^{2}}\label{eq:UVG},
\end{equation}
with $C_{IX}$ any of the LECs of Eq.~(\ref{eq:VLO}) in the
case of the $X(3872)$, or the relevant ones for any other $J^{PC}$
sector.  We take cutoff values $\Lambda = 0.5-1$
GeV~\cite{HidalgoDuque:2012pq,Nieves:2012tt}, where the range is
chosen such that $\Lambda$ will be bigger than the wave number of the
states, but at the same time it will be small enough to preserve HQSS and
prevent that the theory might become sensitive to the specific details
of short-distance dynamics.\footnote{However, as will be shown later on, the situation is more complicated in the two-body $d$-wave hadronic decays.} The dependence of the results on the cutoff,
when it varies within this window, provides a rough estimate of the
expected size of sub-leading corrections. Bound states correspond to
poles of the $T$-matrix below threshold on the real axis in the first
Riemann sheet~(RS) of the complex energy, while the residues at the pole
give the $s$-wave couplings of the state to each channel ($D^0 \bar
D^{*0}$ and $D^+ D^{*-}$ in the case of the $X(3872)$
resonance\footnote{For instance, in the case of the $X(3872)$, we have
\begin{eqnarray*}
\left(g_0^X\right)^2 &=& \lim_{E\to M_\X} \left[ E-M_\X \right] \times T_{11}(E)~,\\
g_0^Xg_c^X &=& \lim_{E\to M_\X} \left[ E-M_\X \right] \times T_{12}(E)~,
\end{eqnarray*}
where $T_{ij}$ are the matrix elements of the $T$-matrix solution of the UV regularized  LSE.}).

The LECs $C_{0X}$ and $C_{1X}$ can in principle be determined~\cite{HidalgoDuque:2012pq}
from  $M_{X(3872)}=(3871.69 \pm 0.17)$ MeV (mass average quoted by the
PDG~\cite{Agashe:2014kda}) and the isospin violating ratio of the decay amplitudes for the $X(3872) \to J/\psi\pi\pi$ and $X(3872) \to
J/\psi\pi\pi\pi$, $R_{X(3872)}=0.26\pm
0.07$~\cite{Hanhart:2011tn}. The ratio is given by (see Eq.~(80) of
Ref.~\cite{HidalgoDuque:2012pq})
\begin{equation}
R_{X(3872)}= \frac{\hat{\Psi}_{\rm n}-\hat{\Psi}_{\rm
    c}}{\hat{\Psi}_{\rm n}+\hat{\Psi}_{\rm c}}
\end{equation}
where $\hat{\Psi}_{\rm n,c}$ give the average of the neutral and
charged wave function components in the
vicinity of the origin, and are related to
the LECs introduced in Eq.~(\ref{eq:VLO}) by~\cite{Gamermann:2009uq}
\begin{equation}
\frac{\hat{\Psi}_{\rm n}}{\hat{\Psi}_{\rm c}} = \frac{1-G_{\rm
    2}\left(C_{0X}+C_{1X}\right)/2}{G_{\rm
    2}\left(C_{0X}-C_{1X}\right)/2} = \frac{G_{\rm 1}\left(C_{0X}-C_{1X}\right)/2}{1-G_{\rm 1}\left(C_{0X}+C_{1X}\right)/2}
\end{equation}
where the neutral and charged loop functions, $G_{\rm 1}=G_{D^0\bar
  D^{*0}}=G_{D^{*0}\bar D^{0}}, G_{\rm 2}=G_{D^+D^{*-}}
=G_{D^{*+}D^{-}}$,  are defined in Eq.~(\ref{eq:gloop}), and should be
evaluated at the $X(3872)$ pole mass.  We  use $m_{D^{0}} = (1864.84 \pm 0.07)~\text{MeV}$, $m_{D^{+}} =(1869.61 \pm 0.10)~\text{MeV}$, $m_{D^{*0}} =(
2006.96 \pm 0.10)~\text{MeV}$ and $m_{D^{*+}} =(2010.26 \pm 0.07)~\text{MeV}$~\cite{Agashe:2014kda}. Note that  $m_{D^{0}}+ m_{D^{*0}}= (3871.80 \pm 0.12)~\text{MeV}$, and  the uncertainty in the value of this lowest
threshold affects the precision of  the $X(3872)$ binding energy. We have  taken into account this effect by adding in quadratures the PDG error of the $\X$ mass and that of the neutral channel threshold and assign
this new error  to the mass of the resonance, that now reads  $M_{X(3872)}=(3871.69 \pm 0.21)~\text{MeV}$. For the LECs, we obtain:
\begin{equation}
\label{eq:ces}
 C_{0X} = -1.70^{+0.03}_{-0.07}~(-0.731^{+0.006}_{-0.015})\, {\rm fm}^{2}, \qquad  C_{1X} = -0.09^{+0.54}_{-0.41}~(-0.38^{+0.12}_{-0.10})\, {\rm
fm}^{2}~,
\end{equation}
for $\Lambda = 0.5(1.0)$ GeV. Errors, at the 68\% confidence level (CL), have
been obtained from a Monte Carlo (MC) simulation assuming uncorrelated
Gaussian distributions for the two inputs
($M_{X(3872)},R_{X(3872)}$). In the simulation, we have rejected MC
samples for which the $\X$ turned out to be unbound, since the scheme
of Ref.~\cite{HidalgoDuque:2012pq} only allows to determine the
properties of the resonance when it is bound.

\subsection{\boldmath $X_2(4013)$: $J^{PC}=2^{++}$, charm sector}
\label{sec:X2charm}
\begin{figure}[tbh]
\begin{center}
\makebox[0pt]{\includegraphics[width=0.5\textwidth]{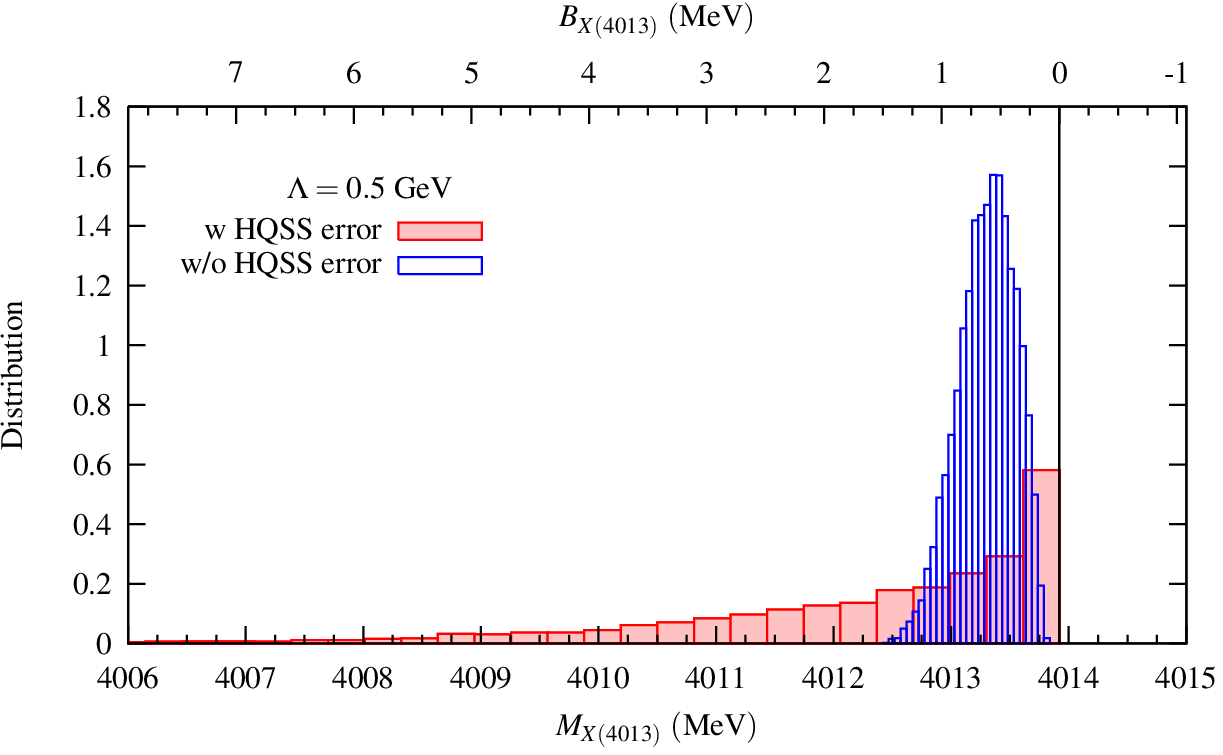}
\includegraphics[width=0.5\textwidth]{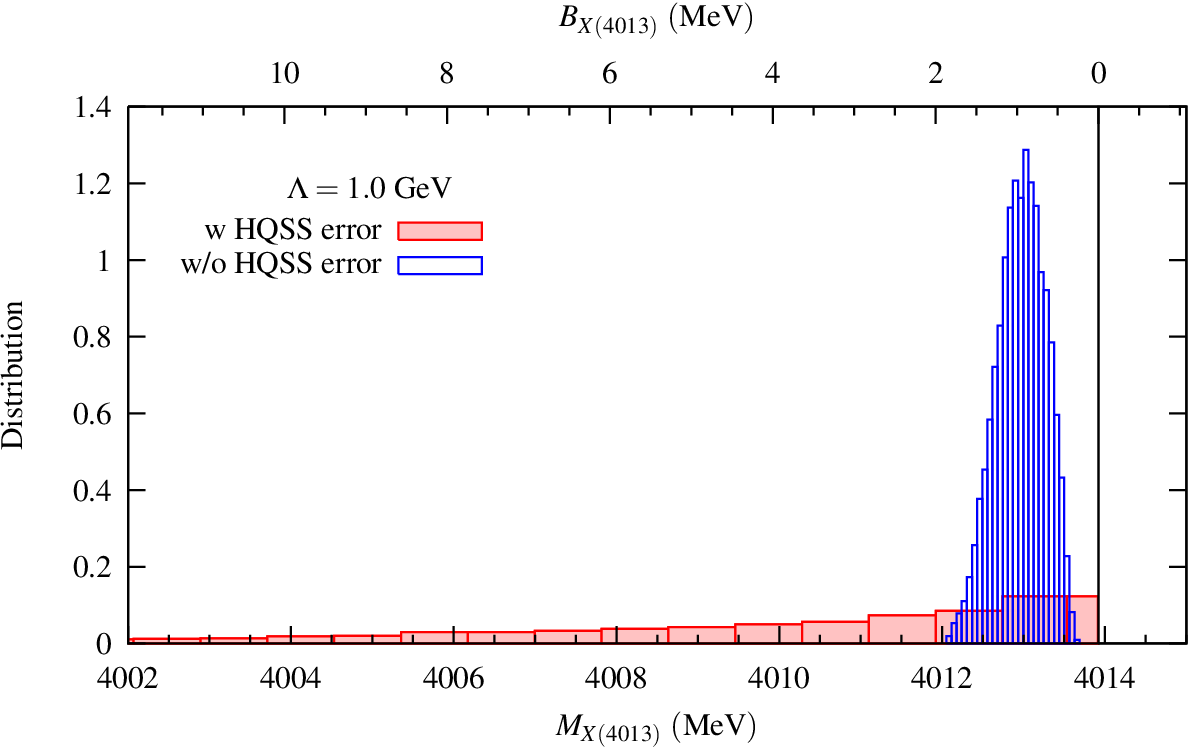}}
\end{center}
\caption{$X_2$ binding energy histograms obtained from the interaction
  of Eq.~(\ref{eq:VLOX2}) using  LECs distributions determined
  from the $\X$ resonance inputs (blue) or using  LECs
  distributions additionally modified to  account for the HQSS
  systematic error (red). Left and right plots
  correspond to UV cutoffs of 0.5 and 1 GeV, respectively.  MC sample
  ($C_{0X} + C_{1X}, C_{0X} - C_{1X})$ pairs fitted to the  input
  ($M_{X(3872)},R_{X(3872)}$) distributions are first generated and are used to evaluate the
  $X_2$ mass. $\X$ mass trials above threshold are rejected. To evaluate the red shaded histogram, and to account for
   the HQSS 20\% uncertainty in the
  $2^{++}$ interaction, each of the members of any MC sample ($C_{0X} +
  C_{1X}, C_{0X} - C_{1X})$ pair is  multiplied by  independent
  N$(\mu=1,\sigma=0.2)$ Gaussian
  distributed random quantities $r_{\pm}$. }\label{fig:mass-histo}
\end{figure}

HQSS predicts that the $s$-wave $D^*\bar D^*$ interaction in the
$2^{++}$ sector is, up to corrections suppressed by the charm quark
mass, identical to that in the $X(3872)$ sector ($1^{++}$) and given
by Eq.~(\ref{eq:VLO}) \cite{Nieves:2012tt,HidalgoDuque:2012pq}. Thus, in the $2^{++}$ sector, the potential in
the ($D^{* 0} \bar D^{*0}$), ($D^{*+} D^{*-}$) coupled channel
space reads (see also Appendix~\ref{sec:4Hpigamma-4H})
\begin{eqnarray}
\label{eq:PotX2}
V_{2^{++}} = \frac{1}{2} \left( \begin{array}{cc}  C_{0X} + C_{1X} & C_{0X} -
C_{1X} \\ C_{0X} - C_{1X} & C_{0X} + C_{1X}     \end{array} \right) +  {\cal
O}(q/m_c), \label{eq:VLOX2}
\end{eqnarray}
with the same structure and involving the same LECs that in the $X(3872)$
channel. Besides, in the above equation $m_c\sim 1.5$ GeV is the charm
quark mass and $q\sim \Lambda_\text{QCD}$, a scale related to the  light degrees of
freedom. Taking $\Lambda_\text{QCD}\sim 300$ MeV
\cite{Agashe:2014kda},  corrections of  the order of 20\% to the
interaction predicted by HQSS cannot be discarded\footnote{The two type of uncertainties
(HQSS subleading corrections and errors inherited from the inputs) affecting the
determination of the LECs in the $2^{++}$ sector are statistically
uncorrelated, and should be accordingly added up. To that end, we use
MC techniques as explained in the caption of
Fig.~\ref{fig:mass-histo}.}.  
Nevertheless, it seems
natural to expect a $2^{++}$ $D^*\bar D^*$ loosely bound state ($X_2$), the HQSS
partner of the $X(3872)$, and  located in the vicinity of the  $D^{* 0} \bar
D^{*0}$ threshold ($\sim 4014$ MeV)~\cite{Nieves:2012tt, HidalgoDuque:2012pq, Guo:2013sya}.
This is illustrated in the $X_2$ binding energy distributions
depicted in Fig.~\ref{fig:mass-histo}. Neglecting the ${\cal O}(q/m_c)$
corrections to the LECs, and using those obtained from the
$X(3872)$ resonance, we find a clear signal (blue histograms) of a
weakly bound state. However, the case is less robust when the latter
corrections are taken into account.  Thus, because of the additional 20\%
HQSS uncertainty, the area below the red shaded $\Lambda= 0.5$ GeV ($\Lambda= 1$
GeV) histogram is only 0.77 (0.68). This means that approximately a 23\% (32\%)
of $\X$ events [($M_{X(3872)},R_{X(3872)}$) MC samples]
do not produce a $X_2$ pole in the first RS, since the strength of
the resulting interaction in the $2^{++}$ sector would  not be attractive enough
to bind the state, though a virtual state in the second RS will be
generated instead. Given the existence of the $\X$ as a $D \bar D^*$
molecule, if the $X_2$ resonance exists, we would expect its mass (binding
energy) to lie most likely in the interval $[2m_{D^{*0}}, 4006~\text{MeV}]$ ($[0,8]~\text{MeV}$), as shown in Fig.~\ref{fig:mass-histo}.
Note that the discussion in Ref.~\cite{Guo:2013sya} was simpler, because there
we worked in the isospin symmetric limit and used the averaged masses of the
heavy mesons, which are larger than those of the physical $D^0$ and $D^{*0}$
mesons.

For later use, we  also  need the couplings of the $X_2$ to its
neutral ($D^{*0}\bar{D}^{*0}$) and charged ($D^{*+}D^{*-}$) components,
$g_0^{X_2}$ and $g_c^{X_2}$, respectively.  They turn out to be  slightly
different because the $X_{2}$ resonance is an admixture of isospin 0 and 1,
since its binding energy is much smaller than the  energy difference between
the two thresholds~\cite{Gamermann:2009uq,HidalgoDuque:2012pq}. Considering  the
HQSS uncertainties, we find:
\begin{align}
10^2 g^{X_2}_0 =  1.4^{+1.1}_{-0.4} & ~~( 1.5^{+1.1}_{-0.4})~~
{\rm{MeV^{-1/2}}}~,\label{eq:gneutral} \\
10^2 g^{X_2}_c =  1.5 ^{+1.4}_{-0.2}&  ~~ (1.3^{+1.3}_{-0.3})~~
{\rm{MeV^{-1/2}}}~,\label{eq:gcargado}
\end{align}
for $\Lambda = 0.5(1.0)$ GeV.

\subsection{\boldmath $X_{b2}$: $J^{PC}=2^{++}$, bottom sector}
\label{sec:xb2}

\begin{figure}[tbh]
\begin{center}
\makebox[0pt]{\includegraphics[width=0.5\textwidth]{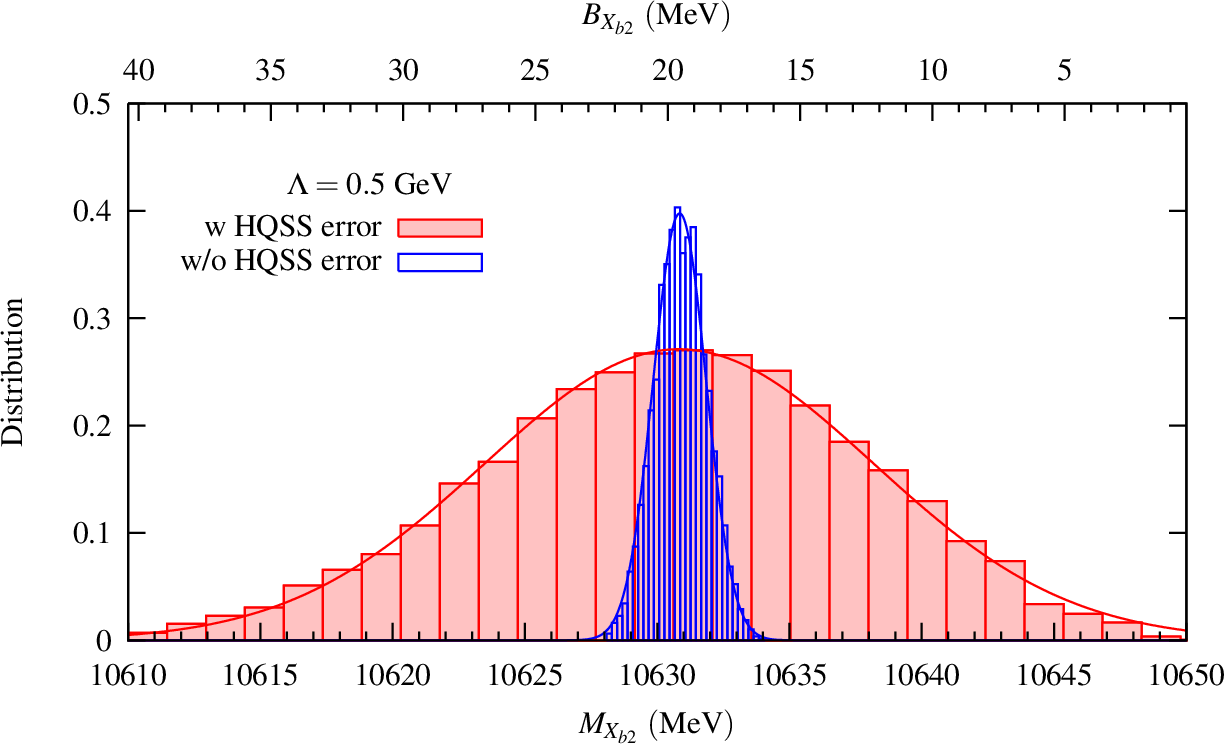}
\includegraphics[width=0.5\textwidth]{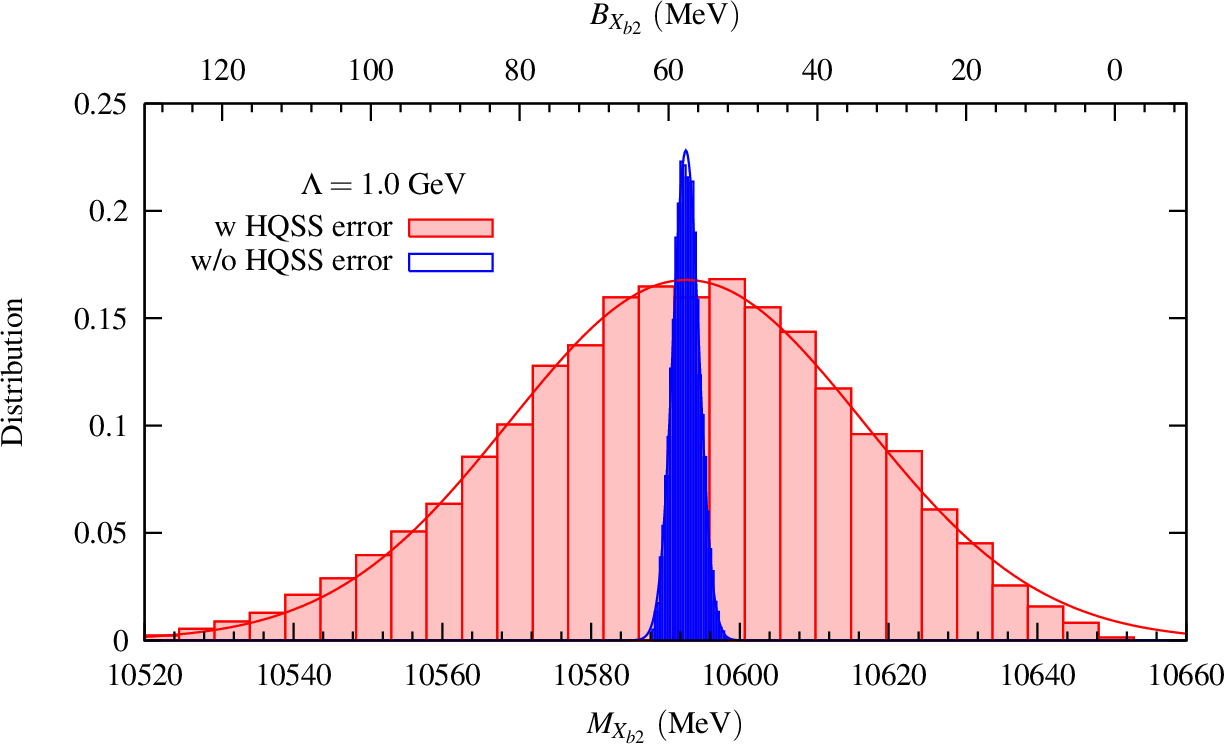}}
\end{center}
\caption{Same as in Fig.~\ref{fig:mass-histo} but in the bottom
  sector. To better appreciate the distribution details, the
  $\Lambda =$ 0.5 (1) GeV red histogram, which includes the 20\% HQSS
  error, has been multiplied by a factor of 5
  (10). }\label{fig:mass-histo-b}
\end{figure}

Owing to the heavy flavor symmetry, the LO $2^{++}$ $B^*\bar B^*$ interaction is
given by Eq.~(\ref{eq:PotX2}) as well, and thus we should also expect a $2^{++}$
 $B^*\bar B^*$ bound state ($X_{b2}$), the HQSFS partner of the
 $X(3872)$, located close to the $B^* \bar B^*$ threshold ($\sim 10650$
 MeV)~\cite{Guo:2013sya}. The $X_{b2}$ binding energy distributions are shown in
 Fig.~\ref{fig:mass-histo-b}
for the two UV cutoffs employed in this work. We have used the same masses for
the neutral and charged mesons, $m_B = (m_{B^0}+m_{B^+})/2 = 5279.42~\text{MeV}$
and $m_{B^*}=5325.2~\text{MeV}$. Note that, according to the
PDG~\cite{Agashe:2014kda},
  $|(m_{B^{*0}}-m_{B^0}) - (m_{B^{*+}}-m_{B^+}) |< 6$ MeV   CL=95.0\%,
 and  $m_{B^0}-m_{B^+} =  (0.32 \pm 0.06)~\text{MeV}$, from where we might
  expect isospin breaking effects for the $B^*$ mesons to be significantly
smaller than  in the charm sector. In Ref.~\cite{Guo:2013sya}  we
found that the   binding energy of the $X_{b2}$ state is
significantly larger than that of its counterpart in the charm sector
($X_2$), around a few tens of MeV.
Thus we do not expect any significant isospin breaking effects and the $X_{b2}$
resonance would be a pure isoscalar ($I = 0$) state.

As can be seen in Fig.~\ref{fig:mass-histo-b}, in this case we have a
robust prediction even when HQSS uncertainties (20\%) are taken into
account. We obtain the mass and the coupling from the residue at the pole for
$\Lambda = 0.5 ~(1.0)~\text{GeV}$:\footnote{There appear small differences in the central value of the resonance mass with respect to value quoted in ~\cite{Guo:2013sya} due to small differences in the
used hadron masses.}
\begin{equation}
E_{X_{b2}} =   10631^{+7}_{-8} \left(10594^{+22}_{-26}  \right)  {\rm{MeV}},
\qquad
%\label{eq:mass_Xb2},
%\end{equation}
%
%and the residue at the pole leads to:
%
%\begin{equation}
10^2 g^{X_{b2}} =  5.9^{+2.9}_{-1.9}  \left(
6.4^{+2.8}_{-2.0}\right) {\rm{MeV^{-1/2}}}.
\label{eq:g_Xb2}
\end{equation}
This bound state, being isoscalar, equally couples to the neutral and
charged components and, therefore: $g^{X_{b2}}_{\rm 0} =
g^{X_{b2}}_{\rm c} = \frac{1}{\sqrt{2}} g^{X_{b2}}$. Our predictions
in Eq.~\eqref{eq:g_Xb2}, both for the mass
and the $B^* \bar B^*$ coupling of the resonance show some dependence
on the UV cutoff, which is to some extent diminished when HQSS
uncertainties are taken into account. Nevertheless, this $\Lambda$
dependence might hint to non-negligible sub-leading corrections (among
others, pion exchange and coupled channel
effects~\cite{Nieves:2012tt}, which can be larger here than in the
charm sector due to the larger binding energy and larger meson masses). We will
compute the decay widths for both UV regulators, and the spread of results will account for this source of
uncertainty.

\section{\boldmath The hadronic $X_{2}$ and $X_{b2}$ decays}
\label{section:hadron-decays}

The quantum numbers, $J^{PC} = 2^{++}$, of these resonances constrain their
possible decay channels. In this work, for hadronic decays we only consider the
decays into two heavy hadrons: $X_{2} \to D \bar{D}$ and $X_{2} \to D
\bar{D}^{*} (D^{*}\bar{D})$, and the analogous processes $X_{b2} \to B \bar{B}$
and $X_{b2} \to B \bar{B}^{*} (B^{*}\bar{B})$. We expect that these $d$-wave
decay modes should largely saturate the widths of these states.
Because the $D\bar D$ couples in a $d$-wave to the $2^{++}$ system, its
contribution to the mass renormalization of the $X_2$ is of higher order (see Appendix~\ref{app:pc}).
We thus did not include the $D\bar D$ as a coupled channel in the $T$-matrix, but
treat it perturbatively. This means that the transitions are mediated by the
exchange of a pion.
The relevant $\pi P^{(*)}  P^{(*)}$ vertices are taken from the LO Lagrangian of heavy meson chiral perturbation
theory~\cite{Grinstein:1992qt,Wise:1992hn,Burdman:1992gh,Yan:1992gz} (see
Appendix~\ref{sec:appHHpi}). At LO, besides the pion decay constant,
$f_{\pi}=92.2$ MeV, there appears only one additional $D^*D\pi$ coupling~($g$).
We take $g=0.570 \pm 0.006$ as inferred from the new value of $\Gamma=(83.4 \pm
1.8)~\text{keV}$ for the $D^{*+}$ decay width quoted by the
PDG~\cite{Agashe:2014kda}.
This is mostly determined by the recent BABAR Collaboration
measurement~\cite{Lees:2013zna} of this width, which is approximately  a factor
12 times more precise than the previous value, $\Gamma = (96 \pm 4 \pm 22)~\text{keV}$ by the CLEO Collaboration~\cite{Anastassov:2001cw}. Thus,  we end up with an
uncertainty of the order 1\% for $g$. Though the hadronic $X_{2}$ and $X_{b2}$
widths evaluated in this section will be proportional to $g^4$, this source of
error ($\sim 4\%$) will be much smaller than others and it will be ignored in
what follows.

\subsection{Charm decays}
\label{sec:charm-dec}

\begin{table}[tb]
\begin{center}
\begin{tabular}{l|cc|cc}
 &  \multicolumn{2}{c|}{without pion-exchange FF} & \multicolumn{2}{c}{with
    pion-exchange FF}\\
 &  ~$\Lambda = 0.5 ~{\rm{GeV}}$~ & ~$\Lambda = 1 ~{\rm{GeV}}$~ &
  ~$\Lambda = 0.5 ~{\rm{GeV}}$~ & ~$\Lambda = 1 ~{\rm{GeV}}$~ \\
\hline
$\Gamma (X_{2} \to D^{+} D^{-})$  [MeV] & $3.3^{+3.4}_{-1.4}$
&$7.3^{+7.9}_{-2.1}$ & $0.5^{+0.5}_{-0.2}$ & $0.8^{+0.7}_{-0.2}$\\
$\Gamma (X_{2} \to D^{0} \bar{D}^{0})$  [MeV]  & $2.7^{+3.1}_{-1.2}$ & $5.7^{+7.8}_{-1.8}$
& $0.4^{+0.5}_{-0.2}$ & $0.6^{+0.7}_{-0.2}$ \\
\hline
$\Gamma (X_{2} \to D^{+} D^{*-})$  [MeV]  &
$2.4^{+2.1}_{-1.0}$ & $4.4^{+3.1}_{-1.2}$ & $0.7^{+0.6}_{-0.3}$ & $1.0^{+0.5}_{-0.2}$\\
$\Gamma (X_{2}  \to D^{0} \bar{D}^{*0})$  [MeV] &
$2.0^{+2.1}_{-0.9}$ &$3.5^{+3.5}_{-1.0}$ &  $0.5^{+0.6}_{-0.2}$ &$0.7^{+0.5}_{-0.2}$  \\
\end{tabular}
\end{center}
\caption{ $X_{2}(4013) \to D \bar{D},  D \bar{D}^*$ decay widths using
  different  UV Gaussian  regulators for  the $D^*\bar  D^* X_2$  form
  factor and with/without including a pion-exchange vertex form factor (FF) in
  each of the  $D^*D\pi$  and $D^*D^*\pi$  vertices  in  the three-point  loop
  function. The  decay widths  of the $X_{2}(4013)  \to \bar  D D^{*}$
  modes are the same thanks  to $C$-parity. Uncertainties are obtained
  from  a  Monte  Carlo  simulation using  the  $X_2$  binding  energy
  histograms displayed  in Fig.~\ref{fig:mass-histo} (red  shaded) and
  the    $g_0^{X_2}$    and    $g_c^{X_2}$    couplings    given    in
  Eqs.~(\ref{eq:gneutral})   and   ~(\ref{eq:gcargado}).   Note   that the
  procedure  takes  into  account  20\%  HQSS  uncertainties  and  the
  correlations between $X_2$ masses (binding energies) and $g_0^{X_2}$
  and $g_c^{X_2}$  couplings.  Errors  on the  widths provide  68\%
  CL intervals.  }
\label{tab:Decaywidths_hadron}
\end{table}
%-------------------------------------------------------------------------------

\subsubsection{ $X_{2}(4013) \to D \bar{D}$}
\label{subsect:DDbar_decays}

\begin{figure}[tbh]
\begin{center}
\includegraphics[width=0.8\textwidth]{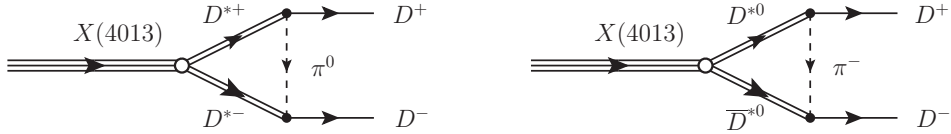}
\end{center}
\caption{Feynman diagrams for the $X_{2}(4013) \to D^{+} D^{-}$ decay.
Diagrams for the $X_{2}(4013) \to D^{0} \bar{D}^{0}$  transition are similar,
with the appropriate changes of the exchanged pion charges.}\label{Decay_DDbar}
\end{figure}

We will first consider the $X_{2}(4013) \to D^{+} D^{-} (D^{0}
\bar{D}^{0})$ decay, which proceeds through the Feynman diagrams depicted
in Fig.~\ref{Decay_DDbar}. We treat charm mesons non-relativistically,
and neglect $p_{D^*,\bar D^*}/m_{D^*}$ terms
and the temporal components in  the $D^*,\bar D^*$ propagators. We
obtain for the $X_{2}(4013) \to D^{+} D^{-}$ process, in the resonance rest frame
and with  $q$ and $k$ the 4-momenta of the $D$ and $\bar D$  final
mesons ($\vec{q}=-\vec{k}$, $q^0+k^0= M_{X_2}$),
\begin{eqnarray}
\nonumber{}
&& -i\, \mathcal{T(\lambda)}_{D^+D^-} \nonumber\\
&=&-\frac{N g^{2}}{f_{\pi}^{2}}
\epsilon_{ij}(\lambda)\left\{ g^{X_2}_c
\int \frac{d^{4}l} {\left(2\pi\right)^{4} }
\frac{l^{i}\,l^{j}  }
 {  \left[ (l+q)^{2} - m_{D^{*+}}^{2}+i\varepsilon\right] \left[(k-l)^{2} -
 m_{D^{*-}}^{2}+ i\varepsilon\right] \left(l^{2} - m_{\pi^0}^{2}
 +i\varepsilon\right) }\right. \nonumber \\
&& \left. +  2 g^{X_2}_0
\int \frac{d^{4}l} {\left(2\pi\right)^{4} }
\frac{l^{i}\,l^{j}  }
 {  \left[ (l+q)^{2} - m_{D^{*0}}^{2}+i\varepsilon\right] \left[(k-l)^{2} -
 m_{\bar D^{*0}}^{2}+ i\varepsilon\right] \left(l^{2} -
   m_{\pi^-}^{2}+i\varepsilon\right) }\right\} \nonumber \\
&=& i\frac{N g^{2}}{f_{\pi}^{2}}
\epsilon_{ij}(\lambda)\left\{ g^{X_2}_c
I^{ij}(m_{D^{*+}},m_{\pi^0}; M_{X_2},q^\mu \,)
+ 2 g^{X_2}_0 I^{ij}(m_{D^{*0}},m_{\pi^-}; M_{X_2},q^\mu \,) \right\}\label{eq:ddpi1},
\end{eqnarray}
where $\epsilon_{ij}(\lambda)$ is the symmetric spin-2 tensor with $\lambda$
denoting the polarization of the $X_2$ state\footnote{ For
  instance, it could be given by
\begin{equation*}
\epsilon_{ij}(\lambda) = \sum_{\lambda_1,\lambda_2}
(1,1,2|\lambda_1,\lambda_2, \lambda) \epsilon^i_{\lambda_1}\epsilon^j_{\lambda_2}
\end{equation*}
with $\epsilon^i_0= (0,0,1)$, and $\epsilon^i_{\pm 1}= \mp
(1,\pm i,0)/\sqrt{2}$, and $(j_1,j_2,j|m_1,m_2,m)$ a Clebsch-Gordan coefficient.} and $N=\sqrt{8
M_{X_2} m^2_{D^{*}}}\left(\sqrt{m_{D}m_{D^{*}}}\right)^{2}$ accounts
for the normalization of the heavy meson
fields\footnote{We use a non-relativistic normalization for the heavy mesons,
which differs from the traditional relativistic one by a factor
$\sqrt{M_H}$.} and some additional factors needed when the couplings
$g^{X_2}_{c,0}$, as determined from the residues at the pole of the EFT
$T$-matrix, are used for the $X_2D^*\bar D^*$ vertex. For the neutral and
charged pion masses, we have used the
values quoted by the PDG~\cite{Agashe:2014kda}  and  heavy meson isospin averaged masses to compute $N$.
Besides, $I^{ij}$ is a three-point loop function, the detailed
evaluation of which is relegated to the
Appendix~\ref{app:three-loopfunction-H}.\footnote{In the computation of
  $I^{ij}$, we are consistent with the former approximations, and
  we  use non-relativistic charm meson propagators.}~The loop is seemingly logarithmically divergent. However, since the $X_2$
polarization is traceless,
the divergent part which comes with a Kronecker
delta does  not contribute. This is because the decay occurs in a
$d$-wave, thus the loop momentum is converted to external momenta, and the
remaining part of the integral is convergent. Nevertheless, we will
include two different form factors in the computation of the three-point loop
function. One is inherited from the UV
regularization/renormalization procedure sketched in
Eq.~(\ref{eq:UVG}) and employed to make the LSE $T$-matrix finite. In addition,
we will include a second form factor to account for the large virtuality of the
pion in the loop. We will discuss this at length below and in
Appendix~\ref{app:three-loopfunction-H}.

Analogously, the $X_{2}(4013) \to D^{0} \bar D^{0}$ amplitude is,
\begin{equation}
-i\, \mathcal{T(\lambda)}_{D^0\bar D^0} = i\frac{Ng^{2}}{f_{\pi}^{2}}
\epsilon_{ij}(\lambda)\left\{ 2g^{X_2}_c
I^{ij}(m_{D^{*+}},m_{\pi^+}; M_{X_2},q^\mu \,)
+  g^{X_2}_0 I^{ij}(m_{D^{*0}},m_{\pi^0}; M_{X_2},q^\mu \,) \right\}.
\label{eq:ddpi2}
\end{equation}

The two-body decay width in the $X_2$ rest-frame reads~\cite{Agashe:2014kda}:
%
%\begin{equation}
%d\Gamma = \frac{1}{32 \pi^{2} } \left| \bar{\mathcal{T}}_{\rm{total}} \right|
% ^{2} \frac{\left| \vec{q_{}} \right| }{M_{X}^{2}} d\Omega
%\end{equation}
\begin{equation}
\label{Eq:decaywidth2body}
\frac{d\Gamma_a}{d\Omega(\hat q)} =  \frac{1}{5} \sum_{\lambda}\left| \mathcal{T(\lambda)}_a\right| ^{2} \frac{\left|
\vec{q}\, \right|}{32 \pi^2 M_{X_2}^{2}}, \qquad a= D^+D^-,\
  D^0\bar D^0~.
\end{equation}
The sum over the $X_2$ polarizations can be easily done in the c.m. frame,
\begin{eqnarray}
\sum_{\lambda} \epsilon_{mn}(\lambda) \epsilon^*_{i j}(\lambda) &=& \frac12 \left[
  \delta_{m i} \delta_{n j} + \delta_{n i}\delta_{m j} - \frac 23
  \delta_{m n}\delta_{i j}\right], \quad m,n,i,j=1,2,3~. \label{eq:e2}
\end{eqnarray}
As discussed in Appendix~\ref{app:three-loopfunction-H}, the
three-point loop function has a tensor structure of the type
\begin{equation}
I^{ij}(\vec{q}\,) =  I_0(\vec{q}^{\,2})\, q^{i}q^{j} + I_{1}(\vec{q}^{\,2})\, \delta^{ij}
\left|\vec q\,\right|^{2}~.
\end{equation}
The $I_{1}$ term carries the UV divergence, which however
does not contribute to the width, because it vanishes after
the contraction with the traceless spin-2 polarization
tensor, as mentioned above. Therefore,  only the $I_{0}$ term is relevant, which is free of UV
divergences. Moreover, the contraction of
$I^{ij}(\vec{q}\,)\,I^{mn}(\vec{q}\,)$ with $\sum_{\lambda}
\epsilon_{mn}(\lambda) \epsilon^*_{i j}(\lambda)$ [Eq.~\eqref{eq:e2}] leads to
a factor of $2\vec{q}^{\,4}$/3. Thus, the integration over the solid angle
$d\Omega(\hat q)$ trivially gives $4\pi$, and the width scales like $|\vec{q}\,|^5$ as expected for a
$d$-wave process.

Our predictions for the $X_{2}(4013) \to D^{+} D^{-}, D^{0}
\bar{D}^{0}$ decays are compiled
in  Table~\ref{tab:Decaywidths_hadron}. If we look at the first two
columns of results in the table, we find widths of the order
of a few MeV, with asymmetric errors that
favour larger values. This is mostly due to the similar asymmetry of the
uncertainties quoted for the $g_0^{X_2}$ and $g_c^{X_2}$ couplings in
  Eqs.~(\ref{eq:gneutral}) and ~(\ref{eq:gcargado}).

\begin{figure}[tb]
\begin{center}
\includegraphics[width=0.49\textwidth]{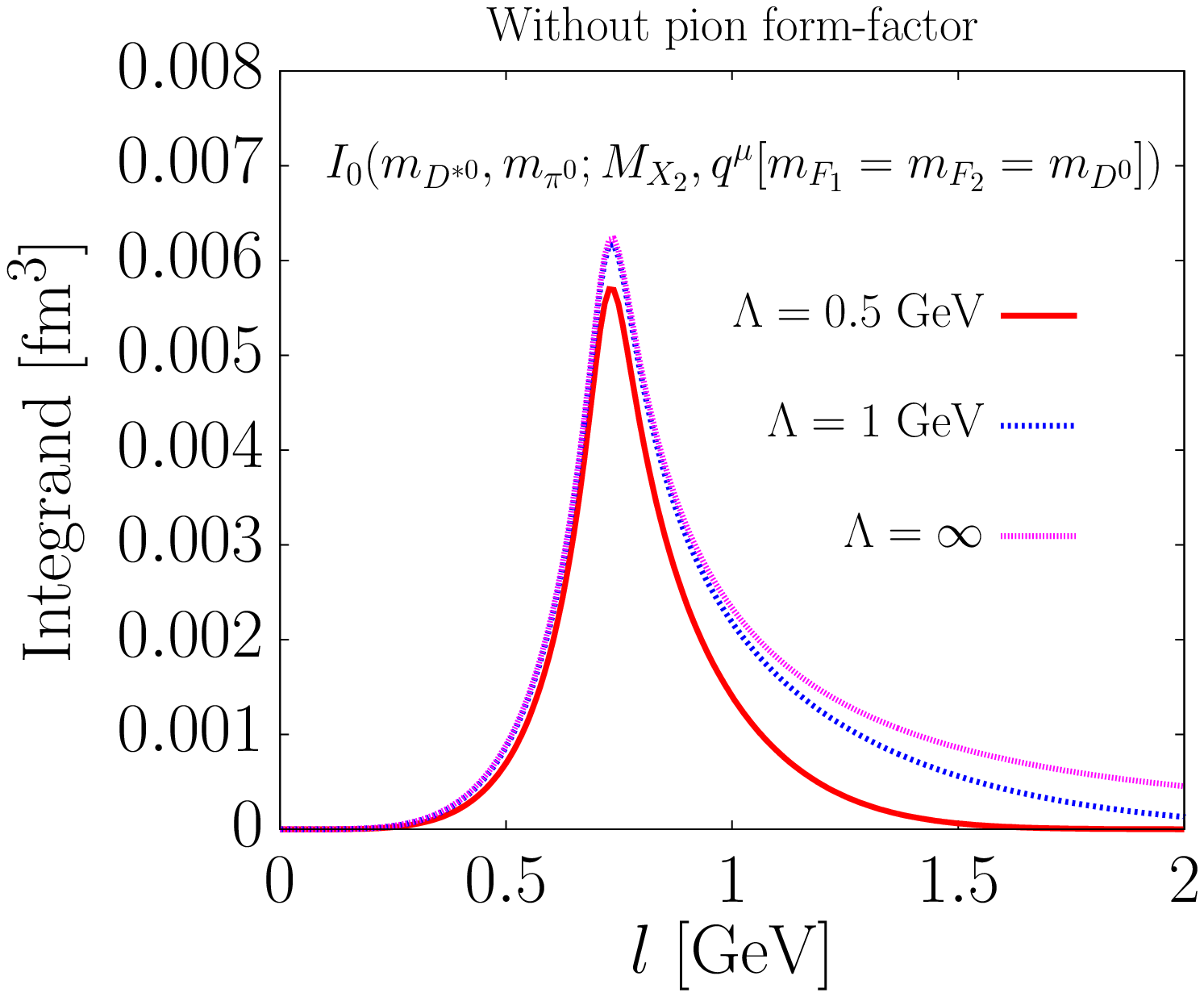}
\includegraphics[width=0.49\textwidth]{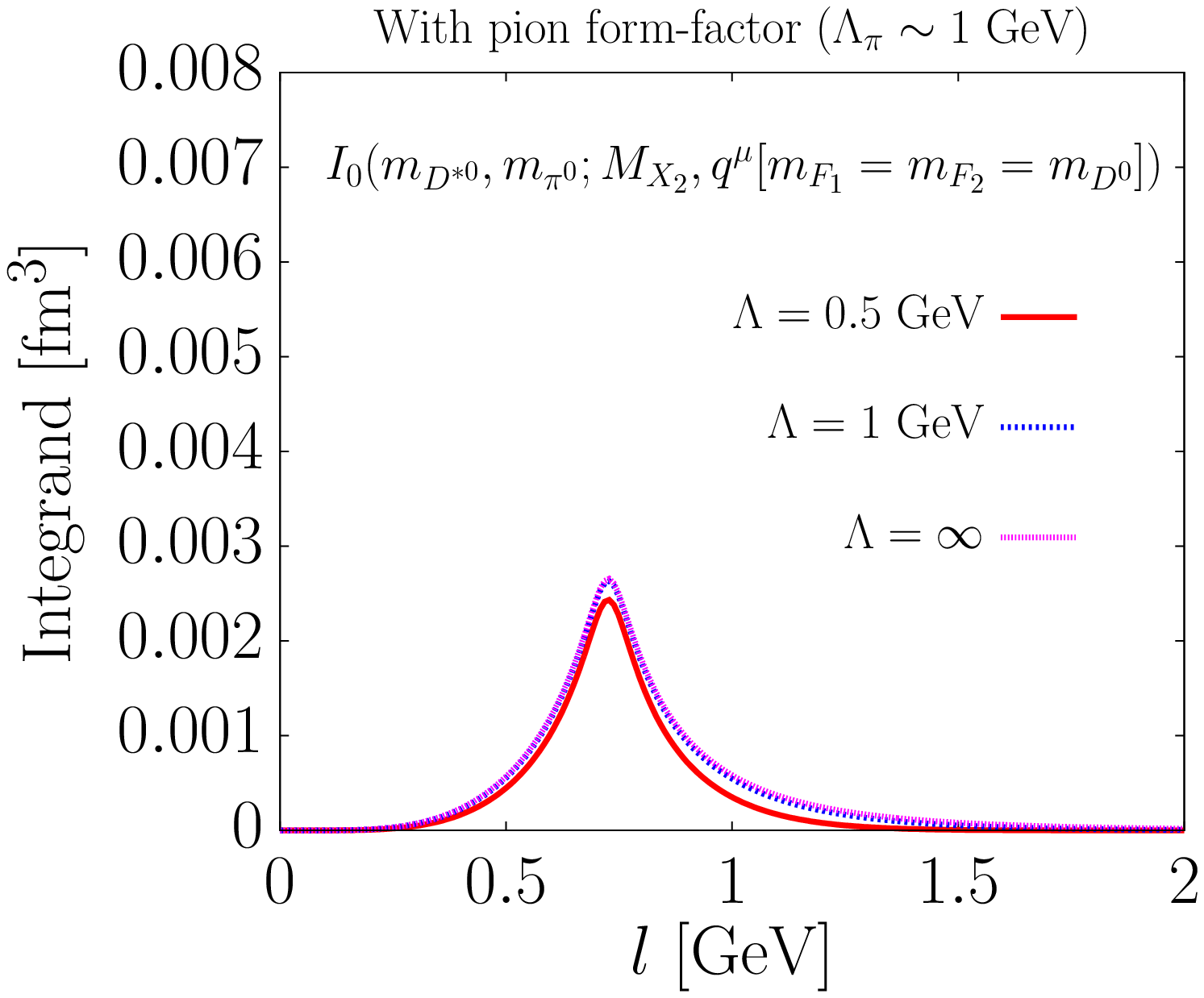}
\end{center}
\caption{Dependence of the $I_0(m_{D^{*0}},m_{\pi^0}; M_{X_2},q^\mu[m_{F_1}=
  m_{F_2}=m_{D^0}])$ integrand [see Eqs.~\eqref{eq:i0}-\eqref{eq:ff}]  on the
pion loop momentum $|\vec l\,|$. For the $X_2$ mass we have used $4013~\text{MeV}$. Results  with and
without the inclusion of the pion form factor [Eq.~\eqref{eq:ff}]
squared are presented in the right and
left plots, respectively. In both cases, three different choices of the
Gaussian regulator [Eq.~\eqref{eq:ffgauss}] in the
$D^*\bar D^* X_2$ vertex have been considered.}\label{fig:Integrands}
\end{figure}

Our scheme is based on a low-energy EFT, in which the momenta should be smaller
than a hard scale which serves as a momentum cutoff [see Eq.~\eqref{eq:UVG}].
The high-momentum modes are out of control in the low-energy EFT.
Therefore in the computation of the width, we include a Gaussian regulator at
the $D^*\bar D^* X_2$ vertex, as discussed in Eq.~(\ref{eq:ffgauss}).  The
cutoff should be the same as the one used in generating the $X_2$ as it is
related to the same unitary cut in the $D^*\bar D^*$ system. In
Fig.~\ref{fig:Integrands}, we display, as an example,  the dependence of the
$I_0(m_{D^{*0}},m_{\pi^0}; M_{X_2},q^\mu[m_{F_1}=
  m_{F_2}=m_{D^0}])$ integrand [see Eqs.~(\ref{eq:i0})-(\ref{eq:ff})]  on the
pion loop momentum. In the left plot we see that in spite of including the
Gaussian $X_2$ form factor, large momenta above 1 GeV provide a sizable
contribution to the integral ($\simeq$ 14\%, 30\% and 45\% for $\Lambda=$ 0.5
GeV, 1 GeV and $\infty$, respectively), which is an unwanted feature within the
low-energy EFT scheme and signals a sizeable short-distance contribution.
Indeed, the momentum of the exchanged pion, peaks at around 750~MeV, which is a
somehow large value in the sense that the hard scale for the chiral expansion
which controls the pionic coupling is $\Lambda_\chi\sim1$~GeV. We see that
below the peak, the curves for both cutoff values are very close to each other,
and they are also close to the curve corresponding to the case without
any regulator. This is the region where the low-energy expansion works and
thus model-independent conclusions can be made. The curves start deviating from
each another after the peak, that is in the region with a pion momentum
$\gtrsim\Lambda_\chi$. Because the loop integrals are not completely dominated
by momentum modes well below $\Lambda_\chi$, the widths of interest will bear
an appreciable systematic uncertainty. This is reflected in the fact that the widths in the second column
in Table \ref{tab:Decaywidths_hadron} are larger than those in the first column by a factor around 2.\footnote{Note that the coupling constants obtained with both cutoffs are
similar, see Eqs.~\eqref{eq:gneutral} and \eqref{eq:gcargado}, and thus the
difference should come mainly from the loop integration.}

On the other hand, the fact that the pion could be quite far off-shell should be
reflected in the $D^* D \pi$ vertex, which should be corrected, similarly as it
is done in the case of the $NN\pi$ one. Thus, to give an estimate of the
hadronic decay widths, in the spirit of the Bonn
potential~\cite{Machleidt:1987hj}, we have included a monopole pion-exchange
vertex form factor, with a hadron scale of the order of 1 GeV, in each of the
$D^* D \pi$ vertices [Eq.~(\ref{eq:ff})]. Its effect on the internal pion
momentum dependence of the $I_0(m_{D^{*0}},m_{\pi^0}; M_{X_2},q^\mu[m_{F_1}=
  m_{F_2}=m_{D^0}])$ integrand  is shown in the right plot of
Fig.~\ref{fig:Integrands}. The large pion momenta contribution ($|\vec{l}\,| >
1$ GeV), which is not reliable in a low-energy EFT calculation, is reduced now
to 6.5\%, 13\% and 16\% for $\Lambda=$ 0.5 GeV, 1 GeV and $\infty$,
respectively. This makes also more appropriate the non-relativistic treatment of
the charmed mesons adopted here. Besides, the dependence of the width on the UV
Gaussian regulator is significantly softer, though the widths are further
reduced by almost an order of magnitude.

We believe that the most realistic estimates are those obtained with the
inclusion of the pion-exchange form factor and the spread of results compiled
in  Table~\ref{tab:Decaywidths_hadron} give a conservative estimate
of the systematic uncertainties, beyond the mere existence of the
$X_2(4013) $ state, as discussed in Sect.~\ref{sec:X2charm}. We
remind here that because of the additional 20\% HQSS uncertainty,
approximately a 23\% (32\%) for $\Lambda=0.5$~GeV ($\Lambda=1$~GeV) of the $\X$
events [($M_{X(3872)},R_{X(3872)}$) MC samples] do not produce the $X_2$
as a bound state pole, since the strength of the resulting interaction in the
$2^{++}$ sector is not attractive enough to bind the $D^*\bar D^*$.

 Nevertheless, assuming the existence of the $X_2$ state, and in view
 of the results given in Table~\ref{tab:Decaywidths_hadron}, we
 estimate the $X_2\to D\bar D$ partial width
 (including both the charged and neutral channels) to be
\begin{equation}
  \Gamma(X_2\to D\bar D) = \big(1.2 \pm {\underbrace{0.3}_{\rm
        sys\,(\Lambda)}}\, ^{+1.3}_{-0.4}\big)~\text{MeV}, \label{eq:X2DD}
\end{equation}
where the first error accounts for the dependence on the UV Gaussian
regulator used in the $D^*\bar D^* X_2$ vertex, while the second one
is obtained from the uncertainties given in
Table~\ref{tab:Decaywidths_hadron}. This latter error includes both
some additional systematic (HQSS violations) and statistical
($X(3872)$ input used to fix the properties of the $X_2$ resonance)
uncertainties.  Notice that, as discussed above, the calculation is
probably already beyond the valid range of the EFT due to the large
contribution of high-momentum modes. We thus have adopted a more
phenomenological strategy and used the pion-exchange form factor with
a cutoff of 1~GeV to make an estimate of the decay widths. The values
presented in Eq.~\eqref{eq:X2DD} refer only to the last two columns in
Table~\ref{tab:Decaywidths_hadron} that include the effect of the
pion-exchange form factor. Thus, and considering both sets
  of errors, we cover the range 0.5--2.8 MeV, which accounts for
  all values including errors quoted in the
  table.

%%%%%%%%%%%%%%%%
%-------------------------------------------------------------------------------

\subsubsection{$X_{2}(4013) \to D \bar{D}^{*} (D^{*}\bar{D})$}

\begin{figure}[htb]
\begin{center}
\includegraphics[width=0.8\textwidth]{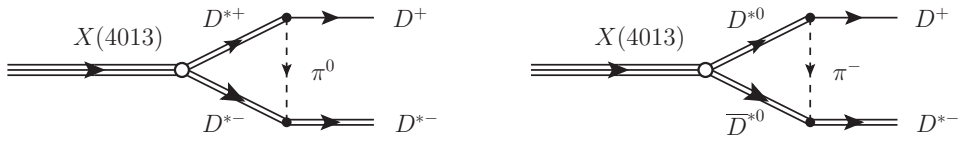}
\end{center}
\caption{Feynman diagrams for the $X_{2}(4013) \to D^{+} D^{*-}$
  decay. Diagrams for the $X_{2}(4013) \to D^{0} \bar{D}^{*0}$  transition are similar,
with the appropriate changes of the exchanged pion charges.}\label{Decay_DDStarbar}
\end{figure}

Here, we will study the $D^{+}D^{*-}$, and $D^{0}\bar{D}^{*0}$ channels, which
proceeds through the Feynman diagrams depicted in Fig.~\ref{Decay_DDStarbar}. This is
also a $d$-wave decay so that both angular momentum and parity are conserved. The decay widths are expected to be comparable to those found for
the $X_{2}(4013) \to D \bar{D}$ decays,  despite the phase space is considerably
more reduced.  This extra enhancement is caused by the extra multiplicity due to
the spin of the final $D^{*}(\bar{D}^{*}$) meson.
%
% -------------------------------------------------------------------------------
%-------------------------------------------------------------------------------

As commented before, we treat charm mesons non-relativistically and
obtain the decay amplitude for the $X_{2}(4013) \to D^{+} D^{*-}$ process as
\begin{eqnarray}
\mathcal{T(\lambda,\lambda_*)}_{D^+D^{*-}} &=& i\frac{N^* g^{2}}{f_{\pi}^{2}}
\epsilon_{ij}(\lambda)\epsilon_{mnj}\epsilon^{*n}(\lambda_*)\Big\{ g^{X_2}_c
I^{im}(m_{D^{*+}},m_{\pi^0}; M_{X_2},q^\mu \,)   \nonumber \\
&&  + 2 g^{X_2}_0 I^{im}(m_{D^{*0}},m_{\pi^-}; M_{X_2},q^\mu \,) \Big\}~,\label{eq:ddstpi1}
\end{eqnarray}
in the resonance rest frame. Here, $q$ is the 4-momenta of the $D^+$ meson,
and $\epsilon^n (\lambda^*)$ is the polarization vector
of the final $D^{*-}$ meson with helicity $\lambda_*$, $N^*=\sqrt{8
M_{X_2} m^2_{D^{*}}}  \sqrt{m_{D}m_{D^{*}}^3}$ and $\epsilon_{imn}$ is the 3-dimensional
Levi-Civita antisymmetric tensor.
Analogously, the $X_{2}(4013) \to D^{0} \bar D^{*0}$ amplitude is,
\begin{eqnarray}
 \mathcal{T(\lambda,\lambda_*)}_{D^0D^{*0}} &=& i\frac{N^* g^{2}}{f_{\pi}^{2}}
\epsilon_{ij}(\lambda)\epsilon_{mnj}\epsilon^{*n}(\lambda_*)\Big\{ 2 g^{X_2}_c
I^{im}(m_{D^{*+}},m_{\pi^+}; M_{X_2},q^\mu \,)   \nonumber \\
&&  +  g^{X_2}_0 I^{im}(m_{D^{*0}},m_{\pi^0}; M_{X_2},q^\mu \,) \Big\}.\label{eq:ddstpi2}
\end{eqnarray}
The two-body decay width in the $X_2$ rest-frame in this case
reads~\cite{Agashe:2014kda}:
%
%\begin{equation}
%d\Gamma = \frac{1}{32 \pi^{2} } \left| \bar{\mathcal{T}}_{\rm{total}} \right|
% ^{2} \frac{\left| \vec{q_{}} \right| }{M_{X}^{2}} d\Omega
%\end{equation}
\begin{equation}
\label{Eq:decaywidth2body-bis}
\frac{d\Gamma_a}{d\Omega(\hat q)} =  \frac{1}{5} \sum_{\lambda,\lambda_*}\left| \mathcal{T(\lambda,\lambda_*)}_a\right| ^{2} \frac{\left|
\vec{q}\, \right|}{32 \pi^2 M_{X_2}^{2}}, \qquad a= D^+D^{*-},
  D^0\bar D^{*0}~.
\end{equation}
The sum over the $\bar D^*$ and
$X_2$ polarizations can be easily done, and we get
\begin{equation}
\sum_{\lambda} \epsilon_{ij}(\lambda) \epsilon^*_{rs}(\lambda)
\sum_{\lambda_*} \epsilon^{*n}(\lambda_*)
\epsilon^p(\lambda_*)\epsilon_{mnj} \epsilon_{lps}   = \frac16 \left[
  7 \delta_{m l} \delta_{i r} + 2 \delta_{il}\delta_{rm} - 3
  \delta_{l r}\delta_{mi}\right], \quad i,l,m,r=1,2,3~. \label{eq:e2e1}
\end{equation}
The above tensor structure should be contracted with
$I^{im}(\vec{q}\,)\,I^{rl}(\vec{q}\,)$. We see that the sum over
polarizations of Eq.~(\ref{eq:e2e1}) is orthogonal to $\delta_{im}$ and
$\delta_{rl}$, which guarantees also here that only the UV finite
$I_{0}$ term of the three-point loop function is relevant.  The
contraction leads to a $\vec{q}^{\,4}$ factor,\footnote{In the $D\bar
  D$ mode, studied in Sect.~\ref{subsect:DDbar_decays} a factor of
  $2\vec{q}^{\,4}/3$ is obtained instead. Thus, neglecting the $D-D^*$
  mass difference, because the loop integrals are the same,
  we would find
\begin{equation}
\left| \mathcal{T}_{D\bar{D}^{*} (D^{*}\bar{D})}  \right|^{2} \simeq \frac{3}{2}  \left| \mathcal{T}_{D\bar{D}}\right|^{2}
\end{equation}
This extra factor 3/2 due to the spin-1 polarization vector produces
an enhancement of the $D\bar{D}^{*}$ decay mode with respect to the
$D\bar{D}$ one, which partially compensates the smaller available phase
space.} and thus the width scales like $|\vec{q}\,|^5$, as expected
for a $d$-wave decay.

Results for the $X_{2}\to D\bar{D}^{*}$ decay widths are also compiled
in Table \ref{tab:Decaywidths_hadron}. We only show predictions for
the $X_{2}(4013) \to D^{+} D^{*-} ( D^{0} \bar{D}^{*0})$ decays,
because being the $X_2$ an even $C$-parity state, its decay modes into
the charge conjugated final states have the same decay widths. In what
respects to the effect of the form factors, the discussion runs in
parallel to that in the Sect.~\ref{subsect:DDbar_decays},
though the effect of the pion-exchange form factor is significantly smaller
here (a factor 4 or 5 at most).  As expected, the widths are
comparable to those found for the $X_{2}\to D\bar{D}$ decays. Finally,
we estimate the partial $X_2\to D\bar D^*\,(D^* \bar D) $ width
(including both the charged and neutral channels as well as the
charge-conjugated modes) to be
\begin{equation}
  \Gamma(X_2\to D\bar D^*) +\Gamma (X_2\to D^* \bar D) = \big(2.9 \pm {\underbrace{0.5}_{\rm
        sys\,(\Lambda)}}\, ^{+2.0}_{-1.0} %\bigcap
)~\text{MeV}, \label{eq:ddsdt-resul}
\end{equation}
where the errors have been estimated as in Eq.~(\ref{eq:X2DD}). The
above result, together with that obtained previously for the $D\bar D$
channel, leads to a total $X_2$ width of the order of 2-8 MeV,
assuming its existence.

%----------------------------------------------------------------------------

\subsubsection{Charm decays: further analysis of uncertainties}

\begin{table}[tb]
\begin{center}
\begin{tabular}{l|cccc|cccc}
& FF$\pi$1 & FF$\pi$2& FF$\pi$1 & FF$\pi$2 &FF$\pi$1 & FF$\pi$2& FF$\pi$1 & FF$\pi$2\\
& FF$X_2-$G~~ & FF$X_2-$G~~ & FF$X_2-$L~~ & FF$X_2-$L~~& FF$X_2-$G~~ & FF$X_2-$G~~ & FF$X_2-$L~~ & FF$X_2-$L~~\\
~~$\Lambda%_{X_2 D^*\bar D^*}
~[{\rm GeV}]$~ &  ~$0.5$~ & ~$0.5 $~ &
  ~$0.4$~ & ~$0.4 ~$~ &  ~$1$~ & ~$1 $~ &
  ~$0.8$~ & ~$0.8 ~$~\\
\hline
$\Gamma (X_{2} \to D^{+} D^{-})$    & $0.5^{+0.5}_{-0.2}$ & 0.6
 & 0.5  & 0.6  & $0.8^{+0.7}_{-0.2}$ & 01.0 & 0.9 & 0.9 \\
$\Gamma (X_{2} \to D^{0} \bar{D}^{0})$   & $0.4^{+0.5}_{-0.2}$ &
0.5 & 0.4 & 0.4 & $0.6^{+0.7}_{-0.2}$ & 0.7  & 0.6  & 0.6  \\
\hline
$\Gamma (X_{2} \to D^{+} D^{*-})$  & $0.7^{+0.6}_{-0.3}$ & 0.9 (1.2)
 & 0.7& 0.8 (1.1) & $1.0^{+0.5}_{-0.2}$ & 1.2 (1.6)  & 1.0  &
1.1 (1.5) \\
$\Gamma (X_{2}  \to D^{0} \bar{D}^{*0})$   &  $0.5^{+0.6}_{-0.2}$
& 0.7 (0.9) & 0.5  & 0.6 (0.8) & $0.7^{+0.5}_{-0.2}$ & 0.8
(1.2)& 0.7& 0.8 (1.1)\\
\end{tabular}
\end{center}
\caption{ $X_{2}(4013) \to D \bar{D}, D \bar{D}^*$ decay widths (in
  MeV units) using different UV regularization schemes for the $D^*\bar D^* X_2$
  vertex and pion-exchange form factors. FF$X_2-$G and FF$X_2-$L stand
  for the results obtained with Gaussian (Eq.~(\ref{eq:UVG})) and
  Lorentzian (Eq.~(\ref{eq:fflor})) regularized local
  interactions, respectively. On the other hand, the widths in the
  columns FF$\pi$1 and FF$\pi$2 were obtained inserting the pion form
  factor of Eq.~(\ref{eq:ff}) and $F(\vec{l}^{\,\, 2},\Lambda_\pi) =
  e^{- \vec{l}^{\,2}/\Lambda_\pi^2}$ in each of the two $D^*
  D^{(*)}\pi$ vertices, respectively.  In the latter case, we take
  $\Lambda_\pi = 1.2$ GeV, as determined in the QCDSR
  calculation of Ref.~\cite{Navarra:2001ju} for the $D^*D\pi$
  coupling. In the $D \bar{D}^*$ decay mode, we also show results (in
  brackets) obtained when a larger cutoff, $\Lambda_\pi=1.85$ GeV, is
  used in the $D^*D^*\pi$ vertex, as obtained  in the QCDSR
  study carried out in Ref.~\cite{Carvalho:2005et} for this
  coupling. The results presented in Table
  \ref{tab:Decaywidths_hadron} correspond to the [FF$X_2-$G \&
  FF$\pi$1] columns and for the rest of choices we only provide
  central values. }
\label{tab:Decaywidths_hadron-uncert}
\end{table}
%-------------------------------------------------------------------------------

The uncertainties on the results compiled in Table
\ref{tab:Decaywidths_hadron} account both for reasonable estimates of HQSS
corrections, as well as for the statistical errors on the
inputs used to fix the LEC's that determine the properties (mass and
$D^*\bar D^*$ couplings) of the $X_2$ resonance.
Moreover we are using an EFT to describe these decays, which means that
there is an intrinsic uncertainty that can be determined systematically.
We indicated the size of this error
in Eqs.~(\ref{eq:X2DD}) and (\ref{eq:ddsdt-resul}).
For obtaining the EFT uncertainty we combined the predictions obtained
for two different UV Gaussian cutoffs ($\Lambda=0.5$ and 1 GeV) in the
$D^*\bar D^* X_2$ vertex, and used the spread of values to quantify this error.
The idea is that the residual dependence of the results on the cutoff
should provide an insight into the size of sub-leading corrections.

Now, we try to further test the robustness of the systematic errors quoted in
Eqs.~(\ref{eq:X2DD}) and (\ref{eq:ddsdt-resul}). To that end, we have
examined:
\begin{itemize}

\item The dependence of our results on the functional form of the
  UV regulators, both in the $D^*\bar  D^* X_2$ and $D^*D^{(*)}\pi$ vertices.

\begin{itemize} \item We have a contact theory with a Gaussian regulator
and a cutoff $\Lambda$ between 0.5 and 1 GeV. This theory, though very
simple, is the ${\rm LO}$ of an EFT expansion describing
the low energy dynamics
of heavy hadrons (see Ref.~\cite{Valderrama:2012jv} for details).
Within the EFT we can expand observable quantities as a power series of the type
\begin{eqnarray}
A = \sum_{\nu} \hat{A}_{\nu} \left( \frac{p}{\Lambda_M} \right)^{\nu}
\, ,
\end{eqnarray}
where $p$ is the momenta of the hadrons and $\Lambda_M$ the scale
at which hadrons stop behaving as point-like particles
(about the $\rho$ mass).
A ${\rm LO}$ calculation only keeps the first term in the series above.
Hence it should have a relative error of order $(p/\Lambda_M)$.
We stress that this is expected to be so regardless of the exact regulator
employed (gaussian, lorentzian, etc.), provided that the cutoff is
at least of the order of $\Lambda_M$.
The reason for that is that the calculations we show are
renormalizable: once the counter-terms are fixed\footnote{See for
  instance the
  discussion of Eqs.~(23) and (24) of Ref.~\cite{Gamermann:2009uq}.}, they
only contain negative powers of the cutoff $\Lambda$ when we expand on the large
cutoff limit. Hence, the uncertainty in the calculation is of order
$(p /\Lambda)$. By taking $\Lambda$ of the order of $\Lambda_M$,
observables are guaranteed to have a cutoff uncertainty
of the order $(p / \Lambda_M)$, equivalent to the EFT uncertainty.

There are several methods for making a more precise estimation of the EFT error:
the one we use in Table~\ref{tab:Decaywidths_hadron} is to vary the cutoff
around values of the order of $\Lambda_M$ (hence the choice of the
$0.5-1 \,{\rm  GeV}$ cut-off window).
EFT predictions for two different cutoffs differ by a factor
of $(p /\Lambda_M)$ and that is why we chose this particular
way of assessing the errors.

Alternatively, one could use different regulators or form factors to assess
the size of this error.
This idea also gives a cross-check of the previous error estimates based
on varying the cut-off.
We have employed a different regulator to check that our former
estimation of the EFT errors is correct and to show that the particular
regulator employed is not important. We closely follow here
the discussion of Sect. VII of Ref.~\cite{Gamermann:2009uq} and
consider a Lorentz form for the regulator
\begin{equation}
 \left<\vec{p}\,|V|\vec{p}\,'\right> = C_{IX} \left
  [\frac{\Lambda^2}{\Lambda^2+\vec{p}\,^{2}}\right]\left
  [\frac{\Lambda^2}{\Lambda^2+\vec{p}\,'^{\,2}} \right] \, , \label{eq:fflor}
\end{equation}
with two values of the cutoff, namely $\Lambda=0.4$ and 0.8 GeV, which were
obtained by multiplying the Gaussian cutoffs by a factor of
$\sqrt{2/\pi}$~\cite{Gamermann:2009uq}.
The resulting\footnote{With the Lorentzian regularized local potential, we
re-obtain the counter-terms $C_{0X}$ and $C_{1X}$ from the $X(3872)$
inputs, which are then used to find the mass of the $X_2$
resonance and its couplings  to the $D^{*0}\bar D^{*0}$ and
$D^{*-} D^{*+}$ meson pairs. These physical quantities hardly
change, because the $X_2$ is a very loosely bound state and its
dynamics is very little sensitive to the details of the $D^*\bar D^*$
interaction at short distances.}
widths are presented in Table~\ref{tab:Decaywidths_hadron-uncert}
and turn out to be rather insensitive to the form of the regulator
(this is to be understood within the limits of the expected EFT uncertainty).

\item Next we have studied the dependence of the widths on the pion
  form factor that accounts for the off-shellness of the pion in the
  mechanisms depicted in Figs.~\ref{Decay_DDbar} and
  \ref{Decay_DDStarbar}. To that end we used the results from the
  QCD sum rule (QCDSR) calculations performed in
  Refs.~\cite{Navarra:2001ju,Carvalho:2005et}. The first of these
  two works considers the $D^*D\pi$ vertex: it was found the
  form factor is harder if the off-shell meson is heavy, implying that
  the size of the vertex depends on the exchanged meson. This means
  that a heavy meson will see the vertex as point-like, whereas the
  pion will see its extension. The authors of Ref.~\cite{Navarra:2001ju}
  find an on-shell value $g=0.48\pm 0.05$ (note the different
  definition used in this reference), around 1-2 sigmas below the
  value of 0.57 used in this work. In addition, they adjust their
  results for off-shell pions to a Gaussian form
  $e^{l^2/\Lambda_\pi^2}$, with $l^\mu$ the pion four momentum, and
  find $\Lambda_\pi=1.2$ GeV. This form-factor\footnote{To use the
    form factor of Ref.~\cite{Navarra:2001ju} in the computation of
    the widths, we have approximated the pion four momentum squared by
    $-\vec{l}^{\,2}$, which is sufficiently accurate because the
    dominant part of the integration comes from regions where the two
    virtual $D^*$ and $\bar D^*$ mesons are almost on-shell. In this
    region, the energy of both heavy light vector mesons is
    approximately $M_{X_2}/2$ which coincides with that of the
    external heavy mesons, and hence $l^0$ is much smaller than
    $|\vec{l}\,|$. }, in the
  region of interest for this work, turns out to be in a good
  agreement with that used to obtain the results presented in
  Table~\ref{tab:Decaywidths_hadron}. This can be seen in the new
  results showed in Table~\ref{tab:Decaywidths_hadron-uncert} and
  obtained with this new pion off-shell form-factor
  (FF$\pi$2). Deviations from our previous estimates of the
  $X_{2}(4013) \to D \bar{D}, D \bar{D}^*$ decay widths are both much
  smaller than the (HQSS \& exp) uncertainties quoted in
  Table~\ref{tab:Decaywidths_hadron}, and well comprised within the
  systematic uncertainty generated when the $D^*\bar D^* X_2$ cutoff
  varies in the 0.5-1 GeV window.

  In the $X_2 \to D\bar D^*$ decay, there also appears the $D^*
  D^*\pi$ coupling in one of the vertices, see
  Fig.~\ref{Decay_DDStarbar}.  The off-shell behavior of this vertex
  might differ from that of the $D^* D\pi$ one. This coupling was
  studied using QCDSR in \cite{Carvalho:2005et} where, translating the
  definition used therein to the one used here, it was found an
  on-shell value $g=0.56\pm 0.07$ in good agreement with the HQSS
  expectations. The off-shell behavior was described by a Gaussian,
  as in the case of the $D^*D\pi$ vertex, but with a significantly
  larger cutoff, $\Lambda_\pi$= 1.85 GeV. This significant difference
  is somehow surprising from the HQSS point of view, and we should
  note that the QCDSR actual calculations in \cite{Carvalho:2005et}
  were carried out for significantly larger values of $l^2> 4$ GeV$^2$
  than in the case of the $D^*D\pi$ coupling analyzed in
  Ref.~\cite{Navarra:2001ju}. Nevertheless, we used this softer
  dependence for the $D^*D^*\pi$ vertex and re-computed the $D\bar
  D^*$ widths. Results are shown in brackets in
  Table~\ref{tab:Decaywidths_hadron-uncert}. Changes are now larger,
  and in general are of the order of 50\%, though they could be still
  accommodated within the HQSS and EFT uncertainties already
  considered in our original calculations. The large momenta of the
  external mesons, that can even exceed 0.5 GeV, make it possible that
  the short distance details of the decay mechanisms could be
  relevant. This is the weakest point in our scheme. The reason is
  that the EFT uncertainty is expected to be 0.5 GeV$/\Lambda_M
  \gtrsim 1/2$, as the variations of the $D\bar D^*$ widths in
  Table~\ref{tab:Decaywidths_hadron-uncert} seem to confirm, and the
  calculation is really on the limit of validity of this kind of
  description and should only be considered as a reasonable estimate.

\end{itemize}

\item The contribution of decay mechanisms driven by the exchange of
  shorter range mesons (heavier) than the pion.

Since the momenta of the external charmed mesons can exceed 0.5~GeV, one might
think that  shorter range contributions such as the $\rho$ and $\omega$
exchanges could be sizable, and
even comparable to those of the diagrams depicted in Figs.~\ref{Decay_DDbar} and
\ref{Decay_DDStarbar} for the exchange of a pion.
We will focus on the $X_2 \to D\bar D$ decay mode,
where the momenta of the external mesons is the largest and
we will begin by studying the effects of the exchange of a $\rho$ meson.
If we use the phenomenological $D^*D\rho$ Lagrangian given in Eq.~(3e)
of Ref.~\cite{Oh:2000qr}, we find that the amplitudes of this new
contribution can be obtained from those driven by pion exchange,
and given in Eqs.~(\ref{eq:ddpi1}) and (\ref{eq:ddpi2}),
by replacing $m_{\pi^0}$ and $m_{\pi^+}$
by $m_{\rho^0}$ and $m_{\rho^+}$ in the loop integrals and
\begin{eqnarray}
\frac{g^2}{f_\pi^2} \to -\frac{m_D}{m_{D^*}} g^2_{D^*D\rho}
\end{eqnarray}
where we have neglected  $|\vec{q}\,|^2/m_{D^{(*)}}^2$ terms, with
$\vec{q}$ the c.m. three-momentum of the external $D$ and $\bar D$
mesons. The coupling constant $g_{D^*D\rho}$ has been computed in
various schemes \cite{Oh:2000qr,Li:2002pp, Wang:2007mc, Bracco:2007sg,
  Rodrigues:2010ed} (ordinary and light cone QCDSR, vector dominance model,
SU(4), etc.)
and it varies in the range $\left[2.8\pm 0.1, 4.3\pm 0.9 \right]$ GeV$^{-1}$
(see Table 5 of Ref.~\cite{Rodrigues:2010ed}).
Taking $g_{D^*D\rho}\sim 5$ GeV$^{-1}$, in the highest part of the
interval of calculated values, we would have
$g^2_{D^*D\rho}/(g^2/f_\pi^2)\sim 2/3$, while a direct calculation of
the loop integrals shows that those calculated using the $\rho$ mass
instead of the pion mass are around a factor of two smaller.
Altogether, this indicates that the absolute values of the $\rho$-exchange amplitudes
are about a factor of three smaller than those driven by the pion exchange.
If one uses $g_{D^*D\rho}\sim 3$ GeV$^{-1}$, now in the lowest part of the
interval of values,  the $\rho$ contribution will be, at the level of
amplitudes, around eight times smaller than those considered in the
present work. In any case, these effects are smaller than  the
 HQSS and EFT uncertainties quoted in
 Table~\ref{tab:Decaywidths_hadron}, and therefore it seems justified
 to neglect them. On the other hand, the $\omega$-exchange contributions
 are even smaller, around a factor of three, because this meson is
 neutral, and it cannot generate mechanisms where a light charged
 meson is exchanged.

To estimate the size of the $\rho$ exchange contribution in the $X_2\to D\bar D^*$
decay mode, in addition to the $D^*D\rho$ vertex, we have used the  
phenomenological $D^*D^*\rho$ Lagrangian given in Eq.~(1f) of 
Ref.~\cite{Oh:2000qr}.  The amplitudes of this new
mechanism can be obtained from those driven by pion exchange,
and given in Eqs.~(\ref{eq:ddstpi1}) and (\ref{eq:ddstpi2}),
by replacing $m_{\pi^0}$ and $m_{\pi^+}$
by $m_{\rho^0}$ and $m_{\rho^+}$, and $P_2(x) \to (P_2(x) + x
|\vec{q}\,|/l)$ in the computation of the loop integral $I_0$ (see
Eq.~(\ref{eq:i0}) of  Appendix \ref{app:three-loopfunction-H})
and using the appropriate coupling constants
\begin{eqnarray}
\frac{g^2}{f_\pi^2} \to \sqrt{\frac{m_D}{m_{D^*}}} g_{D^*D\rho}\frac{g_{D^*D^*\rho}}{m_{D^*}}
\label{eq:phase}
\end{eqnarray}
where we have neglected as before the $|\vec{q}\,|^2/m_{D^{(*)}}^2$
terms. (At this point, the phase in Eq.~(\ref{eq:phase}) is
  arbitrary, because the phase convention used in \cite{Oh:2000qr} and
  in this work for the heavy meson fields is different.) The $g_{D^*D^*\rho}$ coupling is set to 2.52 in
\cite{Oh:2000qr}, while it is estimated to be $4.7\pm 0.2$ in 
\cite{Bracco:2007sg,Bracco:2011pg} (QCDSR). On the other hand, there exist some partial
cancellations in the loop integral in
this case, and it turns out to be at least a factor of 5 smaller than
that calculated for the pion driven
contribution. Thus, and having in mind the previous discussion for the
expected range of values of the $g_{D^*D\rho}$ coupling, we 
conclude that the $\rho-$exchange contribution can be safely neglected in
the $X_2\to D\bar D^*$ decay mode.
\end{itemize}

In view of the results discussed in this subsection, we should acknowledge that as a result
of the contribution from highly virtual pions, which is certainly in
the limit of applicability of the low-energy EFT, the hadronic decay widths bear a large systematic
uncertainty. Nevertheless, we have given arguments to be reasonably
convinced that the results quoted in Eqs.~(\ref{eq:X2DD}) and
(\ref{eq:ddsdt-resul}) provide sensible estimates for
the $X_{2}(4013) \to D \bar{D}, D \bar{D}^*$  widths.

In the next subsection, we will study these hadronic decays in the
bottom sector. There, the considerations  are parallel to those
discussed here for the charm sector.

\subsection{Bottom decays}
\label{section:BottomSector}

Thanks to the heavy flavor symmetry, the results of the
previous subsection can be trivially extended to the bottom sector. There,  we
have a robust prediction, even when HQSS uncertainties (20\%) are taken into account,
for the $X_{b2}$ resonance (see Fig.~\ref{fig:mass-histo-b}). Moreover, all sort
of non-relativistic approximations adopted in the current scheme are now more
suited, since the range of variation of the internal pion momentum in the loops
is similar to that shown in Fig.~\ref{fig:Integrands} for the charm sector.

On the other hand, as discussed in Sect.~\ref{sec:xb2}, we do not
expect any significant isospin breaking effects and the $X_{b2}$
resonance would be a pure $I = 0$ state, with equal coupling to its
neutral and charged components. For simplicity, we will also neglect
the tiny difference between $B^0$ and $B^{\pm}$ masses, and we will
use a common mass $m_B=(m_{B^0}+m_{B^{\pm}})/2=5279.42~\text{MeV}$. Yet, for
the pion mass that appears in the loop integral, we take the isospin
averaged value $m_\pi= (2m_{\pi^{\pm}}+m_{\pi^0})/3$. Note that the relevant
internal pion momentum is around 750 MeV. With all these
approximations, we find
\begin{eqnarray}
\Gamma(X_{b2}\to B\bar B) &=& \frac{3g^4(g^{X_{b2}})^2}{5\pi f_\pi^4}\frac{m_B^2
  m_{B^*}^4}{M_{X_{b2}}}\,|\vec{q}\,|^5\, \Big (I_0(m_{B^*},m_\pi; M_{X_{b2}},q^\mu[m_{F_1}=
  m_{F_2}=m_{B}])\Big)^2, \\
%&&\nonumber \\
\Gamma(X_{b2}\to B\bar B^*) &=& \frac{9g^4(g^{X_{b2}})^2}{10\pi f_\pi^4}\frac{m_B
  m_{B^*}^5}{M_{X_{b2}}}\,|\vec{q}\,|^5\, \Big (I_0(m_{B^*},m_\pi; M_{X_{b2}},q^\mu[m_{F_1}=m_B,
  m_{F_2}=m_{B^*}])\Big)^2,~~~~~
\end{eqnarray}
for any charge channel ($B^+B^-$, $B^0\bar B^0$, $B^+B^{*-}$, $B^0\bar
B^{*0}$) or charge conjugation mode ($B^*\bar B$). Our results for
these decay widths are presented in Table~\ref{tab:Decaywidths_bottom_hadron}.
We notice in passing that following heavy flavor symmetry we use
the same value of $g=0.570\pm0.006$ in the charm and bottom decays.
%Yet lattice QCD
%calculations~\cite{Detmold:2012ge} indicate that the $1/m_Q$ corrections
%to $g$ are not random: the value of $g$ in the $m_Q = \infty$ limit
%is predicted to be $\sim 0.45$, smaller than $g \sim 0.57$
%in the charm sector.
It agrees very well with a recent lattice calculation with relativistic bottom
quarks which gives $g_b=0.569\pm0.076$~\cite{Flynn:2013kwa} (we have added
the systematic and statistic errors in quadrature). Yet, lattice
calculations with static heavy quarks tend to give smaller values, see
Ref.~\cite{Bernardoni:2014kla} and references therein.
For instance, the ALPHA Collaboration presented a precise computation with the result of
$g_\infty=0.492\pm0.029$~\cite{Bernardoni:2014kla}.
Thus we expect that the decay widths
of Table~\ref{tab:Decaywidths_bottom_hadron}
slightly overestimate the real ones.

For the $B\bar B$ mode we find a pronounced dependence on the Gaussian
cutoff $\Lambda$ employed in the dynamical generation of the resonance. This is
inherited from the strong dependence of the $X_{b2}$ mass on this UV
cutoff, as discussed in Eq.~(\ref{eq:g_Xb2}), which affects the
available phase space for the decay. With all these shortcomings, we
expect a partial width in the 1-10~MeV range, when both charge modes
are considered.

In the $B \bar B^{*}$ decay mode, the impact of the Gaussian regulator is
even larger, because it turns out that for $\Lambda = 1$ GeV, the
central value of the resonance mass $M_{X_{b2}} = 10594^{+22}_{-26}~\text{MeV}$
is located below the threshold $m_{B} + m_{B^{*}} \sim 10604~\text{MeV}$. Thus,
in that case, the decay will be forbidden.
For $\Lambda = 0.5$ GeV, we estimate a width also in the 4-12 MeV
range, when the four possible decay modes ($B^+B^{*-}$, $B^0\bar
B^{*0}$, $B^-B^{*+}$, $\bar B^0 B^{*0}$)  are considered.

\begin{table}[tb]
\begin{center}
\begin{tabular}{l|cc|cc}
 &  \multicolumn{2}{c|}{without pion-exchange FF} & \multicolumn{2}{c}{with
    pion-exchange FF}\\
 &  ~$\Lambda = 0.5 ~{\rm{GeV}}$~ & ~$\Lambda = 1 ~{\rm{GeV}}$~ &
  ~$\Lambda = 0.5 ~{\rm{GeV}}$~ & ~$\Lambda = 1 ~{\rm{GeV}}$~ \\
\hline
$\Gamma (X_{b2} \to B \bar B)$  [MeV] &
$26.0^{+1.0}_{-3.3}$ & $8^{+15}_{-7}$ & $4.4^{+0.1}_{-0.4}$ & $0.7^{+1.4}_{-0.6}$  \\
\hline
$\Gamma (X_{b2} \to B \bar B^{*})$  [MeV]  &
$7.1^{+3.4}_{-3.7}$ &$-$ &  $2.0^{+0.9}_{-1.0}$ & $-$
\end{tabular}
\end{center}
\caption{
$X_{b2} \to B \bar{B}, B \bar{B}^{*}$ decay widths (here $B \bar
  B^{(*)}$ refers to  any of the neutral or charged modes, but it is not the sum of both).  The results are given for
  different treatments of the three-point loop function. The errors
  have been obtained using   the  Monte Carlo analysis explained in Table
\ref{tab:Decaywidths_hadron}, but now considering the $X_{b2}$ mass
histograms displayed in Fig.~\ref{fig:mass-histo-b} and the coupling
given in Eq.~(\ref{eq:g_Xb2}). The decay width of the
  $X_{b2 }\to \bar B{B}^{*}$ mode is the  same because of charge
  conjugation symmetry. }
\label{tab:Decaywidths_bottom_hadron}
\end{table}

\section{\boldmath The $X_{2}$ and $X_{b2}$ radiative decays}
\label{sec:radia}
In this section, we will study the $X_2 \to D \bar D^* \gamma $ and
$X_{b2} \to \bar B  B^* \gamma $ decays. The  interaction of the
photon with the $s$-wave heavy mesons contains two contributions
that correspond to the magnetic couplings  to the light  and
heavy quarks~\cite{Amundson:1992yp} (see also
Appendix~\ref{sec:appHHgamma}).
Both terms are needed to understand the observed electromagnetic
branching fractions of the $D^{*+}$ and $D^{*0}$ because a cancellation
between the two terms accounts for the very small width of the
$D^{*+}$ relative to the $D^{*0}$ \cite{Agashe:2014kda}. Actually, one
finds~\cite{Amundson:1992yp,Hu:2005gf}:
\begin{eqnarray}
\Gamma(D^{*0} \to D^0 \gamma) &=& \frac{\alpha}{3}
\frac{m_{D^0}}{m_{D^{*0}}}\left( \beta_1 + \frac2{3m_c}\right)^2
E_\gamma^3, \quad \beta_1= \frac23 \beta - \frac{g^2m_K}{8\pi f^2_K}-
\frac{g^2m_\pi}{8\pi f^2_\pi}, \label{eq:radiaBs1}\\
\Gamma(D^{*+} \to D^+ \gamma) &=& \frac{\alpha}{3}
\frac{m_{D^+}}{m_{D^{*+}}}\left( \beta_2 + \frac2{3m_c}\right)^2
E_\gamma^3, \quad \beta_2= -\frac13 \beta + \frac{g^2m_\pi}{8\pi
  f^2_\pi}, \label{eq:radiaBs2}
\end{eqnarray}
where $E_\gamma$ is the photon energy, $m_c$ the charm quark mass and
$\alpha\sim 1/137.036$ the fine-structure constant. In the
non-relativistic constituent quark model $\beta =1/m_q\sim 1/330$
MeV$^{-1}$, where $m_q$ is the light constituent quark mass. Heavy
meson chiral perturbation theory allows one to improve upon this approximation
by including corrections from loops with light
Goldstone bosons, which give ${\cal O}(\sqrt{m_q})$ corrections
\cite{Amundson:1992yp}.

Using isospin symmetry to relate $\Gamma(D^{*0}\to D^0 \pi^0)$ and
$\Gamma(D^{*+}\to D^0 \pi^+)$, correcting by the slightly different
available $p$-wave phase space in each of the two decays,  and taking into account the
experimental $D^{*0}$ and $D^{*+}$ widths and radiative branching
fractions quoted by the PDG~\cite{Agashe:2014kda}, we find:
\begin{equation}
\Gamma(D^{*0} \to D^0 \gamma) = (22.7 \pm 2.6)~\text{keV}, \qquad
\Gamma(D^{*+} \to D^+ \gamma) = (1.33 \pm 0.33)~\text{keV}.
\end{equation}
These values differ from those used in Ref.~\cite{Hu:2005gf} because of the
recent accurate BABAR measurement of the $D^{*+}$ decay width,
mentioned in Sect.~\ref{section:hadron-decays}, which is around  10\%
smaller than the previous CLEO one used in Ref.~\cite{Hu:2005gf}.  Fixing
the charm quark mass to $m_c=1.5~\text{GeV}$, we fit the parameter $\beta$ to the
above experimental values and find $\beta^{-1} = (293\pm 11)~\text{MeV}$.

In what follows, we will study decays of the type $X_2\to P   \bar P^*
\gamma$, being $P$ and $ P^*$ pseudoscalar and vector heavy-light
mesons, respectively. Let us define $p_1^\mu$, $p_2^\mu$ and $p_3^\mu$  as the
four vectors of the final photon, pseudoscalar and vector mesons,
respectively. Besides, let us define the invariant masses
$m^2_{ij}=p_{ij}^2=(p_i+p_j)^2$, which satisfy
\begin{equation}
m^2_{12}+m^2_{13}+m^2_{23}= M_{X_2}^2+m_1^2+m_2^2+m_3^2 =
M_{X_2}^2+m_P^2+m_{P^*}^2 .
\end{equation}
Since, as we will see, the Feynman amplitudes  depend only on the
invariant masses $m_{12}^2$ and $m_{23}^2$  of the final
$\gamma P$ and $P\bar P^*$ pairs, respectively,
we can use the standard form for the Dalitz plot~\cite{Agashe:2014kda}
\begin{equation}
d\Gamma = \frac{1}{(2\pi)^3}\frac{1}{32 M_{X_2}^3} \overline{|\mathcal{T}|}^2
dm_{23}^2 dm_{12}^2 ,
\end{equation}
with $\overline{|\mathcal{T}|}^2$ the absolute value squared of the decay
amplitude with the initial and final polarizations being averaged and summed,
respectively. Thus, we readily obtain
\begin{eqnarray}
\Gamma &=& \frac{1}{(2\pi)^3}\frac{1}{32 M_{X_2}^3}
\int^{M_{X_2}^2}_{(m_{P}+m_{P^*})^2} dm_{23}^2
\int^{(m^2_{12})_{\rm max}}_{(m^2_{12})_{\rm min}}d m^2_{12}
\overline{|\mathcal{T}|}^2 ,
\end{eqnarray}
where for a given value of $m^2_{23}$, the range of $m^2_{12}$ is determined by its
values when $\vec{p}_P$ is parallel or anti-parallel to
$\vec{p}_{\gamma}$~\cite{Agashe:2014kda}:
\begin{eqnarray}
(m^2_{12})_{\rm max} & = & (E^*_\gamma+E^*_P)^2 -(p^*_\gamma-p^*_P)^2 ,
\nonumber \\
(m^2_{12})_{\rm min} & = & (E^*_\gamma+E^*_P)^2 -(p^*_\gamma+p^*_P)^2 ,
\end{eqnarray}
with $E^*_P = (m_{23}^2+m_{P}^2-m_{P^*}^2) /2m_{23}$ and
$E^*_{\gamma} =
(M^2_X-m_{23}^2) /2m_{23}$ the energies
of the $P$ meson and photon in the $m_{23}$ c.m. frame, respectively, and
$p^*_{P,\gamma}$ the moduli of the corresponding 3-momenta.

Because of parity conservation, this is a $p$-wave decay and hence
the photon  momentum  appears always
in the amplitudes. In the $X_2$  rest frame, it is given by
%
%\begin{equation}
$ |\vec{p}_\gamma| = E_\gamma = {M_{X_2}^2-m_{23}^2}/({2M_{X_2}}) $.
%\end{equation}

\subsection{\boldmath $X_2(4013) \to D \bar D^* \gamma $}

\begin{figure}[tb]
\begin{center}
\includegraphics[width=0.65\textwidth]{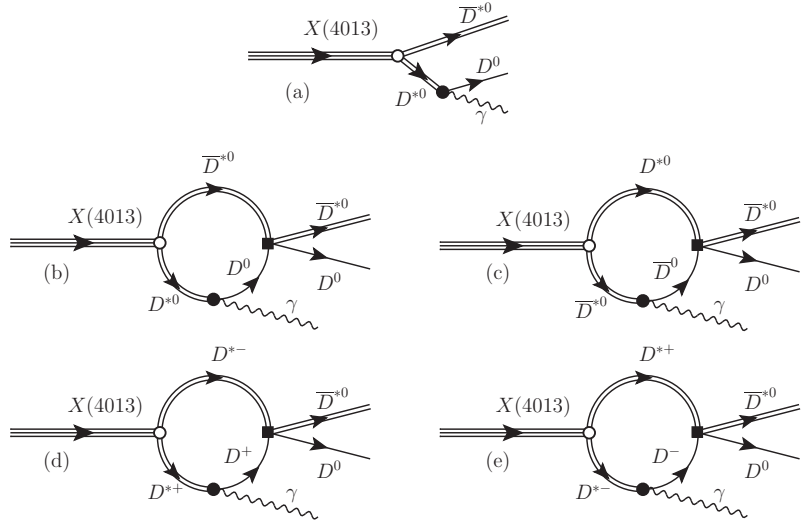}
\caption{Feynman diagrams for the $X_{2}(4013) \to D^{0}
\bar{D}^{*0} \gamma$ decay. Diagrams for the $D^{+} D^{*-}\gamma$
transition are similar. }\label{fig:Decay_gamma}
\end{center}
\end{figure}

We will first consider the $X_{2}(4013) \to D^{0} \bar{D}^{*0} \gamma$
decay, which proceeds according to the Feynman diagrams depicted in
Fig.~\ref{fig:Decay_gamma}. This decay can take place directly through
the radiative transition of the constituent $D^{*0}$ as shown in
Fig.~\ref{fig:Decay_gamma}(a), which is the tree level
approximation. However, there are other mechanisms driven by the
$D\bar D^*$ FSI. After emitting the photon, the vector meson
$D^{*0}$ transits into the $D^0$, and it can interact with the
other constituent in the $X_2$ as shown in
Fig.~\ref{fig:Decay_gamma}(b). There is a third (c) mechanism in which
the photon is emitted from the $\bar D^{*0}$ meson, and the virtual
$D^{*0}\bar D^0$ rescatter into $D^{0}\bar D^{*0}$. Finally,
Fig.~\ref{fig:Decay_gamma}(d) and (e) present another possibility,
namely the decay can also occur through the charged $D^{*+}D^{*-}$
component of the $X_2$ resonance, and the virtual charged $D^+D^{*-}$
and $D^{*+}D^{-}$ pairs then rescatter into $D^{0}\bar
D^{*0}$. Because the $X_2$ has a well defined charge parity~($+$), the decay
width into the charge conjugated mode $\bar D^{0}D^{*0}\gamma$ is the same.
The Feynman diagrams contributing to the $D^{+}
D^{*-}\gamma$ decay mode are similar with obvious replacements. 

One could also construct other FSI mechanisms by replacing the $D^*D\gamma$ vertices in
   Fig.~\ref{fig:Decay_gamma} with the $D^* D^*\gamma $ ones. These
   contributions are small and will be discussed below.

\subsubsection{Tree Level Approximation}

The Feynman amplitude of the mechanism depicted in
Fig.~\ref{fig:Decay_gamma}(a) reads (as in the previous sections, we treat the charm mesons
non-relativistically)
\begin{equation}
-i\mathcal{T(\lambda,\lambda_*,\lambda_\gamma)}_{D^{0} \bar{D}^{*0}
  \gamma} = g^{X_2}_0(m_{12}) \sqrt{4\pi \alpha}N_\gamma
  \left(\beta_1+\frac{2}{3m_c}\right) \frac{\epsilon_{ijm}
  \epsilon^{jn}(\lambda)\epsilon^{*n}(\lambda_*)
  \epsilon^{*i}_\gamma(\lambda_\gamma)p^m_\gamma }
  {2m_{D^{*0}}\left(m_{12}-m_{D^{*0}}+ i\varepsilon\right)} , \label{eq:radia1}
\end{equation}
with  $m_{12}$  the invariant mass of the final
$\gamma D^0$ pair. Besides, $\epsilon^i (\lambda_\gamma)$ is the polarization vector
of the final photon with helicity $\lambda_\gamma$, $\vec{p}_\gamma$
is its three momentum and $N_\gamma=\sqrt{8
M_{X_2} m^2_{D^{*}}}  \sqrt{m_{D}m_{D^{*}}}$ accounts
for the normalization of the heavy meson
fields and the $X_2D^{*0}\bar D^{*0}$ coupling. Finally,
\begin{equation}
g^{X_2}_0(m_{12}) = g^{X_2}_0 \times
e^{-(\vec{p}_{12}^{\,2}-\gamma^2)/\Lambda^2}=  g^{X_2}_0 \times
e^{-m_{D^{*0}}(m_{D^{*0}}-m_{12})/\Lambda^2}  \label{eq:offshell}
\end{equation}
with $\vec{p}_{12}^{\,2} \simeq m_{D^{*0}} (M_{X_2}-m_{D^{*0}}-m_{12})$
  and $\gamma^2= m_{D^{*0}}(M_{X_2}-2m_{D^{*0}})<0$. The Gaussian form
    factor is inherited from the $D^{(*)} \bar D^{(*)}$ EFT UV
    renormalization scheme.

We have neglected the $ D^{*0}$ width in the above
propagator because, since it is quite small, its inclusion only leads to small numerical
variations in the  decay rate which are certainly smaller than uncertainties
induced by the errors in the coupling $g_0^{X_2}$ and
the mass of the $X_2(4013)$ resonance. Similarly, the use of the non-relativistic $D^{*0}$ propagator instead of
$\left(m_{12}^2-m^2_{D^{*0}} + i\varepsilon\right)^{-1}$ leads also to
numerically negligible differences, as compared to the HQSS corrections.
The sum over the $\bar D^{*0}$, $\gamma$ and
$X_2$ polarizations can be easily done, and we get
\begin{equation}
\overline{|\mathcal{T}|}^2_{D^{0} \bar{D}^{*0}} = \frac{16 \pi \alpha
    M_{X_2}m_{D^*}
    m_D}{3}\left(g^{X_2}_0(m_{12})\right)^2\frac{\left(\beta_1+\frac{2}{3m_c}\right)^2}{\left(m_{12}-m_{D^{*0}}+
  i\varepsilon\right)^2}\, \vec{p}_\gamma^{\,2} .
\end{equation}

The amplitude for $D^{+} \bar{D}^{*-}
  \gamma$ decay is readily obtained from Eq.~(\ref{eq:radia1})
making the obvious replacements: $g^{X_2}_0 \to g^{X_2}_c$, $\beta_1
\to \beta_2$ and  $(m_{D^0}, m_{D^{*0}}) \to (m_{D^+}, m_{D^{*+}})$.
Performing the phase space integration, we find at tree level (assuming the
existence of the $X_2$ state)
\begin{eqnarray}
\label{eq:restree}
    \Gamma(X_2(4013)\to D^0\bar D^{*0}\gamma)_\text{tree} &=&
18_{-6}^{+2}
\left( 16_{-9}^{+2} \right) ~\text{keV} , \\
    \Gamma(X_2(4013) \to D^+ D^{*-}\gamma)_\text{tree} &=&
0.10_{-0.05}^{+0.10}
\left( 0.09_{-0.03}^{+0.06} \right) ~\text{keV} ,\label{eq:restree-bis}
\end{eqnarray}
where the values outside and inside the parentheses are obtained with
$\Lambda=0.5$ and 1~GeV, respectively. The errors account for the
uncertainty, both in the mass of the $X_2$ state and in its couplings
$g^{X_2}_{0,c}$, derived from the $X(3872)$ input ($M_{\X}$ and the
ratio $R_{\X}$ of the decay amplitudes for the $\X\to J/\psi\rho$ and
$\X\to J/\psi\omega$ decays) and the HQSS corrections, as explained in
the the caption of Table \ref{tab:Decaywidths_hadron}. We have
neglected additional uncertainties stemming from the error on $\beta$ ($\simeq
3\%)$, because it is totally uncorrelated to those discussed above,
and it is much smaller than those affecting for instance the
$g^{X_2}_{0,c}$ couplings.  The neutral mode width is much larger than
the charged one, thanks to the bigger magnetic
$D^*D\gamma$ coupling and a larger available phase space.

In analogy with the discussion of Eqs.~(20) and (21) in
Ref.~\cite{Guo:2014hqa} for the $\X\to D^0 \bar D^0\pi^0$ decay, in
the $X_2$ radiative processes the relative distance of the $D^*\bar
D^*$ pair can be as large as allowed by the size of the $X_2(4013)$
resonance, since the final state is produced by the one body decay of
the $\bar D^*$ meson instead of by a strong two body transition.
Thus, the radiative $D\bar D^{*}\gamma$ decays might provide details
on the long-distance part of the resonance wave function. For instance, the
$d\Gamma/d|\vec{p}_{\bar D^{*0}}|$ [$d\Gamma/d|\vec{p}_{
    D^{*-}}|$] distribution is related to the $X_2(4013)$
wave-function $\Psi(\vec{p}_{\bar D^{*0}})$ [$\Psi(\vec{p}_{
    D^{*-}})$]~\cite{Guo:2014hqa}. This is in
sharp contrast to the $D\bar D$ and $D \bar D^*$ decay modes studied
in the previous sections, which turned out to be strongly sensitive to
short distance dynamics of the resonance, as revealed by the notorious dependence on the UV form factors.

 However, all these considerations are affected by the $D\bar D^*$ FSI
 effects to be discussed next.

\subsubsection {{$D\bar{D^*}$} FSI Effects}
To account for the FSI effects, we include in the analysis the $D
\bar{D}^*\to D\bar D^*$ and $D^* \bar{D} \to D\bar D^*$ $T$-matrices,
which are obtained by solving the LSE [Eq.~\eqref{eq:lse-app}] in
coupled channels with the $V_{D^{(*)}\bar D^{(*)}}$ potential given in
Eq.~\eqref{eq:potDD*}. Some isospin symmetry breaking effects are
taken into account because the physical masses of the neutral ($D\bar
D^*$) and charged ($D^+D^{*-}$) channels are used in Eq.~\eqref{eq:lse-app}.
 The $X_2\to D^0\bar D^{*0}\gamma$ decay amplitude for the
mechanisms depicted in Fig.~\ref{fig:Decay_gamma}(b) and (c) reads
\begin{eqnarray}
-i\mathcal{T(\lambda,\lambda_*,\lambda_\gamma)}^{\rm FSI\, (b+c)}_{D^{0} \bar{D}^{*0}
  \gamma}&=&g^{X_2}_0 \sqrt{4\pi \alpha}N_\gamma \left(\beta_1+\frac{2}{3m_c}\right)
\epsilon_{ijm}\epsilon^{jn}(\lambda)\epsilon^{*n}(\lambda_*)\epsilon^{*i}_\gamma(\lambda_\gamma)p^m_\gamma
\nonumber\\
&\times & 4m_Dm_{D^*}\, \widehat T_{00\to 00}(m_{23})\,
 J(m_{D^{*0}},m_{D^{*0}},m_{D^{0}},\vec{p}_\gamma)~,
 \label{eq:radia2}
\end{eqnarray}
where $m_{23}$ is the invariant mass of the final
$D^0 \bar D^{*0}$ pair,
\begin{equation}
\widehat T_{00\to 00}(m_{23}) \equiv T_{D^0\bar D^{*0}\to D^0\bar
D^{*0}}(m_{23})+ T_{D^{*0}\bar D^0\to D^0\bar D^{*0}}(m_{23}),
\end{equation}
and the three-point loop integral function
$J\left(M_{1},M_{2},M_{3},\vec{p}_\gamma\right)$ is given in the
Appendix~\ref{sec:app-rad-loop}.
The integral is convergent,
however, for consistency it is evaluated
using the same UV regularization scheme as that employed in the
$D^{(*)} \bar D^{(*)}$ EFT. In sharp contrast with the hadronic decays studied above, the momenta
 involved in these integrals are rather low.

On the other hand, we see that in the (b)+(c) contribution there
appears the combination $T_{D^0\bar D^{*0}\to D^0\bar D^{*0}}(m_{23}) + T_{D^{*0}\bar D^0\to D^0\bar D^{*0}}(m_{23})$.
 In the isospin limit, when the mass differences
 between the neutral ($D^0\bar D^{*0}$) and charged ($D^+ D^{*-}$)
 channels are neglected, we will find $\widehat T_{00\to 00}= (T^{I=0}_{C=-1}+T^{I=1}_{C=-1})/2$.
From  Eqs.~(\ref{eq:potDD*}) and (\ref{eq:lse-app}), we find
the $C$-parity odd isospin amplitudes,\footnote{Here and for simplicity
  we do not explicitly write the on-shell UV Gaussian form factors [see
  Eqs.(\ref{eq:lse-app}) and (\ref{eq:ffparaLSE})].}
\begin{equation}
  \left[T^I_{C=-1}\right]^{-1}(m_{23}) = C_{IZ}^{-1}+ G_{D\bar
    D^*}(m_{23}), \qquad I=0,1 \ , \label{eq:lsebis}
\end{equation}
with $G_{D\bar D^*}\simeq G_{D^0 \bar D^{*0}}\simeq G_{D^+ \bar
  D^{*-}}$ a common loop function. Note that the kernel of this LSE is fixed by the isoscalar ($C_{0Z}$) and isovector ($C_{1Z}$)
$C ({\rm charge~ conjugation})=-1$ terms of  $V_{D^{(*)}\bar
  D^{(*)}}$. This is a trivial consequence of the conservation of this
symmetry, taking into account that the $X_2$ and the photon are even and odd $C$-parity
states, respectively.

The (d) and (e) FSI contributions of
Fig.~\ref{fig:Decay_gamma} are similar, with obvious replacements. We find
\begin{eqnarray}
-i\mathcal{T(\lambda,\lambda_*,\lambda_\gamma)}^{\rm FSI\, (d+e)}_{D^{0} \bar{D}^{*0}
  \gamma}&=& g^{X_2}_c \sqrt{4\pi \alpha}N_\gamma \left(\beta_2+\frac{2}{3m_c}\right)
\epsilon_{ijm}\epsilon^{jn}(\lambda)\epsilon^{*n}(\lambda_*)\epsilon^{*i}_\gamma(\lambda_\gamma)p^m_\gamma
\nonumber\\
&\times & \left\{4m_Dm_{D^*} \widehat T_{+-\to 00}(m_{23})\right\}
 J(m_{D^{*+}},m_{D^{*+}},m_{D^{+}},\vec{p}_\gamma),
 \label{eq:radia3}
\end{eqnarray}
where
\begin{eqnarray}
\widehat T_{+-\to 00}(m_{23}) \equiv T_{D^+ D^{*-}\to D^0\bar D^{*0}}(m_{23})+
T_{D^{*+} D^-\to D^0\bar D^{*0}}(m_{23}),
\end{eqnarray}
and in the isospin limit, we would have  $\widehat T_{+-\to 00}=
(T^{I=0}_{C=-1}-T^{I=1}_{C=-1})/2$.

For the $X_2(4013) \to D^+ D^{*-}\gamma$ decay, the FSI
contribution is
\begin{eqnarray}
-i\mathcal{T(\lambda,\lambda_*,\lambda_\gamma)}^{\rm FSI}_{D^{+} \bar{D}^{*-}
  \gamma}&=&  \sqrt{4\pi \alpha}N_\gamma
\epsilon_{ijm}\epsilon^{jn}(\lambda)\epsilon^{*n}(\lambda_*)\epsilon^{*i}_\gamma(\lambda_\gamma)p^m_\gamma
\nonumber\\
&\times & 4m_Dm_{D^*} \left\{ g^{X_2}_c \left(\beta_2+\frac{2}{3m_c}\right)
\left[ \widehat T_{+-\to
    +-}(m_{23})J(m_{D^{*+}},m_{D^{*+}},m_{D^{+}},\vec{p}_\gamma)
  \right] \right. \nonumber \\
&+& \left. g^{X_2}_0 \left(\beta_1+\frac{2}{3m_c}\right)
\left[ \widehat T_{00\to +-}(m_{23})
J(m_{D^{*0}},m_{D^{*0}},m_{D^{0}},\vec{p}_\gamma) \right]\right\} ,
\label{eq:radia4}
\end{eqnarray}
with
\begin{eqnarray}
\widehat T_{+-\to +-}(m_{23}) &=& T_{D^+ D^{*-}\to D^+ D^{*-}}(m_{23})+ T_{D^{*+} D^-\to
  D^+ D^{*-}}(m_{23}) , \\
\widehat T_{00\to +-}(m_{23}) &=& T_{D^0\bar D^{*0}\to D^+ D^{*-}}(m_{23})+ T_{D^{*0}\bar D^0\to
  D^+ D^{*-}}(m_{23}) = \widehat T_{+-\to 00}(m_{23}) .
\end{eqnarray}
and $\widehat T_{+-\to +-}(m_{23})= \widehat T_{00\to 00}(m_{23})$ in the isospin
limit.

Taking into account isospin corrections, induced by the meson mass differences,
all the needed $T$-matrices ($\widehat T_{00\to 00}, \widehat T_{+-\to 00}$ and
$\widehat T_{+-\to +-}$) can be calculated by solving the coupled channel LSE,
Eq.~\eqref{eq:lse-app}, with the $V_{D^{(*)}\bar D^{(*)}}$ potentials of
Eq.~\eqref{eq:potDD*}, as mentioned above. Thanks to the conservation of
$C$-parity, the FSI corrections will depend only on $C_{0Z}$ and $C_{1Z}$.
Indeed, one finds
\begin{equation}\label{eq:Tmat-FSI-OddP}
\left( \begin{matrix}
\widehat T_{00\to 00} & \widehat T_{+-\to 00} \cr
\widehat T_{00\to +-} & \widehat T_{+-\to +-}\cr
\end{matrix} \right)^{-1} = \widehat{\cal F}_\Lambda^{-1}(E) \cdot
\left\{ \left( \begin{matrix}
  \frac{C_{0Z}+C_{1Z}}{2} & \frac{C_{0Z}-C_{1Z}}{2} \cr
\frac{C_{0Z}-C_{1Z}}{2} & \frac{C_{0Z}+C_{1Z}}{2}
\end{matrix} \right)^{-1} - \left( \begin{matrix} G_{D^0 \bar D^{*0}}
  & 0 \cr
0 & G_{D^+ \bar  D^{*-}}\end{matrix} \right)\right \}
\cdot \widehat {\cal F}_\Lambda^{-1}(E)~,
\end{equation}
with $\widehat{\cal F}_\Lambda(E)= {\rm Diag}\left( f^{\rm neu}_\Lambda(E),
f^{\rm ch}_\Lambda(E)\right)$, where the Gaussian form factors are
defined after Eq.~(\ref{eq:ffparaLSE}).

The $Z_b(10610)$ observed in Ref.~\cite{Belle:2011aa} carries electric charge,
and its neutral partner was also reported by the Belle
Collaboration~\cite{Adachi:2012im}. It lies within a few MeV of the $B\bar B^*$
threshold and it is tempting to speculate about it as a hadronic molecule. Belle also
reported the discovery of a second exotic electrically charged bottomonium
state~\cite{Belle:2011aa}, $Z_b(10650)$ in the vicinity of the $B^*\bar B^*$
threshold. For both the $Z_b(10610)$ and $Z_b(10650)$ states, $J^P=1^+$ are
favored from angular analyses.

Within our scheme, we assume that the  $Z_b(10610)$ resonance is  an isovector $\left(B\bar B^*+B^* \bar
B\right)/\sqrt{2}$ $s$-wave bound state with $J^{PC}=1^{+-}$ \cite{Guo:2013sya}. Note, HQSS
predicts the interaction of the $B^* \bar B^*$ system with $I = 1$, $J^{PC} =
1^{+−}$ quantum numbers to be identical to that of the $B\bar B^*$
pair in the $Z_b(10610)$ sector. Thus,
HQSS naturally explains~\cite{Bondar:2011ev} the approximate
degeneracy of the $Z_b(10610)$ and $Z_b(10650)$. Taking for the  $Z_b(10610)$
binding energy $(2.0 \pm 2.0)~\text{MeV}$~\cite{Cleven:2011gp}, we could fix
a third linear combination of the LECs that appear in the
LO Lagrangian of Eq.~(\ref{eq:LaLO})
\begin{eqnarray}
C_{1Z}\equiv C_{1A}-C_{1B}&=&-0.75^{+0.10}_{-0.17}~(-0.30^{+0.02}_{-0.04})\,
{\rm fm}^{2}
\label{eq:c1Zvalue}
\end{eqnarray}
for $\Lambda = 0.5(1.0)$ GeV. Errors have been obtained with a
procedure similar to that described in the discussion of Eq.~(\ref{eq:ces}) and used
in the case of the $X(3872)$. Assuming heavy quark flavor symmetry,
the above value of $C_{1Z}$ could be also used in the charm sector,
subject to corrections of the order of ${\cal O} (\Lambda_{\rm QCD}/m_c)\simeq
20\% $ that we take into account. Therefore, we predict the existence of
the isovector charmonium partners of the $Z_b(10610)$ and the
$Z_b(10650)$, though as virtual states in the second Riemann
sheet~\cite{Guo:2013sya}, which probably correspond to the recently observed
charged charmonium-like\footnote{The simple EFT scheme
  employed in this work does not allow for finite width
resonances, only for virtual or bound states. The merit of that EFT is
actually making a connection between the 
$Z_b(10610/10650)$ and $Z_c(3900/4025)$ resonances on the basis
of heavy flavor symmetry, not so much predicting the exact location. In that
simple theory, we treat the $Z_b(10610)$ as a stable bound state and that is
the reason why $Z_c$'s are predicted to be  virtual states. 
Had we used a more complete EFT that
takes into account the finite width of the $Z_b(10610)$ and includes
next-to-leading corrections, then we would have predicted
the $Z_c$'s as  resonances and located them 
 more accurately in the complex plane.} $Z_c(3900)$ and $Z_c(4025)$
states~\cite{Ablikim:2013mio,Liu:2013dau, Xiao:2013iha, Ablikim:2013emm,Ablikim:2013wzq}. These resonances lie close to the
$D\bar D^*$ and $D^*\bar D^*$ thresholds, respectively, while
$J^P=1^+$ quantum numbers are favored from some angular analyses.

Due to the presence of the $Z_c(3900)$  close to threshold,
one should expect the loop (FSI) mechanisms depicted in
Fig.~\ref{fig:Decay_gamma} to be important since $T^{I=1}_{C=-1}$
must have a pole. However, the value of
$C_{0Z}=(C_{0A}-C_{0B})$ is still unknown. It is not determined by the
inputs deduced from the $X(3872)$ and $Z_b(10610)$ states used in
the present analysis. Depending on the  value of $C_{0Z}$, there can be an
isoscalar $J^{PC}=1^{+-}$ $D\bar D^*$ $s$-wave bound state or not. For
instance, considering the case for $\Lambda=0.5$~GeV and taking the central
value for $C_{0Z}\sim -2.5$~fm$^2$ one finds a
bound state pole in the $D\bar D^*$ system bound by around 10 MeV; if
$C_{0Z}\sim -1.5$~fm$^2$, there will be a $D\bar D^*$ bound state
almost at threshold; if
the value of $C_{0Z}$ is larger, there will be no bound state pole any more.
Therefore, the information of $C_{0Z}$ will be crucial in understanding the
$D\bar D^*$ system and other exotic systems related to it through heavy quark
symmetries~\cite{Guo:2013sya,Guo:2013xga}. Conversely, as we will see, the $X_2$ radiative decay width could be used
to extract information on the fourth LEC, $C_{0Z}$, thanks to the FSI effects.

\begin{figure}[tb]
\begin{center}
\makebox[0pt]{\includegraphics[width=0.45\textwidth]{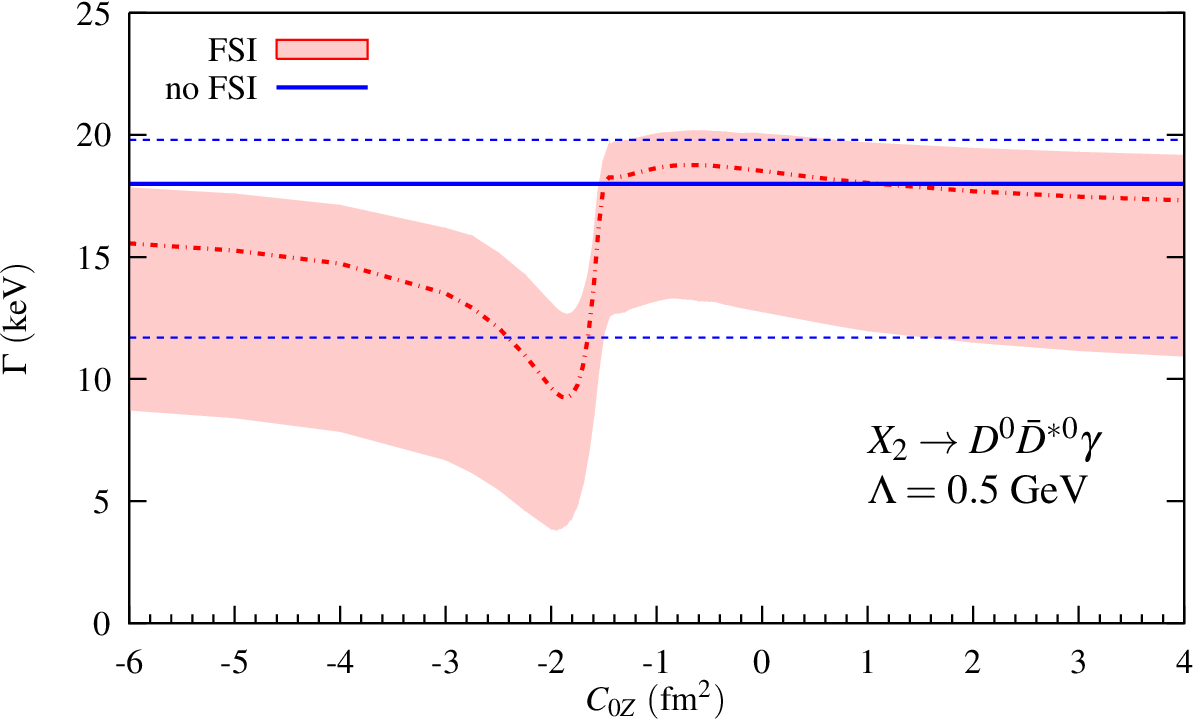}
\hspace{0.3cm}\includegraphics[width=0.45\textwidth]{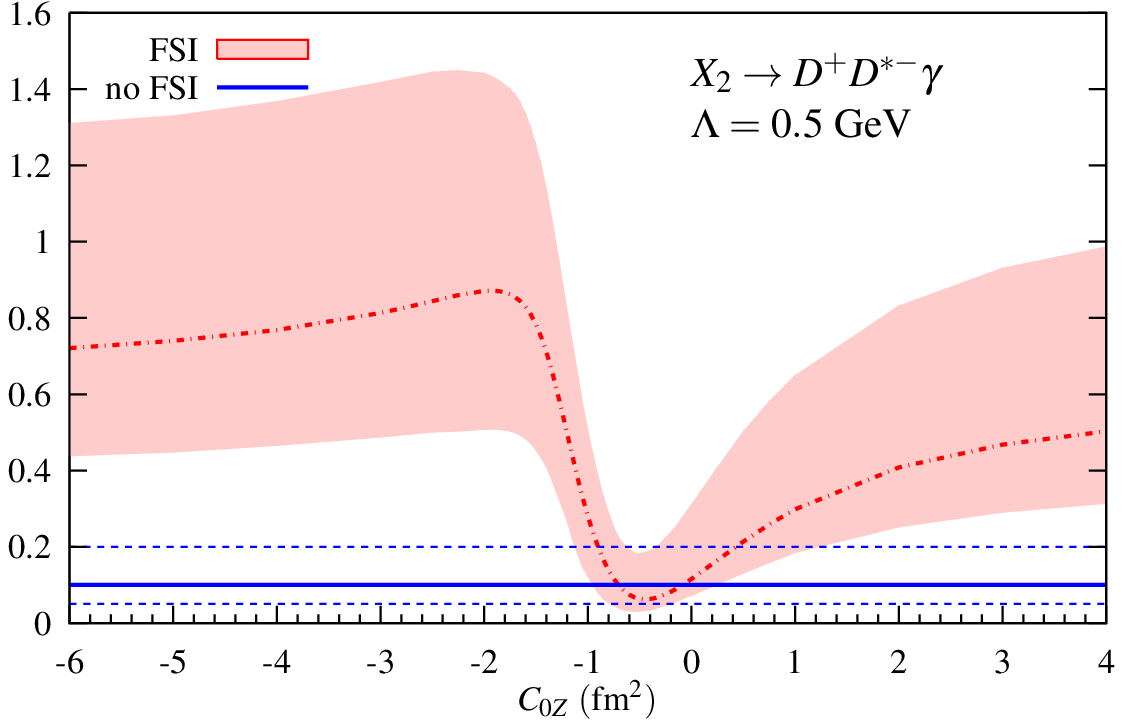}}\\
\vspace{0.3cm}
\makebox[0pt]{\includegraphics[width=0.45\textwidth]{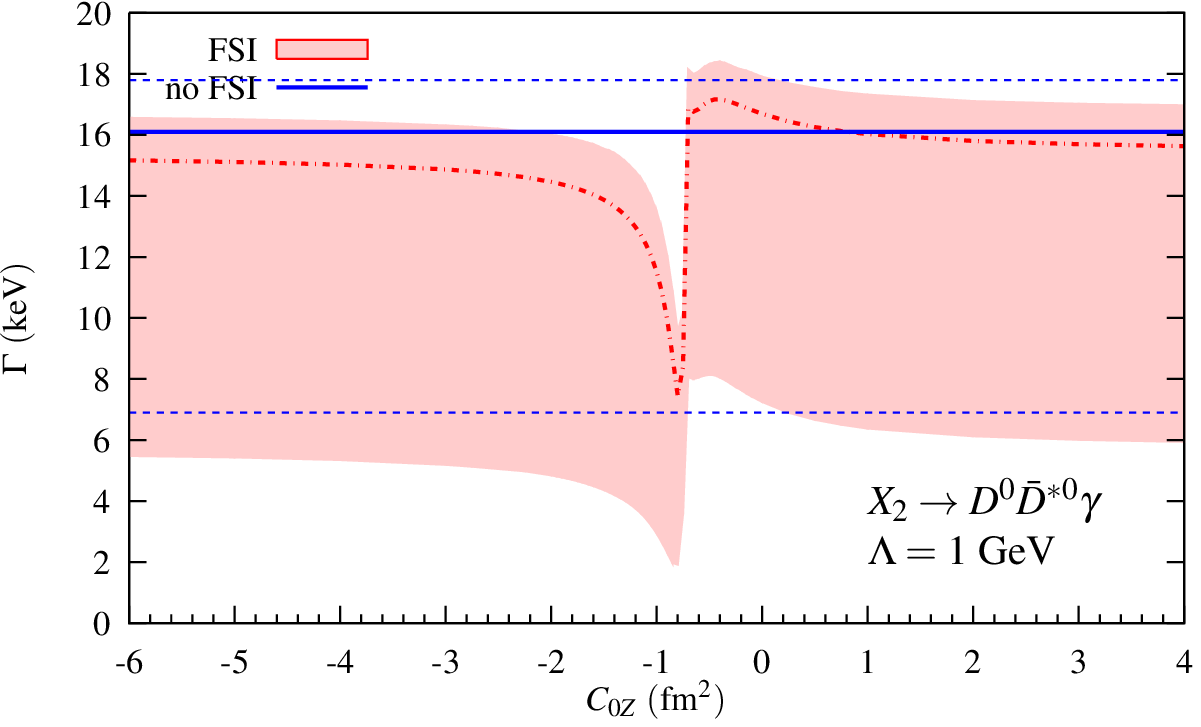}
\hspace{0.3cm}\includegraphics[width=0.45\textwidth]{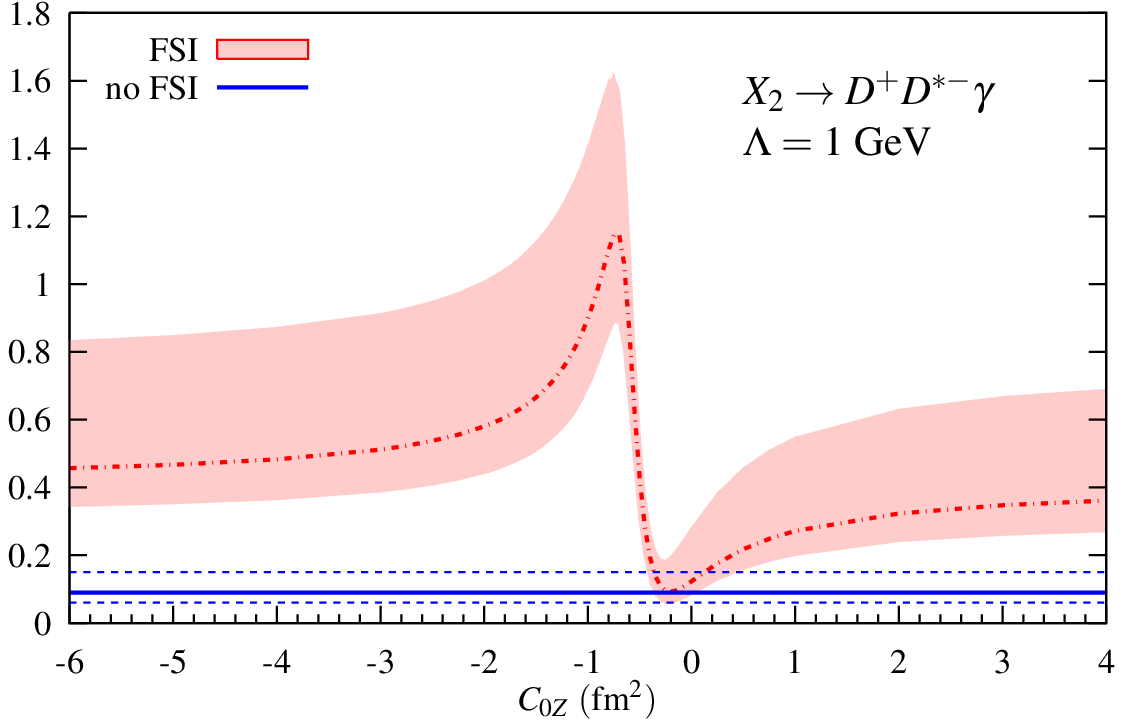}}
\end{center}
\vspace{-5mm}
\caption{Dependence of the $X_2(4013) \to D^0\bar
  D^{*0}\gamma$ and the $X_2(4013) \to D^+ D^{*-}\gamma$ partial decay
      widths on the low-energy constant $C_{0Z}$. The UV cutoff is set
      to $\Lambda=0.5$ GeV (1 GeV) in the top (bottom) panels.  The
      red error bands contain the $D\bar D^*$ FSI effects, while the three
      horizontal blue lines stand for the tree level predictions of
      Eqs.~(\ref{eq:restree}) and (\ref{eq:restree-bis}). Besides the uncertainties on the mass
      and the couplings of the $X_2$ resonance, the errors on $C_{1Z}$
      quoted in Eq.~(\ref{eq:c1Zvalue}), together with the expected
      20\% heavy quark flavor symmetry corrections when it is used in
      the charm sector, are also accounted in the 68\% CL bands
      displayed in the panels.  The red dash-dotted (full calculation)
      and solid blue (tree level) lines stand for the results obtained with the
      central values of the parameters.}\label{fig:fsi-charm}
\end{figure}

To investigate the impact of the FSI, in Fig.~\ref{fig:fsi-charm} we
show the dependence of the partial $X_2(4013) \to D \bar D^{*}\gamma$
decay widths on $C_{0Z}$. For comparison, the tree-level results are
also shown in the same plots.  As expected, for the decay into the $D^0\bar
D^{*0}\gamma$ channel, the FSI effects turn out to be important, and for some
values of $C_{0Z}$, they dominate the decay width.  The maximum effects of the FSI mechanisms approximately occur
for values of $C_{0Z}$ which give rise to an isoscalar $1^{+-}$ $D\bar
D^*$ bound or virtual state close to threshold.  One can see an
apparent deviation from the tree-level results in this region.  When
$C_{0Z}$ takes smaller values, the binding energy of the bound state
increases and moves apart from threshold; when $C_{0Z}$ takes
larger values, the pole moves deeper into a non-physical RS
and becomes a virtual state further from the threshold. In both
situations, the FSI corrections turn out to be less important. On the other
hand, the FSI corrections are always important in the $D^+ D^{*-}\gamma$
channel. This is because the tree level amplitude involves only the
$D^{*\pm}D^{\pm}\gamma$ magnetic coupling, while FSI brings in the
neutral magnetic coupling, which is much larger than the former
one. This is also the reason, besides phase space, why the tree level
width is much larger in the neutral mode than in the charged one, as
commented above.

   Notice that in the above calculations, we did not include the
   contribution from the coupled-channel FSI $D^*\bar D^*\to D\bar
   D^*$ which can come from replacing the $D^*D\gamma$ vertices in
   Fig.~\ref{fig:Decay_gamma} by the $D^* D^*\gamma $
   ones\footnote{The electric part of the $D^* D^*\gamma $ vertex does
     not contribute to the FSI $X_2$ decay width amplitude when the
     quantum numbers of the final $D^* \bar D^*$ pair are $1^{+-}$,
     with the two heavy mesons in relative $s-$wave. Thus, we are left
     with the contribution from the magnetic coupling, as in the
     $D^*D\gamma$ case. }. We have checked that this is a good
   approximation when the re-summation of the charmed meson scattering
   is switched off. This is partly because the loop integral defined
   in Eq.~\eqref{3pointEq} takes a much larger absolute value for the
   considered diagrams than for those with the $D^*D^*\gamma$ vertices
   in most regions of the phase space. The only exception is when the
   photon has a very low energy so that the $D^*\bar D^*$ are almost
   on-shell.  However, after integrating over the phase space, this
   region provides a small contribution to the partial decay width
   because of the $p_\gamma^3$ suppression due to relative $p$-wave
   between the photon and the $1^{+-}$ $D\bar D^*$ system.  The
   re-summation would introduce a further complication because of the
   presence of the $Z_c(4025)$ which couples dominantly to the
   $D^*\bar D^*$ in the $1^{+-}$ channel. It contributes mainly to the
   region close to the pole position of the $Z_c(4025)$. Again, this
   corresponds to the region with the very low-energy photon and thus
   it is suppressed due to phase space. Therefore, we expect that the
   neglected contribution discussed here has little impact, and its
   numerical effects should be safely covered by the sizable HQSS
   uncertainty exhibited in the results.

\begin{figure}[t]
\begin{center}
\includegraphics[width=0.54\textwidth]{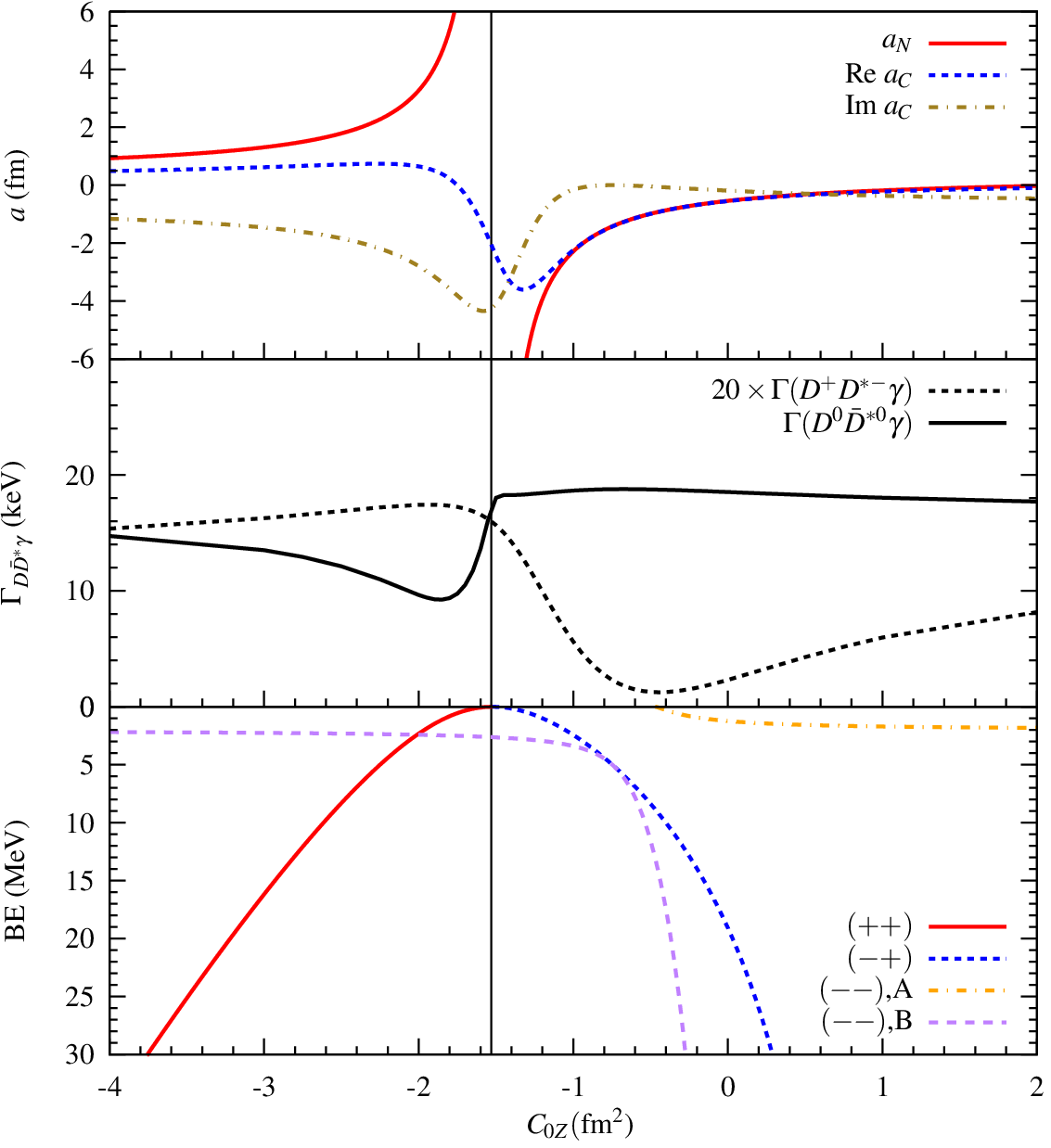}
\end{center}
\vspace{-5mm}
\caption{Dependence on $C_{0Z}$ of several physical quantities predicted in this work. In all cases an UV cutoff $\Lambda=0.5$ GeV is
  employed in the Gaussian regulator. Top: $\widehat T_{00\to
  00}$ and $\widehat T_{+-\to +-}$ scattering lengths $a_N$ (red solid curve)
  and $a_C$ (blue dashed and green dash-dotted curves),
  respectively. They are defined as $a_i=\mu_i \widehat
  T_i(E=M_{1i}+M_{2i})/2\pi$, with $\mu_i$ the corresponding reduced
  mass and $ (M_{1i},M_{2i})= (m_{D^0},m_{D^{*0}})$ and
  $(m_{D^\pm},m_{D^{*\pm}})$ for the neutral and charged channels,
  respectively. The scattering length $a_C$ is complex because the
  neutral threshold is lower than the charged one, and therefore it is
  open. Middle: Central values of the $X_2(4013) \to D^0\bar
  D^{*0}\gamma$ (solid curve) and $X_2(4013) \to D^+ D^{*-}\gamma$
  (dashed curve) partial decay
      widths including FSI effects. Bottom: Position of the poles of
 the odd $C$-parity $D \bar{D}^*\to D\bar D^*$ $T$-matrix, with respect to
 the neutral $(m_{D^0}+m_{D^{*0}})$ threshold. Poles found
in the various RS's are shown (see the text for details). The red solid
curve shows the evolution of the bound state with $C_{0Z}$, while the dashed and
the dash-dotted curves show that of the virtual ones. The vertical
black line marks the value of $C_{0Z}$ for which a $D\bar D^*$ bound state is
      generated at the $D^0\bar D^{*0}$ threshold.}\label{fig:expli}
\end{figure}

To better understand the dependence of the $X_2(4013) \to D \bar
D^{*}\gamma$ decay widths on $C_{0Z}$ in the bottom panel of
Fig.~\ref{fig:expli} we show the pole positions of
 the odd $C$-parity $D \bar{D}^*\to D\bar D^*$ $T$-matrix as functions
 of this LEC and for an UV cutoff of $0.5~\text{GeV}$.  There are two
coupled channels, the neutral one which has the lowest threshold, and
the charged channel. As a consequence, there are three relevant
RS's (among four). The first of them [$(++)$] is the physical one, while
the two non-physical RS's are reached by changing the sign of the
imaginary part of the momentum inside of the loop functions $G$ in
Eq.~\eqref{eq:Tmat-FSI-OddP} of the first channel [$(-+)$] or the momenta
of both channels [$(--)$]. Each of these non-physical RS's are
continuously connected with the physical one on the real axis above the
relevant thresholds. The solid red and dashed blue curves stand for  poles in the
$(++)$ and $(-+)$ RS's, respectively. For sufficiently large negative values of
$C_{0Z}$, there is a bound state (a pole below threshold and located
in the physical RS), which becomes less bound when $|C_{0Z}|$
decreases. For a value of this LEC around $-1.5\ \text{fm}^2$, marked
in Fig.~\ref{fig:expli} with a vertical black line, this state reaches
the threshold. When $C_{0Z}$ is further increased, the pole jumps into
the $(-+)$ RS, becoming thus a virtual state (a pole below threshold and
located in an non-physical RS) and it moves away from the threshold.  In the
medium panel we show the central values of the $X_2(4013)$ radiative
widths for both decay modes as a function of $C_{0Z}$,
which were already presented in Fig.~\ref{fig:fsi-charm}. As can be seen, it is
around this critical value $C_{0Z}=-1.5\ \text{fm}^2$, when the FSI
effects are larger for both decays, due to the vicinity of the pole to
the threshold. This happens regardless of whether it is a bound
or a virtual state, since the presence of the pole in both situations
greatly enhances the odd $C$-parity $D \bar{D}^*\to D\bar D^*$
$T$-matrix near threshold. This can be appreciated in the top panel of
Fig.~\ref{fig:expli}, where the dependence of the $\widehat T_{00\to
  00}$ and $\widehat T_{+-\to +-}$ scattering lengths on $C_{0Z}$ is shown.

Thus we have understood why in the region of values of $C_{0Z}$ around
 $-1.5\ \text{fm}^2$, FSI corrections strongly affect  the
 $D^{0}\bar D^{* 0}\gamma$ decay width: this channel has the
 lowest threshold and the bound or virtual state is located on or
 nearby it. For the charged decay mode, the width exhibits a  maximum
 for values of $C_{0Z}$ also in this region, followed by a clear minimum
 placed now in the vicinity of $C_{0Z} \sim
 C_{1Z}$. Notice that when $C_{0Z} = C_{1Z}$,  $\widehat
 T_{00\to +-}$ vanishes (see Eq.~(\ref{eq:Tmat-FSI-OddP})) and
 therefore the contribution due to the neutral mesons, driven by the
 largest magnetic coupling ($\beta_1$), in the FSI loops disappears. The
 exact position of the minimum is modulated by the further
 interference between  the tree level and the FSI charged loops, which
 are comparable.

 In the bottom panel, we also observe a virtual state pole,
the position of which is rather
insensitive\footnote{The situation is more
   complicated, as can be seen in the plot. There is a narrow region
   of values of $C_{0Z}$ around $[-0.4, -0.2]$ $\text{fm}^2$, where
   the virtual state moves quickly away from threshold, shortly
   reappearing again (orange dash-dotted line) in a position similar
   to the one that it had at the left of $C_{0Z}=-0.4$
   $\text{fm}^2$. More in detail, in a narrow region included in the
   above interval of values of $C_{0Z}$, two poles (virtual) in
   the $(--)$ RS coexist. Among these two poles, the decay width should be
   influenced only by the one closest to threshold, since it
   overshadows the second one placed deeper in the real
   axis. On the other hand, for $C_{0Z}=-0.75$~fm$^2$ the $(--)$ RS
   virtual state (magenta, B) coincides with the one located in the $(-+)$
   RS (blue). This is easy to understand since at this point
   $C_{0Z}=C_{1Z}\equiv C_Z$ and then according to
   Eq.~\eqref{eq:Tmat-FSI-OddP} the off-diagonal interaction term
   vanishes. In this situation, neutral and charged channels
   decouple, the scattering length $a_C$ is real,
   and the determinant $(1-C_Z G_{D^0 \bar D^{*0}})(1-C_Z G_{D^+ \bar
     D^{*-}})$ would vanish when either of the two factors in brackets
   is zero. It
   turns out that the factor $(1-C_Z G_{D^0 \bar D^{*0}})$ vanishes at
   the same energy both in the $(-+)$ and $(--)$ RS's, since by construction
   the $G_{D^0 \bar D^{*0}}$ loop function is identical in both
   unphysical sheets. However, the charged factor $(1-C_Z G_{D^+ \bar
     D^{*-}})$ does not lead to any further pole for $C_{0Z}=C_{1Z}=
   -0.75$~fm$^2$ and $\Lambda=0.5~\text{GeV}$. In the $\Lambda=1~\text{GeV}$ case,
   not shown in  Fig.~\ref{fig:expli}, it happens that for
   $C_{0Z}=C_{1Z}= -0.3$~fm$^2$, there exist identical poles in the
   $(-+)$ and $(--)$, and $(+-)$ and $(--)$ RS's, respectively, whose origin can be
   traced to the above discussion having in mind that now both terms
   in the decomposition of the determinant lead to poles.}
to $C_{0Z}$.  It is originated by the interaction
 in the $I=1$ sector, $C_{1Z}$, and it should be related to the
 $Z_c(3900)$ exotic charmonium-like state reported by the BESIII and
 Belle collaborations. Within our LO EFT scheme, we do not find a
 $D\bar D^*$ bound state, but instead a pole located near threshold in
 a non-physical RS~\cite{Guo:2013sya}.

\newpage

\subsection{\boldmath $X_{b2} \to \bar B  B^* \gamma $}

The expressions found in the charm sector can be readily used here,
having in mind the following correspondence: $D^0 \leftrightarrow B^-, \ D^+ \leftrightarrow \bar B^0, \
\bar D^0 \leftrightarrow \bar B^+, $ and $ D^- \leftrightarrow B^0$. Since the heavy quark flavor symmetry ensures that $g$ and $\beta_{1,2}$ are the
same in the $b$ and $c$ systems (up to corrections of the order
$\Lambda_{\rm QCD}/m_c$), the expressions of Eqs.~(\ref{eq:radiaBs1}) and
(\ref{eq:radiaBs2}) can be used to
predict the widths for the  $B^*$ radiative decays~\cite{Amundson:1992yp},
\begin{eqnarray}
\Gamma(B^{*-} \to B^- \gamma) &=& \frac{\alpha}{3}
\frac{m_{B}}{m_{B^*}}\left( \beta_1 - \frac1{3m_b}\right)^2
E_\gamma^3 =  (0.49 \pm 0.05)~\text{keV} , \\
\Gamma(\bar B^{*0} \to \bar B^0 \gamma) &=& \frac{\alpha}{3}
\frac{m_{B}}{m_{B^{*}}}\left( \beta_2 - \frac1{3m_b}\right)^2 E_\gamma^3=
(0.23 \pm 0.02)~\text{keV},
\end{eqnarray}
where we have taken the value $m_b=4.8~\text{GeV}$ for the bottom quark mass.

As in the study of its hadronic decays, we assume the $X_{b2}$
be a pure $I = 0$ state, with equal coupling to its neutral and charged
components, $g^{X_{b2}}_{\rm 0} = g^{X_{b2}}_{\rm c} = \frac{1}{\sqrt{2}}
g^{X_{b2}}$. The isospin breaking effects for the $B^*$ mesons are expected to
be small and the tiny difference between the $B^0$ and $B^{\pm}$ masses can
safely be neglected as well. In this limit we find at tree level
\begin{eqnarray}
-i\mathcal{T(\lambda,\lambda_*,\lambda_\gamma)}_{ \bar B {B}^{*}
  \gamma}&=& \frac{g^{X_{b2}}(m_{12})}{\sqrt 2} \sqrt{4\pi
  \alpha}N^b_\gamma \left(\beta_a-\frac{1}{3m_b}\right)
\frac{\epsilon_{ijm} \epsilon^{jn}(\lambda)\epsilon^{*n}(\lambda_*)
\epsilon^{*i}_\gamma(\lambda_\gamma)p^m_\gamma } {2m_{B^{*}}
\left(m_{12}-m_{B^{*}}+ i\varepsilon\right)} ,
\label{eq:radia1xb2}
\end{eqnarray}
where $\beta_a= \beta_1 (\beta_2)$ for the $B^- B^{*+}\gamma (\bar B^0 B^{*0}
\gamma)$ mode, $N^b_\gamma=\sqrt{8 M_{X_{b2}} m^2_{B^{*}}}
\sqrt{m_{B}m_{B^{*}}}$, and $m_{12}$ is the invariant mass of the final $\gamma
\bar B$ pair. These amplitudes lead to
\begin{equation}
    \Gamma(X_{b2}\to B^- B^{*+}\gamma)_\text{tree} =
13_{-10}^{+23}  ~\text{eV}, \quad
    \Gamma(X_{b2}\to \bar B^0 B^{*0}\gamma)_\text{tree} =
    6_{-5}^{+10} ~\text{eV},
    \label{eq:restree-b}
\end{equation}
where the values have been obtained
with $\Lambda=0.5$ GeV. We remind  that for $\Lambda = 1$ GeV, the
 central value of the resonance mass $M_{X_{b2}}$
 is located below the threshold $(m_{B} + m_{B^{*}}) \sim 10604~\text{MeV}$ and
 the decay is forbidden. The errors
reflect the uncertainty in the inputs from the $\X$ and the HQSS breaking
corrections, as outlined in the caption of Table
\ref{tab:Decaywidths_bottom_hadron}; they are quite large and are
dominated by those quoted for $g^{X_{b2}}$ in Eq.~(\ref{eq:g_Xb2}). The widths are of the
order of a few eV, significantly  smaller than those of the radiative
decays of the $B^{*0}$ and $B^{*-}$ meson because of the quite reduced
phase space available ($\sim 20$ MeV) for this $p$-wave decay. They are also
orders of magnitude smaller than $\Gamma(X_2\to D^0\bar D^{*0}\gamma)$ as a
result of $\Gamma(D^{*0}\to D^0\gamma)\gg \Gamma(B^*\to B\gamma)$.

\begin{figure}[tbh]
\begin{center}
\makebox[0pt]{\includegraphics[width=0.45\textwidth]{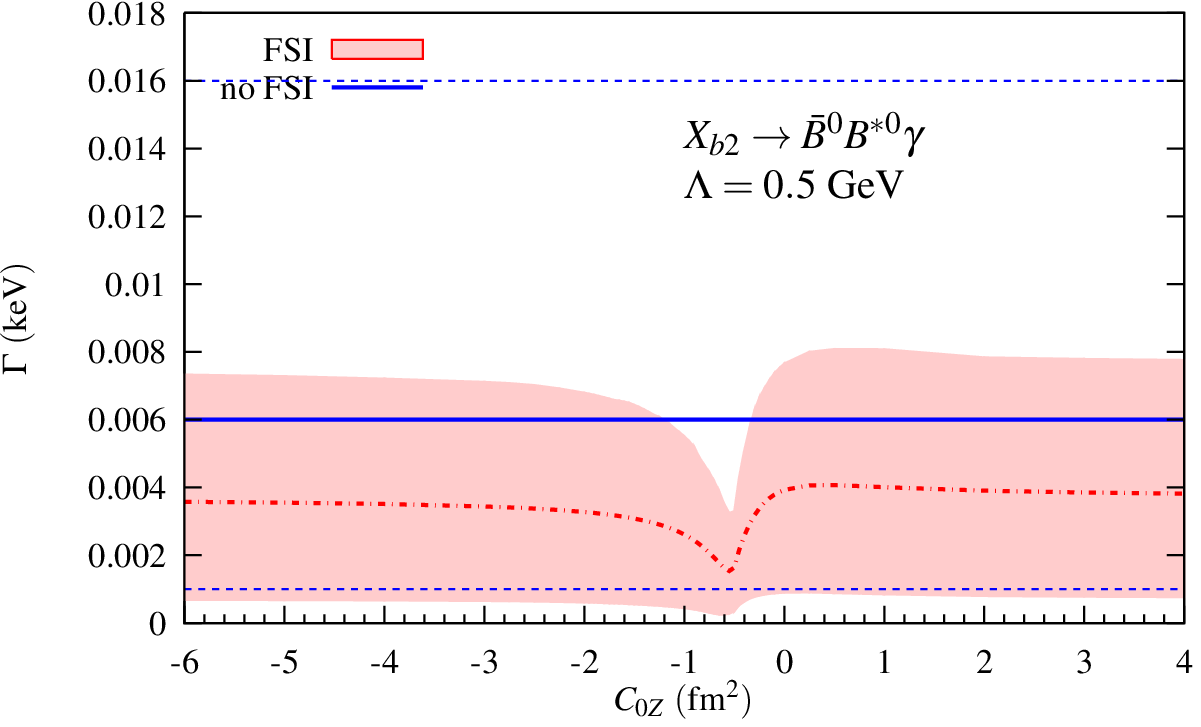}
\hspace{0.3cm}\includegraphics[width=0.45\textwidth]{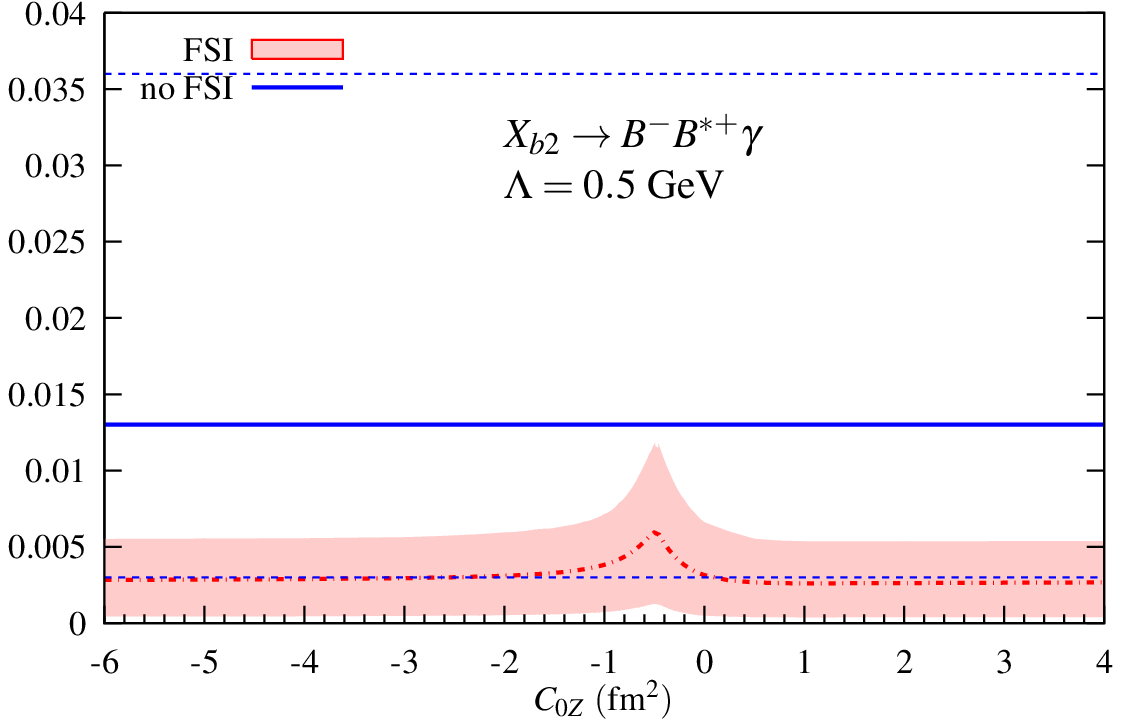}}
\end{center}
\caption{Dependence of the $X_{b2} \to B^- B^{*+} \gamma$ and  $X_{b2}
  \to \bar B^0 B^{*0} \gamma$ partial decay
      widths on the low-energy constant $C_{0Z}$. The
      red error bands contain the $\bar B B^*$ FSI effects, while the three
      horizontal blue lines stand for the tree level predictions of
      Eq.~(\ref{eq:restree-b}).}\label{fig:fsi-bottom}
\end{figure}

The amplitude for the FSI mechanisms is readily evaluated and we find
\begin{eqnarray}
-i\mathcal{T(\lambda,\lambda_*,\lambda_\gamma)}^{\rm FSI}_{\bar B B^{*}
  \gamma}&=& \sqrt{4\pi \alpha}N^b_\gamma \frac{g^{X_{b2}}}{\sqrt 2}
\epsilon_{ijm}\epsilon^{jn}(\lambda)\epsilon^{*n}(\lambda_*)\epsilon^{*i}_\gamma(\lambda_\gamma)p^m_\gamma
J(m_{B^{*}},m_{B^{*}},m_{B},\vec{p}_\gamma)\nonumber\\
&\times & 4m_B m_{B^*} \left\{ \left(\beta_1-\frac{1}{3m_b}\right)
\left[ \frac{T_{C=-1}^{I=0}(m_{23})\pm T_{C=-1}^{I=1}(m_{23})}{2}
  \right]_{\bar B B^*} \right. \nonumber \\
&+& \left.  \left(\beta_2-\frac{1}{3m_b}\right)
\left[  \frac{T_{C=-1}^{I=0}(m_{23})\mp T_{C=-1}^{I=1}(m_{23})}{2}
  \right]_{\bar B B^*}\right\},
\label{eq:radia4xb2}
\end{eqnarray}
where the $+-$ ($-+$) combination stands for the $B^- B^{*+}\gamma$
($\bar B^0 B^{*0}  \gamma$) decay mode and $m_{23}$ is now the invariant
mass of the $\bar B B^*$ pair.  The $C$-parity odd isospin
amplitudes are obtained by solving Eq.~(\ref{eq:lsebis}) using the
bottom sector loop function $G_{\bar B B^*}$.

FSI corrections turn out to be important, as can be appreciated in
Fig.~\ref{fig:fsi-bottom}. This is because  we are generating in
the $T_{C=-1}^{I=1}$ amplitude a bound state [$Z_b(10610)$], almost at
threshold (binding energy $(2.0 \pm 2.0)$~MeV~\cite{Cleven:2011gp}), that
enhances the loop mechanisms, as we discussed in the charm
sector. If we pay attention for instance
to the charged $B^- B^{*+}\gamma$ mode, we
could appreciate a distinctive feature: there appears a
 destructive interference pattern between the tree level and the FSI
amplitudes. Thanks to our MC
procedure where correlations are consistently propagated, we also observe a
reduction of the size in the uncertainties.
Besides the uncertainties on the mass and the couplings of the
$X_{b2}$ resonance, the errors on $C_{1Z}$ quoted in
Eq.~(\ref{eq:c1Zvalue}) are also accounted for in the 68\% CL bands
displayed in the panels. Actually, these latter uncertainties should
have also an important impact on the total CL bands. This is because
variations of $C_{1Z}$ allow for situations  where the pole
 is located precisely at threshold (zero binding energy) or bound by about
4 MeV.  In the first case the FSI contribution should be larger than
that obtained with the central value of $C_{1Z}$, which
correspond to a binding energy of 2 MeV. These big 68\% CL bands makes
hard to disentangle any further dependence on $C_{0Z}$, which in this
case turns out to be quite mild.

\section{Conclusions}
\label{section:Conclusions}

In this work we have studied the hadronic and radiative decays of a molecular
$P^{*}\bar{P}^{*}$ state with quantum numbers $J^{PC} = 2^{++}$ in the charm
($X_2$) and bottom ($X_{b2}$) sectors using an EFT approach. We have considered
the $X(3872)$ resonance as a $J^{PC}=1^{++}$ $D\bar D^*$
hadronic molecule. The $X_2$ and the $X_{b2}$ states will be HQSFS partners of
the  $X(3872)$ with masses and couplings to the $P^{*}\bar{P}^{*}$ heavy meson
pair determined by the properties of the $X(3872)$ resonance.

The hadronic $d$-wave $X_{2} \to D \bar{D}$ and $X_{2} \to D \bar{D}^{*}$
two-body decays are driven via one pion exchange. We observed that as a result
of the contribution from highly virtual pions, which is out of control in the
low-energy EFT, these hadronic decay widths (hence the total width of the $X_2$
as well) bear a large systematic uncertainty.
Even though the momenta involved in these decays probably lie outside the range
of applicability of EFT the calculations are still valuable as a way
to find reasonable estimates of these partial decay widths,
which we expect to almost saturate the $X_2$ decay width. To this end
and in analogy to
the Bonn potential, we have included a monopole pion-exchange form factor,
with a cutoff around 1~GeV, in each
of the $D^* D \pi$ and $D^* D^* \pi$ vertices to suppress the contribution of
large momenta.  We finally estimate the partial widths of both processes
to be of the order of a few MeV. The analysis runs in parallel in the bottom
sector with the assumption that the bare contact terms in the Lagrangian
are independent of the heavy flavor. In this sector, we also find
widths of the order of a few MeV.

We discussed the radiative $X_2\to D\bar D^{*}\gamma$ and $X_{b2} \to \bar B
B^{*}\gamma$ decays as well. The widths  are small, of the order of keV's (eV's)
in the charm (bottom) sectors. Furthermore, they are affected by the $D\bar
D^{*}$ or $B\bar B^{*}$ FSI mechanisms.
FSI effects are large because they are enhanced by the presence of the isovector
$Z_c(3900)$ and $Z_b(10610)$ resonances located near the $D^0 \bar D^{*0}$ and
$\bar B B^{*}$ thresholds, respectively.  In the charm sector, FSI corrections
turn out to be also sensitive to the negative $C$-parity isoscalar  $D \bar D^*$
interaction ($C_{0Z}$).
% We have shown that if there was a near threshold pole in this sector, the
% partial decay width can be very different from the result neglecting the FSI
% effects.
Thus,  future precise measurements of these radiative decay widths might provide
valuable information on this LEC, which cannot be in principle determined from
the properties of the $\X$, $Z_b(10610)$ and $Z_b(10650)$ resonances.
Constraints on this latter LEC are important in order to understand the dynamics
of the $P^{(*)}\bar P^{(*)}$ system.

% \begin{acknowledgments}
\section*{Acknowledgments}

We would like to thank C. Hanhart for discussions about the expected width of
the $X_2$ that prompted us to initiate this work and for valuable comments. We
are also very grateful to U. van Kolck for valuable discussions on the heavy
flavor dependence of the contact terms. We thank U.-G.~Mei\ss{}ner and
E. Oset for  careful
proof readings and comments. C.~H.-D. thanks the support of the JAE-CSIC Program
and the hospitality of the HISKP during his visit. M.~A. acknowledges financial
support from the ``Juan de la Cierva'' program (reference 27-13-463B-731) from
the Spanish Government through the Ministerio de Econom\'ia y Competitividad.
This work is supported in part by the DFG and the NSFC through funds provided to
the Sino-German CRC 110 ``Symmetries and the Emergence of Structure in QCD''
(NSFC Grant No. 11261130311), by the NSFC (Grant No.
11165005), by the Spanish Ministerio de Econom\'\i a y Competitividad and
European FEDER funds under the contract FIS2011-28853-C02-02 and the Spanish
Consolider-Ingenio 2010 Programme CPAN (CSD2007-00042), by Generalitat
Valenciana under contract  PROMETEOII/2014/0068 and by the EU HadronPhysics3
project, grant agreement no. 283286.
% \end{acknowledgments}

\medskip

\appendix

\section{\boldmath Heavy meson Lagrangians: $s$-wave interactions and pionic
and electromagnetic decays}
\label{sec:4Hpigamma}
We collect in this appendix the Lagrangians
used in this work. We use the matrix field $H^{(Q)}$ [$H^{(\bar Q)}$] to
describe the combined isospin doublet of pseudoscalar heavy-mesons
$P^{(Q)}_a=(Q\bar u,Q\bar d)$ [$P^{(\bar Q)}_a=(u \bar Q,d \bar Q )$] fields and
their vector HQSS partners $P^{*(Q)}_a$ [$P^{*(\bar Q)}_a$] (see for
example \cite{Grinstein:1992qt}),
\begin{eqnarray}
H_a^{(Q)} &=& \frac{1+\vsl}2 \left (P_{a\mu}^{* (Q)}\gamma^\mu -
P_a^{(Q)}\gamma_5 \right), \qquad v\cdot P_{a}^{* (Q)} = 0,  \nonumber \\
H^{(\bar Q)}_a &=&  \left (P_{a\mu}^{* (\bar Q)}\gamma^\mu -
P^{(\bar Q)}_a\gamma_5 \right) \frac{1-\vsl}2 , \qquad v\cdot P^{*
  (\bar Q)}_a = 0.
\end{eqnarray}
The matrix field $H^{c}$ [$H^{\bar c}$] annihilates $P$ [$\bar P$]
and $P^*$ [$\bar P^*$] mesons with a definite velocity $v$. The field
$H_a^{(Q)}$ [$H^{(\bar Q)}_a$] transforms as a $(2,\bar 2)$ [$(\bar
  2,2)$] under the heavy spin $\otimes $ SU(2)$_V$ isospin
symmetry~\cite{Grinstein:1992qt}, this is to say:
\begin{equation}
H_a^{(Q)} \to S \left( H^{(Q)} U ^\dagger\right)_a, \qquad
H^{(\bar Q) a} \to \left(U  H^{(\bar Q)}\right)^a S^\dagger.
\end{equation}
Their hermitian conjugate fields are defined by:
\begin{equation}
\bar H^{(Q)a} =\gamma^0 H_a^{(Q)\dagger} \gamma^0, \qquad
\bar H_a^{(\bar Q)} =\gamma^0 \bar H^{(\bar Q)a\dagger} \gamma^0 ,
\end{equation}
and transform as~\cite{Grinstein:1992qt}:
\begin{equation}
\bar H^{(Q)a} \to  \left( U \bar H^{(Q)} \right)^a S^\dagger , \qquad
\bar H^{(\bar Q)}_a \to S\left(\bar H^{(\bar Q)} U^\dagger \right)_a .
\end{equation}

The definition for $H_a^{(\bar Q)}$ also
specifies our convention for charge conjugation, which is $\mathcal{C}P_a^{(Q)}
\mathcal{C}^{-1} = P^{(\bar Q) a} $ and $\mathcal{C}P_{a\mu}^{*(Q)}\mathcal{C}^{-1}
= -P_\mu^{*(\bar Q) a} $, and thus it follows
\begin{equation}
\mathcal{C}H_a^{(Q)} \mathcal{C}^{-1} = c\, H^{(\bar Q)aT}\, c^{-1}, \qquad
\mathcal{C}\bar H^{(Q)a}\mathcal{C}^{-1} = c\, \bar H_a^{(\bar Q)T}\, c^{-1}
\end{equation}
with $c$ the Dirac space charge conjugation matrix satisfying $c\gamma_\mu
c^{-1}=-\gamma_\mu^T$.

\subsection{Quadruple-heavy-meson contact interaction}

\label{sec:4Hpigamma-4H}

At very low energies, the
interaction between a heavy and anti-heavy meson can be accurately
described just in terms of a contact-range potential.
The LO Lagrangian respecting HQSS reads~\cite{AlFiky:2005jd}
\begin{eqnarray}
\label{eq:LaLO}
\mathcal{L}_{4H} & = & C_{A}\,\Tr\left[\bar{H}^{(Q)a}{H}_a^{(Q)} \gamma_{\mu}
\right] \Tr\left[{H}^{(\bar{Q})a} \bar{H}^{(\bar{Q})}_a \gamma^{\mu} \right]
\nonumber\\ &+&
C_{A}^{\tau}\,\Tr\left[\bar{H}^{(Q)a} \vec\tau_a^{\,b}
{H}^{(Q)}_{b} \gamma_{\mu} \right] \Tr\left[{H}^{(\bar{Q})c}
\vec\tau_c^{\,d}\bar{H}^{(\bar{Q})}_{d} \gamma^{\mu} \right]
\nonumber\\
&+& C_{B}\,\Tr\left[\bar{H}^{(Q)a}{H}_a^{(Q)} \gamma_{\mu}\gamma_5
\right] \Tr\left[{H}^{(\bar{Q})a} \bar{H}^{(\bar{Q})}_a \gamma^{\mu}\gamma_5
\right] \nonumber\\
&+&
C_{B}^{\tau}\,\Tr\left[\bar{H}^{(Q)a} \vec\tau_a^{\,b}
{H}^{(Q)}_{b} \gamma_{\mu}\gamma_5 \right] \Tr\left[{H}^{(\bar{Q})c}
\vec\tau_c^{\,d}\bar{H}^{(\bar{Q})}_{d} \gamma^{\mu} \gamma_5\right]
\end{eqnarray}
with $\vec\tau$ the Pauli
  matrices in isospin space, and $C_{A,B}^{(\tau)}$ light flavor independent
  LECs, which are also assumed to be heavy flavor independent.  Note that in our normalization the heavy or anti-heavy meson
  fields, $H^{(Q)}$ or $H^{(\bar Q)}$, have dimensions of
  $E^{3/2}$ (see \cite{Manohar:2000dt} for details). This is because
  we use a non-relativistic normalization for the heavy mesons, which
  differs from the traditional relativistic one by a factor
  $\sqrt{M_H}$.
For later use, the four LECs that appear above  are rewritten into
$C_{0A}$, $C_{0B}$ and $C_{1A}$, $C_{1B}$ which stand for the
LECs in the isospin $I=0$ and $I=1$ channels, respectively. The
relations read
\begin{equation}
C_{0\phi} = C_{\phi} + 3 C_{\phi}^{\tau}, \qquad
C_{1\phi} = C_{\phi} - C_{\phi}^{\tau}, \qquad \text{for}~ \phi = A,B\ .
\end{equation}
The LO Lagrangian determines the contact interaction potential
$V=-\mathcal{L}/4$, which
is then used as kernel of the two body elastic LSE (see  Eq.~(\ref{eq:lse}) and
the related discussion).

The LECs that appear in the $J^{PC}=1^{++}$ and $2^{++}$
sectors [Eqs.~\eqref{eq:VLO} and \eqref{eq:PotX2}] turn out to be
$C_{0X}\equiv C_{0A}+C_{0B}$ and $C_{1X}\equiv C_{1A}+C_{1B}$. The
contact interaction in the $Z_b(10610)$ sector ($I=1$,
$J^{PC}=1^{+-}$) is $C_{1Z}\equiv C_{1A}-C_{1B}$.

On the other hand, the interaction  in the $\left\{D^0\bar D^{*0},
D^{*0}\bar D^{0}, D^+D^{*-}, D^{*+}D^{-}\right\}$ space
reads:\footnote{In the bottom sector, the corresponding basis is:
  $\left\{B^-B^{*+}, B^{*-} B^{+}, \bar B^0B^{*0},  \bar B^{*0}B^{0}\right\}$. }
\begin{eqnarray}
V_{D^{(*)}\bar D^{(*)}} &=& A^T \times {\rm Diag}(C_{0Z},
C_{0X},C_{1Z},C_{1X})\times A\nonumber \\
& = &  \frac12 \left( \begin{matrix} C_{0A}+C_{1A} & -C_{0B}-C_{1B} & C_{0A}-C_{1A}  & C_{1B}-C_{0B} \cr
                            -C_{0B}-C_{1B} & C_{0A}+C_{1A} & C_{1B}-C_{0B} & C_{0A}-C_{1A} \cr
                            C_{0A}-C_{1A} & C_{1B}-C_{0B} & C_{0A}+C_{1A} & -C_{0B}-C_{1B} \cr
                            C_{1B}-C_{0B} & C_{0A}-C_{1A} & -C_{0B}-C_{1B}  &  C_{0A}+C_{1A} \cr
\end{matrix} \right) , \label{eq:potDD*}
\end{eqnarray}
with $C_{0Z}=C_{0A}-C_{0B}$ and the orthogonal matrix $A$ given by:
\begin{equation}
A= \frac12 \left( \begin{matrix} 1 & 1 & 1  & 1 \cr
                            1 & -1 & 1 & -1 \cr
                            1 & 1 & -1 & -1 \cr
                            1 & -1 & -1  &  1 \cr
\end{matrix} \right) .
\end{equation}
Equation~(\ref{eq:potDD*}) trivially follows from the fact that the
$\mathcal{L}_{4H}$ interaction of
Eq.~(\ref{eq:LaLO}) is diagonal in the isospin basis and
the charge conjugation is well defined.\footnote{For instance in the charm
  sector, the $C$-parity states are $[D\bar D^*]_{1,2}= \frac{D\bar
D^*\pm D^*\bar D}{\sqrt2}$ ($1\leftrightarrow + $, $2\leftrightarrow -
$). In our convention, the $C$-parity of these states is independent
of the isospin and it is equal to $\mp 1$.}~The interaction given in
Eq.~(\ref{eq:potDD*}) can be used as the kernel of an  UV finite LSE
to obtain the
$T$-matrix that we use to account for the FSI in the
  radiative decays studied in Section~\ref{sec:radia},
\begin{eqnarray}
\left[T_{D^{(*)}\bar D^{(*)}}(E)\right]^{-1} =  {\cal F}_\Lambda^{-1}(E) \cdot
\left\{ \left[V_{D^{(*)}\bar
    D^{(*)}}\right]^{-1} - \widehat G(E)\right \}
\cdot {\cal F}_\Lambda^{-1}(E) ,
\label{eq:lse-app}
\end{eqnarray}
with the two particle regularized matrix propagator defined as
\begin{eqnarray}
\widehat G(E)&=&{\rm Diag}\left(G_{D^0\bar D^{*0}},
G_{D^{*0}\bar D^{0}}, G_{D^+D^{*-}}, G_{D^{*+}D^{-}} \right), \\
G_{ij}(E) &=& \int \frac{d^3\vec{q}}{(2\pi)^3}
\frac{e^{-2\vec{q}^{\,2}/\Lambda^2}}{E-\vec{q}^{\,2}/2\mu_{ij}-M_i-M_j +   i
\varepsilon}, \label{eq:gloop}
\end{eqnarray}
where trivially $G_{D^0\bar D^{*0}}=G_{D^{*0}\bar D^{0}}$ and  $G_{D^+D^{*-}}
=G_{D^{*+}D^{-}}$. In addition, the on-shell UV Gaussian form factor
matrix reads
\begin{equation}
{\cal F}_\Lambda(E) = {\rm Diag}\left( f^{\rm neu}_\Lambda(E),f^{\rm
   neu}_\Lambda(E), f^{\rm ch}_\Lambda(E), f^{\rm ch}_\Lambda(E)\right)\label{eq:ffparaLSE}
\end{equation}
with $f^{ (a)}_\Lambda(E) = \exp(-\vec{k}_a^{2}/\Lambda^2)$ and
$\vec{k}_a^2= 2\mu_a(E-M_{1a}-M_{2a})$, with  $a={\rm (neu)},\,{\rm (ch)}$.

\subsection{\boldmath $P^{(*)}\bar P^{(*)}\pi$ interactions}
\label{sec:appHHpi}

The relevant term in the LO Lagrangian of the heavy meson chiral perturbation
theory~\cite{Grinstein:1992qt,Wise:1992hn,Burdman:1992gh,Yan:1992gz}
that provides the $D^{*} D \pi$ and $D^{*} D^* \pi$ couplings is
\begin{eqnarray}
{\cal L}_{\pi HH} &=& -\frac{g}{2 f_\pi} \left ( {\rm Tr} \left [\bar H^{(Q)b}
  H^{(Q)}_a \gamma_\mu \gamma_5\right] + {\rm Tr} \left [
  H^{(\bar Q)b}\bar H^{(\bar Q)}_a \gamma^\mu \gamma_5\right] \right)
(\vec\tau \partial_\mu  \vec\phi)^{\, a}_b  + \cdots, \label{eq:LpiHH}
\end{eqnarray}
with $\vec\phi$ a relativistic field that describes the
pion,\footnote{We use a convention such that $\phi= \frac{\phi_x-{\rm
      i} \phi_y}{\sqrt 2}$ creates a $\pi^-$ from the vacuum or
  annihilates a $\pi^+$, and the $\phi_z$ field creates or annihilates
  a $\pi^0$. We adopt the usual convention $\mathcal{C}
  (\vec{\tau}\cdot \vec{\phi})\mathcal{C}^{-1}= (\vec{\tau}\cdot
  \vec{\phi})^T$.} $g$ is the heavy flavor independent
$PP^*\pi$ coupling and $f_{\pi}
= 92.2\,{\rm MeV}$ the pion decay constant. Note that in our
normalization, the pion field has a dimension of energy, while the
heavy meson or antimeson fields $H^{(Q)}$ or $H^{(\bar Q)}$ have
dimensions of $E^{3/2}$, as we already mentioned.

\subsection{\boldmath $HH\gamma$ interactions}
\label{sec:appHHgamma}

The magnetic coupling of the photon to the $s$-wave heavy mesons is
described by the Lagrangian~\cite{Amundson:1992yp,Hu:2005gf}
\begin{eqnarray}
{\cal L}_{\gamma HH} &=& \frac{e\beta}{4} Q_b^a \left ( {\rm Tr} \left [\bar H^{(Q)b}
  H^{(Q)}_a \sigma^{\mu\nu}\right] + {\rm Tr} \left [
  H^{(\bar Q)b}\bar H^{(\bar Q)}_a \sigma^{\mu\nu}\right]
\right)F_{\mu\nu}\nonumber \\
&+& \frac{eQ'}{4m_Q} \left ( {\rm Tr} \left [\bar H^{(Q)a}\sigma^{\mu\nu}
  H^{(Q)}_a \right] + {\rm Tr} \left [
  H^{(\bar Q)a}\sigma^{\mu\nu}\bar H^{(\bar Q)}_a \right]
\right)F_{\mu\nu}+ \cdots  \label{eq:Lpigamma}
\end{eqnarray}
where $F_{\mu\nu}=\partial_\mu A_\nu-\partial_\nu A_\mu$, with $A_\mu$
the photon field
($\mathcal{C} A_\mu(x)\mathcal{C}^{-1} = -A_\mu(x)$),
$Q={\rm Diag}(2/3,-1/3)$ is the light quark charge matrix, and $Q'$ is the
heavy quark electric charge (in units of the proton charge
$e=\sqrt{4\pi\alpha}$). For charm (bottom) $Q'=2/3$
($Q'=-1/3$). Besides, $m_Q$ is the heavy quark mass and $\beta$ is the
parameter introduced in Ref.~\cite{Amundson:1992yp}.  These two terms
describe the magnetic coupling due to the light (preserves HQSS) and
heavy quarks (suppressed by $1/m_Q$),
respectively. Both terms are needed to understand the observed electromagnetic
branching fractions of the $D^{*+}$ and $D^{*0}$ because a cancellation
between the two terms accounts for the very small width of the
$D^{*+}$ relative to the $D^{*0}$~\cite{Agashe:2014kda}.

In the
non-relativistic constituent quark model $\beta =1/m_q\sim 1/330$
MeV$^{-1}$, where $m_q$ is the light constituent quark mass. Heavy
meson chiral perturbation theory provides contributions from Goldstone
boson loops, which give ${\cal O}(\sqrt{m_q})$ corrections to the
decay rates~\cite{Amundson:1992yp}. If these loop corrections are evaluated in
an approximation where heavy hadron mass differences are
neglected, the correction to the above formulas can be
incorporated by making the following replacements~\cite{Amundson:1992yp}
\begin{eqnarray}
\beta Q_{11} &\to& \beta Q_{11} - \frac{g^2m_K}{8\pi f^2_K}-
\frac{g^2m_\pi}{8\pi f^2_\pi}, \\
\beta Q_{22} &\to& \beta Q_{22} + \frac{g^2m_\pi}{8\pi f^2_\pi},
\end{eqnarray}
with $f_K \sim 1.2 f_\pi$.

\section{Validity of the perturbative treatment of the \boldmath $D\bar D$ for the $X_2$}
\label{app:pc}

\begin{figure}[tb]
\centering
\includegraphics[width=0.7\linewidth]{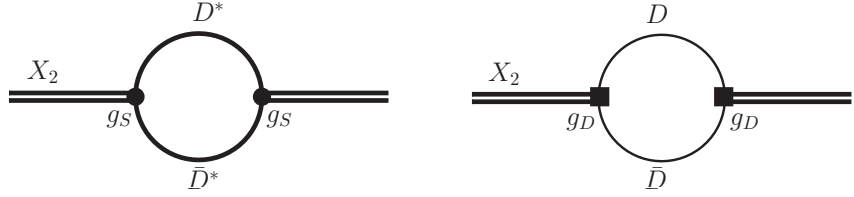}
\caption{ The $X_2$ self-energy diagrams from the $s$-wave $D^*\bar D^*$ and
$d$-wave $D\bar D$, respectively.
\label{fig:selfenergy}}
\end{figure}

In this appendix, we will argue that the $d$-wave $D\bar D$ may be treated
perturbatively in the $2^{++}$ system.
Even though this was already discussed in Ref.~\cite{Nieves:2012tt}, we have
included here a new argument grounded on a different EFT
to make a more compelling case on the smallness of
this contribution to the $X_2$ mass.
We will compare the power counting of the
self-energy diagrams of the $X_2$ from the $d$-wave $D\bar D$ and the
$s$-wave $D^*\bar D^*$ two-point loops, see Fig.~\ref{fig:selfenergy}. If the
$D\bar D$ loop is suppressed in comparison with the $D^*\bar D^*$ one, it will validate the perturbative
treatment of the $D\bar D$. Because in our case the heavy mesons are
non-relativistic, we can apply a velocity counting for the loops analogous to the
power counting of the heavy meson loops in heavy quarkonium
transitions~\cite{Guo:2009wr,Guo:2010ak}.

For the $D^*\bar D^*$ loop, the velocity counting of the self-energy reads as
\begin{equation}
   \Sigma_{D^*\bar D^*} \sim g_S^2 \frac{v^5}{(v^2)^2} = g_S^2\, v,
\end{equation}
where $g_S$ denotes the value of the $s$-wave coupling of the $X_2$ to the
$D^*\bar D^*$, $v$ denotes the velocity of the $D^*$ meson, $v^5$ is for the
loop integral measure since the non-relativistic energy is counted as
$\mathcal{O}(v^2)$, and $1/(v^2)^2$ accounts for the two non-relativistic
propagators.

Similarly, for the $D\bar D$ loop, denoting the velocity of the $D$ meson by
$w$, the velocity counting is given by
\begin{equation}
   \Sigma_{D\bar D} \sim g_D^2 \frac{w^5 w^4}{(w^2)^2} = g_D^2\, w^5,
\end{equation}
where $g_D$ is the $d$-wave coupling constant normalized to have the same
dimension as $g_S$, and the factor of $w^4$ in the denominator comes from the
two $d$-wave vertices.

Therefore, we obtain the ratio
\begin{equation}
r_{D/S} \equiv \frac{\Sigma_{D\bar D} }{\Sigma_{D^*\bar D^*} } =
\frac{g_D^2}{g_S^2} \frac{w^5}{v}.
\label{eq:rds1}
\end{equation}
\begin{figure}[tb]
\centering
\includegraphics[width=0.6\linewidth]{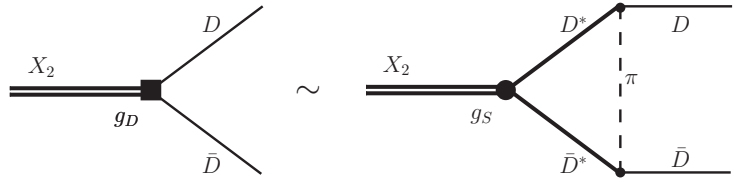}
\caption{ The contact term of the $d$-wave coupling of the $X_2$ to the $D\bar
D$ may be estimated by the one-pion exchange diagram.
\label{fig:x2dd_pc}}
\end{figure}
The question is now how $g_D$ compares with $g_S$. We can estimate $g_D$ by
considering the one-pion exchange diagram considered in this work as
illustrated in Fig.~\ref{fig:x2dd_pc}. Because the $X_2$ is very close to the
$D^*\bar D^*$ threshold, we should count each of the $D^*(\bar D^*)$ propagators
as $1/v^2$. This is equivalent to affirm that the cut due to the $D^*\bar D^*$ in the
triangle diagram in Fig.~\ref{fig:x2dd_pc} is the same as that in the $D^*\bar
D^*$ bubble diagram of the $X_2$ self-energy. Thus, we can count the $D^*$ in
both diagrams in the same way. But the pion propagator should be counted
differently. The reason is that because the $X_2$ couples to the $D\bar D$ in a
$d$-wave, the momenta in the $D^*D\pi$ vertices of the one-pion exchange diagram
should become the momenta of the $D$ and $\bar D$, $q_D= m_D w$, and the pion
momentum is of the same order as we discussed in Sect.~\ref{subsect:DDbar_decays}. This is to
say that the pion propagator should be counted as $1/w^2$ rather than $1/v^2$.
Thus, expressing the content of Fig.~\ref{fig:x2dd_pc} in terms of power
counting gives
\begin{equation}
  g_D w^2 \sim g_S \frac{v^5}{(v^2)^2 w^2} \left(\frac{g}{\Lambda_\chi}\right)^2
  (m_D w)^2 = \left[ g_S g^2 \frac{v}{w^2}
  \left(\frac{m_D}{\Lambda_\chi}\right)^2 \right] w^2,
  \label{eq:gD}
\end{equation}
where $g$ is the axial coupling constant in Eq.~\eqref{eq:LpiHH}, and
$\Lambda_\chi=4\pi f_\pi$ is the hard scale for the chiral expansion.

Numerically, for the case of the $X_2$, we have $w\simeq\sqrt{ (M_{X_2}-2
m_D)/m_D } \simeq0.38$, and $v\simeq\sqrt{(M_{X_2}-2m_{D^*})/m_{D^*}}\sim 0.06$ if we
take 7~MeV as the binding energy (recall that we have the charged $D^*\bar
D^*$ channel explicitly whose threshold is around 7~MeV above the neutral one).
With these values, we use Eqs.~\eqref{eq:rds1} and \eqref{eq:gD}, which leads
to $g_D\sim 0.6\, g_S$, to obtain an estimate of the contribution of the
$d$-wave $D\bar D$ to the $X_2$ self-energy relative to the $s$-wave $D^*\bar D^*$,
\begin{equation}
  r_{D/S} \sim 0.05.
\end{equation}
The above value suggests a high suppression of the $d$-wave $D\bar D$ in
comparison with the $s$-wave $D^*\bar D^*$.
We notice that the power counting of Ref.~\cite{Nieves:2012tt} indicates
that the size of the $D\bar{D}$ loop is ${\rm N^4LO}$
(next-to-next-to-next-to-next-to-leading order),
in line with the velocity power counting arguments.

\section{Three-point loop functions}
\label{app:three-loopfunction}

\subsection{Hadron decays}
\label{app:three-loopfunction-H}
In this section we address the tensor three-point loop function that
appears in the hadron decay amplitudes studied in the
Sect.~\ref{section:hadron-decays}. It is composed
by a pion, and two heavy meson ($P^{*}\bar P^{*})$ propagators.
The integral reads ($\vec{q}=-\vec{k}$, $q^0+k^0= M_{X_2}$)
\begin{eqnarray}
I^{ij}(M,m; M_{X_2},q^\mu \,) &=& i\int \frac{d^{4}l} {\left(2\pi\right)^{4} }
\frac{l^{i}\,l^{j}  }
 { \left[ (l+q)^{2} - M^{2}+i\varepsilon\right] \left[(k-l)^{2} -
 M^{2}+i\varepsilon\right] \left(l^{2} - m^{2}+i\varepsilon\right)
 }\nonumber \\
&\simeq & \frac{i}{4M^2}\int \frac{d^{4}l} {\left(2\pi\right)^{4} }
\frac{l^{i}\,l^{j}  }
 { \left( l^0+q^0-\omega_{h} %(\vec{l}+\vec{q}\,)
 +i\varepsilon\right)
   \left(k^0-l^0-\omega_{h} %(\vec{l}+\vec{q}\,)
   +i\varepsilon\right) \left(l^{2} - m^{2}+i\varepsilon\right) },~~~
\label{eq:b1}
\end{eqnarray}
where $M$ is the mass of the heavy particles in the loop, $m$ is the
mass for the light intermediate particle, $M_{X_2}$ is the total c.m. energy,
and $q$ and $k$ are the external 4-momenta
of the two particles in the final state of masses $m_{F_1}$ and
$m_{F_2}$, respectively. In addition,
\begin{equation}
q^0 = \frac{M_{X_2}^2+m^2_{F_1}-m^2_{F_2}}{2M_{X_2}}, \quad k^0 =
\frac{M_{X_2}^2+m^2_{F_2}-m^2_{F_1}}{2M_{X_2}},
\end{equation}
and  $\omega_{h} %(\vec{l}+\vec{q}\,)
= M+(\vec{q}+\vec{l}\,)^{\,2}/2M $ is the non-relativistic energy of the virtual
heavy mesons. Using Cauchy's theorem to integrate over the virtual pion
energy $l^0$, we obtain:\footnote{If $m_{F_1}=m_{F_2}$, $k^0 =
    q^0= M_{X_2}/2$ and the loop function now reads,
\begin{equation*}
I^{ij}(M,m; M_{X_2}, \vec{q}\,) \simeq \frac1{8M^2} \int \frac{d^3
  l}{(2\pi)^3} \frac{ l^il^j}{\omega
  (M_{X_2}/2-\omega-\omega_h)(M_{X_2}/2-\omega_h)},
\quad M_{X_2}< 2M~.
\end{equation*}}
\begin{equation}
I^{ij}(M,m; M_{X_2},q^\mu \,) \simeq \frac1{4M^2} \int \frac{d^3
  l}{(2\pi)^3} l^il^j \frac{ M_{X_2}-2\omega_h-2\omega}{2\omega
  (k^0-\omega-\omega_h)(q^0-\omega-\omega_h)(M_{X_2}-2\omega_h)}~,
\end{equation}
for $|m^2_{F_1}-m^2_{F_2} |< 2 m M_{X_2}<4Mm$ to guarantee that
the integral in Eq.~(\ref{eq:b1}) is real, and
$\omega(\vec{l}\,)= \sqrt{m^2+\vec{l}^{\,2}}$. The loop integral $I^{ij}$ presents a logarithmic UV divergence.
Indeed, $I^{ij}$ admits a tensor decomposition
\begin{equation}
I^{ij}(\vec{q}\,) =  I_0(\vec{q}^{\,2})\, q^{i}q^{j} + I_{1}(\vec{q}^{\,2})\, \delta^{ij}
\left|\vec q\,\right|^{2} . \label{eq:Iij}
\end{equation}
The $I_{1}$ term presents an
UV divergence, but it does not contribute to the amplitude because
it annihilates the traceless spin-2 polarization tensor. This
means that only the $I_{0}$ term is relevant. It can be computed as:
\begin{eqnarray}
&& I_0(M,m; M_{X_2},\vec{q}^{\,2})  \nonumber\\
&\simeq& \frac1{32M^2\pi^2
\vec{q}^{\,2}} \int_0^{+\infty} \frac{dl\,l^4}{\omega}\int_{-1}^{+1}\!\! dx
P_2(x) \frac{ M_{X_2}-2\omega_h(l,x)-2\omega}{ (k^0-\omega-\omega_h(l,x))(q^0-\omega-\omega_h(l,x))(M_{X_2}-2\omega_h(l,x))},
  ~~~~~~\label{eq:i0}
\end{eqnarray}
with $\omega_h(l,x)= M+(\vec
{l}^{\,2}+\vec{q}^{\,2}+2|\vec{l}||\vec{q}\,|x)/2M$, and  $P_2$ the
Legendre's polynomial of order 2. This term is not
UV divergent because in the limit $l\to + \infty$ all dependence on $x$, besides
$P_2(x)$, disappears and the integration over $x$ gives
zero. The convergence of the integral is greatly enhanced because
$P_2(x)$ is orthogonal to $x$ as well.
Moreover, the same type of arguments guarantees that $I_0(M,m;
M_{X_2},\vec{q}^{\,2})\sim \text{const.}$  in the $\vec{q}^{\,2}\to 0$
limit. Numerically, we use non-relativistic kinematics to compute
$|\vec{q}\,|$ in the evaluation of
$I_0(M,m; M_{X_2},\vec{q}^{\,2})$ in Eq.~(\ref{eq:i0}), {\it i.e.}, $\vec{q}^{\,2} \simeq 2\mu_{F_1F_2}(M_{X_2}-m_{F_1}-m_{F_2})$.
However, to guarantee the appropriate $d$-wave phase space, we use relativistic
kinematics to evaluate $q^i q^j$ in Eq.~(\ref{eq:Iij}) and the
$|\vec{q}\,|$ phase-space factor that appears in
Eqs.~(\ref{Eq:decaywidth2body}) and (\ref{Eq:decaywidth2body-bis}).

For consistency with the scheme adopted in Eq.~(\ref{eq:UVG}),  we
include a Gaussian regulator in the
$P^*\bar P^* X_2$ vertex by multiplying the integrand in Eq.~(\ref{eq:i0})
by  the exponential factor,
\begin{equation}
\frac{e^{-\left( \vec{q} + \vec{l}
    \,\right)^{2}/\Lambda^{2}}}{e^{-\gamma^2/\Lambda^{2}}}=\frac{e^{-\left(\vec
{l}^{\,2}+\vec{q}^{\,2}+2|\vec{l}\,||\vec{q}\,|x \right)/\Lambda^{2}}}{e^{-\gamma^2/\Lambda^{2}}} \label{eq:ffgauss}
\end{equation}
with $0>\gamma^2= M(M_{X_2}-2M)$. We divide by the factor
$e^{-\gamma^2/\Lambda^{2}}$, because it was incorporated in
the  $P^*\bar P^* X_2$ coupling.

In addition, the exchanged pion is highly virtual, and one might include a
vertex form factor of the form
\begin{equation}
  F(\vec{l}^{\,\, 2},\Lambda_\pi) = \frac{\Lambda_\pi^2}{\vec
  {l}^{\,2}+\Lambda_\pi^2}, \qquad \Lambda_\pi \sim 1~{\rm GeV},
  \label{eq:ff}
\end{equation}
in each of the two $\pi P^{(*)} P^{(*)} $ vertices.

\subsection{Radiative decays}
\label{sec:app-rad-loop}
In the computation of the FSI effects on the radiative decays of the
$X_2$ and $X_{b2}$ resonances in the Sect.~\ref{sec:radia}, the
following three-point loop function appeared
\begin{eqnarray}
\label{3pointEq}
J\left(M_{1},M_{2},M_{3},\vec{p}_\gamma\right) &=& i \int{
\frac{d^{4}q}{\left(2\pi\right)^{4}
}   \frac{1}{q^{2}-M_{1}^{2}+i\varepsilon} \frac{1}{\left(P -
q\right)^{2}-M_{2}^{2}+i\varepsilon}
\frac{1}{\left(q - p_{\gamma}\right)^{2}-M_{3}^{2}+i\varepsilon}} \\
&\simeq& \frac{i}{8M_1M_2M_3} \int
\frac{d^{4}q}{\left(2\pi\right)^{4}
}   \frac{1}{q^0-M_{1}-\vec{q}^{\,2}/2M_1+i\varepsilon}
\frac{1}{M_{X_2}-q^0-M_{2}-\vec{q}^{\,2}/2M_2+i\varepsilon}  \nonumber \\
&\times&\frac{1}{q^0 -
  E_\gamma-M_3-(\vec{q}-\vec{p}_\gamma)^2/2M_3+i\varepsilon}\nonumber
\\
&=& \frac{\mu_{12}\mu_{23}}{2M_1M_2M_3} \int
\frac{d^{3}q}{\left(2\pi\right)^{3}
}   \frac{1}{\left(\vec{q}^{\,2}+ c-i\varepsilon\right)}  \frac{1}{\left(\vec{q}^{\,2} -2\mu_{23}\,\vec{q}\cdot
  \vec{p}_\gamma/M_3 +c' -i\varepsilon\right)} ,
\end{eqnarray}
with $P^\mu=(M_{X_2},\vec{0})$ in the rest frame of the $X_2$ and
$\mu_{ij}^{-1}= (M_i^{-1}+M_j^{-1})$. In addition,
$b_{12}=M_1+M_2-M_{X_2}$, $b_{23}=M_2+M_3+E_\gamma-M_{X_2}$,
$c=2\mu_{12}b_{12}$ and $c'=
2\mu_{23}b_{23}+\mu_{23}\vec{p}_\gamma^{\,2}/M_3$.
 Since all the intermediate mesons in the present case are
non-relativistic, the three point loop has been treated
non-relativistically.  This loop integral is
convergent and its analytic expression
can be found in Eq.~(A2) of Ref.~\cite{Guo:2010ak}. However, for
consistency, despite the three-point loop function in
Eq.~(\ref{3pointEq}) being finite, it should be evaluated
using the same UV renormalization scheme as that employed in the
$D^{(*)} \bar D^{(*)}$ EFT. This is accomplished by including in the
integrand of Eq.~(\ref{3pointEq}) a Gaussian form factor,
$F_\Lambda(\vec{q}\,)$ defined as
\begin{equation}
F_\Lambda(\vec{q}\,) =
e^{-\left(\vec{q}^{\,2}-\gamma^2\right)/\Lambda^2}e^{-\left(\vec{q}^{\,2}_{\rm
  cm}-\vec{q}^{\,2}_{\rm on\,shell} \right)/\Lambda^2} .
\end{equation}
Here $\gamma^2=2\mu_{12}(M_{X_2}-M_1-M_2)$, $\vec{q}^{\,2}_{\rm
  on\,shell}= 2\mu_{23}(m_{23}-M_2-M_3)$, with  $m_{23}^2=(P-p_\gamma)^2 = M_{X_2}^2-2M_{X_2}E_\gamma$,  and
\begin{equation}
\vec{q}^{\,2}_{\rm cm} = \frac{M_2(\vec{q}-\vec{p}_\gamma)^2+
M_3\vec{q}^{\,2}}{M_2+M_3} .
\end{equation}
Note that the first exponential factor accounts for the off-shellness
in the $X_2D^{*0}\bar D^{*0}$ coupling, as in Eq.~(\ref{eq:offshell}), while the second one
accounts for the virtuality of the incoming mesons in the $D
\bar{D}^*\to D\bar D^*$ and $D^* \bar{D} \to D\bar D^*$ $T$-matrices.
Note that, after the inclusion of this factors, an analytical expression for the integral cannot be easily obtained, and it needs to be computed numerically.

\end{document}